\documentclass[]{style}
\usepackage{amsmath}  
\usepackage{amsfonts}
\usepackage{graphicx} 
\usepackage{bm}
\usepackage{ulem}
\usepackage{color}
\definecolor{lyellow}{rgb}{0.97,0.94,0.8}
\definecolor{gray}{rgb}{0.4,0.4,0.4}

\newcommand{\I}{\mathrm{i}}

\newcommand{\ffrac}[2]{{\textstyle\frac{{#1}}{{#2}}}}
\newcommand{\half}{\ffrac{1}{2}}
\renewcommand{\vec}[1]{{\boldsymbol #1}}

\begin{document}

\title{Estimating the relevance of predictions from nuclear mean-field
 models}

\author{P.-G.~Reinhard\email{paul-gerhard.reinhard@fau.de}
\\
Institut f\"ur Theoretische Physik II, Universit\"at Erlangen-N\"urnberg,
\\
Staudtstrasse 7, D-91058 Erlangen, German} 


\maketitle

\begin{abstract}
This contribution reviews the present status of the
Skyrme-Hartree-Fock (SHF) approach as one of the leading
self-consistent mean-field models in the physics of atomic nuclei.  It
starts with a brief summary of the formalism and strategy for proper
calibration of the SHF functional. The main emphasis lies on an
exploration of the reliability of predictions, particularly in the
regime of extrapolations. Various strategies are discussed to explore
the statistical and systematic errors of SHF. The strategies are
illustrated on examples from actual applications.
Variations of model and fit data are used to get an idea about
systematic errors. The statistical error is evaluated in
straightforward manner by statistical analysis based on $\chi^2$
fits. This also allows also to evaluate the correlations (covariances)
between observables which provides useful insights into the structure
of the model and of the fitting strategy.
\end{abstract}




\section{Introduction}
\label{sec:intro}

This contribution is devoted to a brief review of the
Skyrme-Hartree-Fock (SHF) approach to nuclear structure and dynamics
with emphasis on a critical analysis of its predictive power. SHF
belongs to the class of nuclear self-consistent mean-field models
which became competitive in the early 1970ies. The first nuclear
models were purely macroscopic of the type of a liquid drop model. The
observation of shell structure led to the invention of an empirical
mean-field model, the nuclear shell model, with which one could
reveal the basic mechanisms of nuclear structure from magic nuclei
over deformation to low-energy collective motion for overviews see the
much celebrated nobel lectures \cite{Boh75aR,Mot75aR,Rai75aR} and
books \cite{Boh99aB}. These findings have confirmed that nuclear
structure and dynamics can be well described in terms of a mean-field.
The mixed macroscopic-microscopic model employing empirical
shell models has been driven to a precise instrument of describing and
predicting nuclear masses, see e.g.  \cite{Mol95a}.  Parallel to these
developments, one has looked for the natural next step, namely for
self-consistent mean-field models of the type of the Hartree-Fock
approximation.  Self consistency eliminates the need for intuitive
explicit shaping of the mean-field potential and is thus applicable to
a much broader range of processes.

The SHF approach resembles Hartree-Fock but uses an effective
interaction, called the Skyrme force. In fact, it is more precise to say that
it uses an effective energy functional derived from the Skyrme
force. The Skyrme force was proposed already in \cite{Sky59a}, had a
break-through for practical applications with the first high-quality
parametrizations coming up \cite{Vau72a}, and has evolved meanwhile to
one of the most widely used standard schemes for self-consistent
nuclear modeling.  There are many more models of that sort which rely on
effective energy functionals. The two strongest competitors of SHF are
the Gogny force \cite{Gog70a,Dec80a} and the relativistic mean-field
(RMF) model \cite{Ser86aR,Rei89aR} for which meanwhile a couple of
variants exists \cite{Vre05aR}.  For a review of all three nuclear
effective energy functionals see \cite{Bender_2003}.  It is
interesting to note that these nuclear effective energy functionals
became fashionable almost at the same time as density functional
theory for electronic systems \cite{Hoh64,Koh65,Per96}. The difference
is that electronic energy functionals have been derived from first
principle and ab-initio calculations while nuclear functionals are
mostly determined by calibration with respect to empirical data.  That
is because the nuclear many-body problem is by orders of magnitude
more involved than that for electrons. In spite of considerable
progress in nuclear ab-initio calculations (for reviews see e.g.
\cite{Dic92aR,Pan97aR,Epe09aR,Mac11aR,Ham13aR}), the case is still not
sufficiently well settled to serve as basis for high-quality
functionals. The standard pathway is since the early days to motivate
the form of the functional by the basic structure of a low-momentum
expansion and to calibrate the parameters of the functional by fits to
empirical data, usually nuclear ground state properties. This raises
the problem how to get an idea about the predictive power of the
effective energy functional and to estimate the uncertainties in
extrapolations to observables outside the pool of fit data.  As we
know, it holds for every physical theory that it can never be
ultimately proven, but only be confirmed stepwise by experience until
one hits the range of validity. This holds even more true for nuclear
mean-field models in their pragmatic mix of theoretical and empirical
input.  It is the main topic of this contribution to present and
discuss the various approaches to estimating the uncertainties within
nuclear self-consistent mean-field models. Thereby we confine the
considerations to the SHF functional as a prototype example.  The
results are transferable to the competitors, the Gogny force and the
RMF.

The paper is outlined as follows:
Section \ref{sec:functional} summarizes the SHF functional as it is
used throughout the paper and sketches its formal motivation.
Section \ref{sec:obs} briefly collects the various observables
employed for calibration of the SHF functional (bulk properties of the
ground states of finite nuclei), for characterization (response
properties of homogeneous nuclear matter), and in a broad range of
further applications.
Section \ref{sec:calibration} introduces the method of least-squares
($\chi^2$) fits and the subsequent statistical analysis delivering
extrapolation uncertainties and correlations (also
coined covariances) between observables.
Section \ref{sec:estim} addresses the problem of error estimates 
from many different perspectives.
Section \ref{sec:results} exemplifies most of the methods compiled in
section \ref{sec:estim} in terms of practical applications to a
variety of different observables.

\bigskip

\section{The SHF energy-density functional}
\label{sec:functional}

\subsection{Basic constituents: Densities and currents}

Mean-field theories describe a system through a set of single particle
(s.p.) wavefunctions $\varphi_\alpha(\vec{r})$ and associated BCS
amplitudes $v_\alpha$ for occupation and the complementing 
$u_\alpha=\sqrt{1-v_\alpha^2}$ for non-occupation.  
These are summarized in the one-body density matrix
\begin{equation}
  \varrho_q(\mathbf{r},\mathbf{r}')
  =
  \sum_{\alpha\in q} w_\alpha v_{\alpha}^2
     \varphi_{\alpha}^{\mbox{}}(\mathbf{r})\varphi_{\alpha}^\dagger(\mathbf{r}')
\label{eq:onebodyd}
\end{equation}
where $q$ labels the nucleon species with $q=p$ for protons and $q=n$
for neutrons. The $w_\alpha$ is a further factor which serves to
terminate the summations, so to say a cutoff for pairing space. It will be
discussed in connection with the pairing functional in section
\ref{sec:pair}.

The Skyrme-Hartree-Fock (SHF) energy-density functional requires the
knowledge of only a few local densities and and
currents. These are,
sorted according to time parity:
\begin{equation}
  \vspace*{-1em}
  \begin{array}{rcll}
  \multicolumn{3}{l}{\mbox{time even:}}
\\[2pt]
   \rho_q\hspace*{-0.5em}&=&\hspace*{-0.5em}\displaystyle
   \mbox{tr}_\sigma\{\varrho(\mathbf{r},\mathbf{r}')\}\big|_{r=r'}
  &\hspace*{-0.5em}\equiv\mbox{density}
\\
   \tau_q\hspace*{-0.5em}&=&\hspace*{-0.5em}\displaystyle
   \mbox{tr}_\sigma\{\nabla_r\nabla_{r'}\varrho(\mathbf{r},\mathbf{r}')\}\big|_{r=r'}
  &\hspace*{-0.5em}\equiv\mbox{kinetic density}
  \\
   \vec{J}_q \hspace*{-0.5em}&=&\hspace*{-0.5em}\displaystyle
   -\I\,\mbox{tr}_\sigma\{\nabla_r\!\times\!\hat{\vec{\sigma}}
         \varrho(\mathbf{r},\mathbf{r}')\}\big|_{r=r'}
  &\hspace*{-0.5em}\equiv\mbox{spin-orbit density}
\\[14pt]
  \multicolumn{3}{l}{\mbox{time odd:}}
\\[2pt]
   \vec{\sigma}_q\hspace*{-0.5em}&=&\displaystyle\hspace*{-0.5em}
   \mbox{tr}_\sigma\{\hat{\vec{\sigma}}\varrho(\mathbf{r},\mathbf{r}')\}\big|_{r=r'}
  &\hspace*{-0.5em}\equiv\mbox{spin density}
  \\
   \vec{j}_q\hspace*{-0.5em}&=&\hspace*{-0.5em}\displaystyle
   \Im{m}\left\{\mbox{tr}_\sigma\{\nabla_r\varrho(\mathbf{r},\mathbf{r}')\}\big|_{r=r'}\right\}
  &\hspace*{-0.5em}\equiv\mbox{current}
\\
   \vec{\tau}_q\hspace*{-0.5em}&=&\hspace*{-0.5em}\displaystyle
   -\mbox{tr}_\sigma\{\hat{\vec{\sigma}}\nabla_r\nabla_{r'}
                       \varrho(\mathbf{r},\mathbf{r}')\}\big|_{r=r'}
  &\hspace*{-0.5em}\equiv\mbox{kinetic spin-dens.}
\\[14pt]
  \multicolumn{3}{l}{\mbox{time mixed:}}
\\[2pt]
   \xi_q 
   \hspace*{-0.5em}&=&\hspace*{-0.5em}
     \sum_{\alpha\in q}w_\alpha  u_{\alpha}v_{\alpha}|\varphi_{\alpha}|^2
  &\hspace*{-0.5em}\equiv\mbox{pairing density}
  \end{array}
\label{eq:rtj}
\vspace*{1em}
\end{equation}
It is advantageous to handle the densities in terms of isospin
$T\in\{0,1\}$ instead of protons $p$ and neutrons $n$. Thus we will
often consider the recoupled forms which read for the local density
\begin{equation}
  \rho_{0}
  \equiv
  \rho
  =
  \rho_p+\rho_n
  \quad,\quad
  \rho_{1}
  =
  \rho_p-\rho_n
  \quad,
\end{equation}
and similarly for the other densities and currents.  The isoscalar
density $\rho_{0}\equiv\rho$ is equivalent to the total density and
the difference $\rho_{1}$ corresponds to the isovector density.
All densities and currents in the collection (\ref{eq:rtj}) 
are real and have definite time parity, except for the pairing
density $\xi$ which is complex and has mixed time parity.

Note that the above collection of densities and currents is richer
than in electronic density-functional theory (DFT) which employs
usually only the local (spin) density $\rho_\sigma(\vec{r})$
\cite{Dre90}. This indicates that nuclear DFT is different
\cite{Mes09a}. This will also become apparent later on in connection
with the calibration of the energy functional, see section
\ref{sec:calibration} which proceeds much different than for
electronic DFT.

\subsection{The composition of the total energy}

Starting point of all self-consistent mean-field theories based on DFT
is an expression for the total energy. This reads for SHF
\begin{equation}
E_\mathrm{total}
=  \int \! d^3 r \; \mathcal{E}_\mathrm{kin}
  + \int \! d^3 r \; \mathcal{E}_\mathrm{Sk}
  + E_\mathrm{Coul}
  + E_\mathrm{pair}
  - E_\mathrm{corr}
  \quad.
\label{eq:Etot}
\end{equation}
The one-body kinetic energy is given by 
\begin{equation}
  \mathcal{E}_\mathrm{kin} 
  =
  \frac{\hbar^2}{2m_p} \tau_p+  \frac{\hbar^2}{2m_n} \tau_n
  \quad.
\label{eq:ekin}
\end{equation}
Keeping it at the exact level implies that full quantum mechanical shell
structure is maintained which means that we deal with DFT at the level of the
Kohn-Sham approach \cite{Koh65,Dre90}.

\subsubsection{The Skyrme energy functional}

Key piece is the SHF energy density:
\begin{subequations}
\label{eq:basfunct}
\begin{equation}
  \mathcal{E}_\mathrm{Sk}
  =
  \mathcal{E}_\mathrm{Sk,even}
  +
  \mathcal{E}_\mathrm{Sk,odd}
  \;,
\end{equation}
\begin{equation}
\label{eq:ESkeven}
\begin{array}{rclcl}
  \mathcal{E}_\mathrm{Sk,even}
  &=&
  \textcolor{white}{+}\colorbox{lyellow}{$C_0^\rho\,\rho_0^2$}
  &+&
  \colorbox{lyellow}{$C_1^\rho\,\rho_1^2$}
  \\
  &&
  +\colorbox{lyellow}{$C_0^{\rho,\alpha}\,\rho_0^{2+\alpha}$}
  &+&
  C_1^{\rho,\alpha}\,\rho_1^2\rho_0^\alpha
  \\
  &&
  +\colorbox{lyellow}{$C_0^{\Delta\rho}\,\rho_0\Delta\rho_0$}
  &+&
  C_1^{\Delta\rho}\,\rho_1\Delta\rho_1
  \\
  &&
  +\colorbox{lyellow}{$C_0^{\nabla J}\,\rho_0\nabla\!\cdot\!\vec{J}_0$}
  &+&
  C_1^{\nabla J}\,\rho_1\nabla\!\cdot\!\vec{J}_1
  \\[3pt]
  &&
  +C_0^{\tau}\,\rho_0\tau_0
  &+&
  C_1^{\tau}\,\rho_1\tau_1
  \\[3pt]
  &&
  +C_0^{J}\,\vec{J}_0^2
  &+&
  C_1^{J}\,\vec{J}_1^2
\end{array}
\end{equation}
\begin{equation}
\label{eq:ESkodd}
\begin{array}{rclcl}
  \mathcal{E}_\mathrm{Sk,odd}
  &=&
  \textcolor{white}{+}C_0^{\sigma}\,\vec{\sigma}_0^2
  &+&
  C_1^{\sigma}\,\vec{\sigma}_1^2
  \\
  &&
  +C_0^{\sigma,\alpha}\,\vec{\sigma}_0^2\rho_0^\alpha
  &+&
  C_1^{\sigma,\alpha}\,\vec{\sigma}_1^2\rho_0^\alpha
  \\
  &&
  +C_0^{\Delta\sigma}\,\vec{\sigma}_0\Delta\vec{\sigma}_0
  &+&
  C_1^{\Delta\sigma}\,\vec{\sigma}_1\Delta\vec{\sigma}_1
  \\
  &&
  +\colorbox{lyellow}{$\textcolor{gray}{C_0^{\nabla J}}
             \,\mathbb{\vec{\sigma}}_0\!\cdot\!\nabla\!\times\!\vec{j}_0$}
  &+&
  \textcolor{gray}{C_1^{\nabla J}}\,\vec{\sigma}_1\!\cdot\!\nabla\!\times\!\vec{j}_1
  \\[3pt]
  &&
  -\textcolor{gray}{C_0^{\tau}}\,\vec{j}_0^2
  &-&
  \textcolor{gray}{C_1^{\tau}}\,\vec{j}_1^2
  \\[3pt]
  &&
  -\half \textcolor{gray}{C_0^{J}}\vec{\sigma}_{0} \!\cdot\! \vec{\tau}_{0} 
  &-&
  \half \textcolor{gray}{C_1^{J}}\vec{\sigma}_{1} \!\cdot\! \vec{\tau}_{1} 
\end{array}
\end{equation}
\end{subequations}
The formal reasoning for this functional will be presented in section
\ref{sec:motivateSHF}. But already here, the building principle is
obvious: The energy density contains all conceivable bi-linear
couplings of densities and currents up to second order in
derivatives. Each term is time even, although it may be composed of a
product of two time-odd currents. The coupling constants are denoted
as $C_T^\mathrm{type}$ with obvious abbreviations for ``type''. Each
coupling constant depends, in principle, on the isoscalar density
$\rho_0$. In practice, one takes a minimalistic approach and augments
only the $\rho_0^2$ term by a minimalistic density dependence $\propto
C_T^{\rho,\alpha}$. This suffices to deliver a good description of
ground state properties and excitations. On the other hand, more
density dependence is hard to determine empirically from nuclear data
because finite nuclei cover a small range of densities due to nuclear
saturation.

There is a hierarchy of importance in the various terms of the
functional (\ref{eq:basfunct}). The leading terms are those
which depend on $\rho_T$ only. These already provide a good description of bulk
matter and an acceptable, although rough, zeroth order description of
nuclei, see section \ref{sec:hierarchy}.  Terms with derivatives add
details which lift the model to the level of a quantitative
description. The minimal set of terms is indicated by yellow (gray)
shading. Already these terms alone provide an excellent description of
bulk properties of finite nuclei (energy, charge radius, charge
surface thickness). 
All further terms are necessary to allow also a good
description of response properties as giant resonances,
polarizability, low-energy vibrations, or fission. A large part of the
discussions in this manuscript deals with exploring the impact of the
various terms on the modeling. 

The part $\mathcal{E}_\mathrm{Sk,odd}$ collecting all coupling with
time-odd currents is inactive in static calculations, e.g., for ground
states of even-even nuclei. They come into play with excitations and
with odd nuclei.  Note that part of the coefficients, those printed in
gray, are taken over from the time-even part. There is no freedom to
chose them differently because identity of these coefficients is
crucial to guarantee Galilean invariance of the functional
\cite{Bender_2003,Engel_1975}. There are choices for the other
coefficients. They are all connected with terms carrying spin and
become active only only for odd nuclei and spin excitations.  We will
postpone a discussion of these spin terms and their coefficients to
section \ref{sec:spin}. All excitation modes in even-even nuclei with
natural parity (giant resonances, low-lying vibrational states,
rotation) access only the terms in $\mathcal{E}_\mathrm{Sk,odd}$ which
are fixed by Galilean invariance.

The functional (\ref{eq:ESkeven}) contains two terms leading to a
spin-orbit potential. The compulsory basic term is the one
$\propto\rho\nabla\!\cdot\!\vec{J}$. Its strength $c_T^{\nabla J}$ is
an independent parameter of the SHF functional (much unlike
relativistic models where the spin-orbit strength is automatically
given by the Dirac structure of the wavefunctions
\cite{Ser86aR,Rei89aR,Rin96aR}). The other term $\propto\vec{J}^2$ is
called tensor spin-orbit term. Its parameter $C_T^J$ is linked to the
parameter $c_T^\tau$ of the kinetic term, see table
\ref{tab:restrict}. In fact, we quote here only a simplified version.
The full tensor term has true tensor structure (not reduced to a
vector $\vec{J}$) and there is a complementary tensor term.  These
tensor terms are ignored in the majority of SHF applications, i.e. one
deals with $c_T^J=0$. This is what we will assume throughout the
following. We refer to \cite{Ben07a,Ben09c} for a very detailed
survey of tensor terms and its effects in a great variety of
observables.

Before continuing with the other contributions to the total energy
(\ref{eq:Etot}), we mention in passing that the SHF functional
(\ref{eq:basfunct}) does not necessarily guarantee unconditional
stability. In particular spin and tensor terms are likely to induce
instabilities in symmetry unrestricted calculations. This important
issue has been discussed over the years in several respects. For
recent extensive studies see, e.g.,
\cite{Ben07a,Ben09c,Dav09a,Hel13a}.

\subsubsection{Coulomb energy}
\label{sec:coul}

The Coulomb energy is treated as
\begin{subequations}
\begin{eqnarray}
  E_\mathrm{Coul}
  &=&
  E_\mathrm{Coul,dir}-E_\mathrm{Coul,ex}
  \quad,
\\
  E_\mathrm{Coul,dir}
  &=&
  \frac{e^2}{2}\int d^3r\,d^3r'
   \frac{\rho_p(\vec{r})\rho_p(\vec{r}')}{|\vec{r}-\vec{r}'|}
  \quad,
\\
   E_\mathrm{Coul,ex}
   &=&
   \frac{3e^2}{4}\left(\frac{3}{\pi}\right)^{1/3}
   \int d^3r[\rho_p(\vec{r})]^{4/3}
   \quad,
\end{eqnarray}
\end{subequations}
where $e^2=1.44\,\mathrm{MeV}\,\mathrm{fm}$.  Coulomb exchange is done
at the level of the local-density approximation \cite{Dre90}.  It is
to be noted that not all mean-field models include Coulomb exchange
(for example, most RMF models do not).  in the present examples, it is
included.  The direct Coulomb energy should employ, in principle, the
charge density $\rho_C$. However, the mere proton density $\rho_p$ is
used in most applications. This makes not much of a difference and is
simpler to handle. We follow this tradition also here.

\subsubsection{Pairing energy}
\label{sec:pair}

Pairing is inevitable in case of open shell nuclei. They have a high
density of almost degenerated states. This gives the residual two-body
interaction a chance to mix these states in order to produce a unique
ground state \cite{Rei84c}. Pairing explores what is called the
particle-particle ($pp$) channel of the effective interaction which is
different from the particle-hole ($ph$) channel responsible for the
mean field \cite{Rei94aR}.  Thus it is justified and customary to use
a separate functional for pairing, namely
\begin{equation}
E_{\rm pair}
  =
  \frac{1}{4} \sum_{q\in\{p,n\}}V_\mathrm{pair,q}
  \int d^3r \xi^2_q
  \left[1 -\frac{\rho_0}{\rho_{0,\mathrm{pair}}}\right]\; ,
\label{eq:ep}
\end{equation}     
where $\rho$ and $\xi$ are the particle and pairing densities as
defined in eq. (\ref{eq:rtj}). The pairing functional (\ref{eq:ep})
contains a continuous switch, the parameter $\rho_{0,\mathrm{pair}}$,
where a pure $\delta$-interaction (DI) is recovered for
$\rho_{0,\mathrm{pair}}\longrightarrow\infty$ also called volume
pairing. The general case is the density dependent
$\delta$-interaction (DDDI).  A typical value near matter equilibrium
density $\rho_{0,\mathrm{pair}}=0.16$ fm$^{-3}$ concentrates pairing
to the surface. This is often denoted as surface pairing.  Allowing
$\rho_{0,\mathrm{pair}}$ to be a free parameter of the model yields a
value $\rho_{0,\mathrm{pair}}\approx0.2$ fm$^{-3}$ which puts the
actual functional somewhere half way between volume and surface
pairing \cite{Kluepfel_2009}.  A fully variational treatment of
pairing yields the Hartree-Fock-Bogoliubov (HFB) equations.  A widely
used and robust simplification is the BCS approximation which suffices
for all well bound nuclei \cite{Rin80aB}.  All following calculations
are done within the BCS scheme.

The space of pairing-active states has to be limited because the
zero-range nature of the pairing functional (\ref{eq:ep}) produces
poor convergence with size of phase space.  In numerical calculations
only a moderately large set of states can be included.  This is
expressed by the phase-space weight $w_\alpha$ in the definition
(\ref{eq:onebodyd}) of the on-body density. Older recipes employ a
sharp cutoff
$w_\alpha(\varepsilon_\alpha)=\theta(\epsilon_\mathrm{cut}-\varepsilon_\alpha)$.
This can be done in connection with large pairing spaces reaching up
to 50 MeV above the Fermi surface \cite{Dob84a}. Smaller spaces
require a smooth cut off for which one often uses
\begin{equation}
  w_\alpha
  =
  \left[1+
    \exp{\left((\varepsilon_\alpha-(\epsilon_F+\epsilon_{\rm cut}))
            /\Delta\epsilon\right)}
  \right]^{-1}
\label{eq:softcut}
\end{equation}
where typically $\epsilon_{\rm cut}=5\,{\rm MeV}$ and
$\Delta\epsilon=\epsilon_{\rm cut}/10$ \cite{Bon85a,Kri90a}. The value 
$\epsilon_\alpha$ is the single particle energy of the state $\alpha$ and
$\epsilon_F$ is the chemical potential. This works very well for all 
stable and moderately exotic nuclei. For better extrapolation ability 
away from the valley of stability, the fixed margin $\epsilon_{\rm cut}$ 
is modified to use a band of fixed particle number $\propto N^{2/3}$ 
instead of a fixed energy band \cite{Ben00c}.

\subsubsection{Correlation energy}
\label{sec:correlenerg}

Finally, we come to the correlation energy. It contains several
contributions
\begin{equation}
   E_\mathrm{corr}
   =
   E_\mathrm{cm}
   +
   E_\mathrm{rot}
   +
   E_\mathrm{vib}
   \quad,
\label{eq:Ecorrel}
\end{equation}
a correction $E_\mathrm{cm}$ for the spurious center-of-mass energy,
another term $E_\mathrm{rot}$ for rotational projection (non-zero only
in deformed nuclei), and a term $E_\mathrm{vib}$ accounting for soft
surface vibrations.  The leading part which is used in practically all
applications is $E_\mathrm{cm}$. There exist several variants for it
as summarized, e.g., in \cite{Bender_2003,Ben00a,Erler_2011}. We use
it here in the form
\begin{subequations}
\label{eq:cmcorr}
\begin{eqnarray}
  E_\mathrm{cm}
  &=&
  \frac{\langle\hat{\vec{P}}_\mathrm{cm}^2\rangle}{2mA}
  \quad,\quad
  \hat{\vec{P}}_\mathrm{cm}=\sum_{n=1}^A\hat{\vec{p}}_n
\end{eqnarray}
\end{subequations}
which
accounts for the effect of center-of-mass projection evaluated in
second order Gaussian overlap approximation \cite{Sch91a}. It
employs, in fact, a two-body operator which makes its application in
variational equations very cumbersome. It is thus only evaluated a
posteriori, i.e. for the given solution of the SHF mean-field
equations.

The rotational correction is, similar as $E_\mathrm{cm}$, an
approximation to rotational projection and looks 
\begin{eqnarray}
  E_\mathrm{rot}
  &=&
  \frac{\langle\hat{\vec{J}}^2\rangle}{2\Theta}
  2g(\langle\frac{\hat{\vec{J}}^2}{4}\rangle)
  \quad,
\label{eq:zperot}
\\
  g(x)
  &=&
  x\partial_x\log{\left(\int_0^1dy e^{-x(1-y^2)}\right)}
  \quad,
\nonumber
\end{eqnarray}
where $\hat{\vec{J}}$ is the operator of total angular momentum and
$\Theta$ the momentum of inertia as evaluated in the ATDHF
approximation (also coined self-consistent Inglis inertia)
\cite{Bender_2003,Kluepfel_2008}.  This rotational correction plays a
crucial role for well deformed nuclei and should be included for them.
The switch factor $g$ serves to limit the correction to deformed
nuclei and to leave (nearly) spherical nuclei untouched
\cite{Kluepfel_2008,Rei78d,Hag02a}.  Similar as the c.m. correction
$E_\mathrm{cm}^\mathrm{(full)}$, it is a two-body operator and thus
only evaluated a posteriori.

Finally, $E_\mathrm{vib}$ becomes noticeable in all nuclei with soft
surface vibrations which are typically the transitional nuclei between
spherical and well deformed ones. Its evaluation is very involved, for
details see \cite{Kluepfel_2008,Ben05a,Ben06a}. We will include the full
correlation energies only once in the introductory overview in section
\ref{sec:bulkprop}. All further examples are confined to nuclei which
are proven to have negligible vibrational-rotational corrections
and need only $E_\mathrm{cm}$.

\subsubsection{A comment on large amplitude motion}
\label{sec:largeamp}

The correlation energies discussed in the previous section
\ref{sec:correlenerg} are, in fact, covering effects beyond mean
field. They are included for symmetry reasons (translation, rotation)
and because the (heavily fluctuating) correlations from low-energy
vibrations cannot be embodied into a smooth energy functional.  All
three of these correlations are associated with low-energy,
large-amplitude collective motion. There emerges now a subtle problem
with extending the SHF functional to these cases. It is by definition
a density functional, well defined only as expectation value over one
mean-field state producing unambiguously one density matrix
(\ref{eq:onebodyd}) and local densities therefrom. But the
generator-coordinate method of large amplitude collective motion
requires energy overlaps between different mean-field states
\cite{Rei87aR}. These are a priori undefined for the SHF functional.
One can motivate a unique extension to compute those overlaps which
works well for center-of-mass projection \cite{Sch91a}.  But it was
figured out later that one runs into subtle problems with the analytical
structure of such an extended definition particularly in cases of
particle-number projection \cite{Doba07a}, but also for rotational
projection. Luckily enough, these problems are still absent when
computing energy overlaps of mean-field states which stay still close
to each other. And only these near overlaps are employed in modeling
large-amplitude collective motion within the Gaussian overlap
approximation to the generate-coordinate method \cite{Rei87aR}.  All
the correlations required here (see section \ref{sec:correlenerg}) are
evaluated in this limit and are thus at the safe side. The problem
persists when full projection is necessary. One solution is to
develop an effective energy functional which is developed
consistently from an effective interaction. Work in this direction is
in progress, see e.g. \cite{Rai14a}.

\subsection{Motivation of the SHF functional}
\label{sec:motivateSHF}

The standard pathway to derive an energy-density functional from
ab-initio calculations is for electronic systems the much celebrated
local-density approximation (LDA) \cite{Dre90} refined by non-local
corrections through the generalized gradient approximation (GGA)
\cite{Per96}. This line of development is also a strong motivation for
the nuclear case. But it cannot be simply copied for nuclear systems,
first because nuclear ab-initio calculations are not yet as reliable
as electronic ones are, and second, because nuclear energy functionals
require more than mere density dependence as can be seen from the SHF
functional (\ref{eq:basfunct}). Although a quantitative derivation
from ab-initio theories is still inhibited, at least the desirable
formal structure of a Skyrme-like effective energy functional can be
motivated by the technique of the density-matrix expansion
\cite{Neg72a,Neg75a}. We sketch here the basic steps. For simplicity,
we concentrate on the spatial part of the expansion and ignore
spin-isospin structure. We also skip explicit vector notation.

Most ab-initio models deliver at the end an effective two-body
interaction for an underlying mean-field calculation in terms of an
involved integral operator, the $T$-matrix $\hat{T}$. An example is
the Brueckner-Hartree-Fock method (BHF) whose $T$-matrix serves
finally as effective force for the Hartree-Fock part, for reviews see
e.g \cite{Dic92aR}.  An effective force as integral operator can also
be extracted from other ab-initio models \cite{Rei94aR} as, e.g., the
unitary correlator method. In any case, the most general total
interaction energy reads
\begin{equation}
  E_{\rm pot}
  = 
  \int\!dx\,dx'dy\,dy' 
  \varrho({x},{x}')\,T({x},{x}';{y},{y}')\,\varrho({y},{y}')
\label{eq:energmicro}
\end{equation}
where $\varrho({x},{x}')$ is the one-body density matrix
(\ref{eq:onebodyd}).  The key point is that $\varrho({x},{x}')$ varies
slowly with $x,x'$ within a typical range of $k_F^{-1}$ where $k_F$ is
the Fermi momentum.  The $T$-matrix, on the other hand, is well
concentrated in space, non-zero only for small differences in all
pairs of coordinates with typical ranges mostly below
$k_F^{-1}$. This suggests a Taylor expansion in all four coordinates
$x,x',y,y'$ about the common center $R=(x+x'+y+y')/4$.  This reads up to
second order
\begin{eqnarray}
  \varrho({x},{x'})
  &\hspace*{-0.5em}\approx&
  \rho(R)
  +(\bar{x}\!-\!R)\nabla\rho\big|_R
  +\half(\bar{x}\!-\!R)^2\Delta\rho\big|_R
\nonumber\\
  &&
  \hspace*{-0.7em}+ \I(x\!-\!x')j\big|_R
  + \frac{1}{2}(x\!-\!x')^2 \big(\tau\!-\!\frac{1}{4}\Delta{\rho} \big)\big|_R
\end{eqnarray}
where we abbreviate $\bar{x}=(x+x')/2$.  A similar expansion holds for
$\varrho({y},{y'})$. We insert this into the interaction energy
(\ref{eq:energmicro}) keep all terms up to second order in derivatives
and recall that the $T$-matrix conserves parity as well as
time-parity. This eliminates all terms of first order. What remains is
just the SHF functional (\ref{eq:basfunct}), of course, without the
spin and spin-orbit terms which had been ignored in this quick
``derivation''. The basic ingredient, the $T$-matrix, is assumed to
stem from homogeneous matter and thus depends on the density for which
it was evaluated. Thus all expansion coefficients $C_T^\mathrm{type}$
in the ansatz (\ref{eq:basfunct}) carry, in principle, a density
dependence. This is as far as we can get with formal reasoning.  As
argued above, we are still lacking sufficient information to fix all
possible density dependencies and associate one only to the leading
zeroth order terms $\propto\rho_0^2$ and $\propto\rho_1^2$.  This
looks at first glance like a somewhat helpless escape. But it turns
out to be a pragmatic and fruitful guess delivering a high-quality
functional for many purposes, as we will see. The reason is that the
density of nuclei gathers predominantly around the equilibrium density
$\rho_0\approx\rho_\mathrm{eq}$ due to strong nuclear saturation.

Altogether, the density-matrix expansion demonstrates how a zero-range
effective interaction emerges naturally from the initially given
involved operator structure. We are dealing with a typical low-energy
or low-$q$ expansion \cite{Dun12aB}.  It requires that the spatial
structure should be sufficiently smooth which means in the present
case that the essential physics of the $T$-matrix is concentrated at
length scales smaller than the typical wavelength $k_F^{-1}$. This is
fine at normal nuclear density.  But one should warned with
extensions to high densities. Sooner or later, effective functionals
of zero range will become inappropriate although the validity of
effective functionals exceeds often the range of such safe estimates.

\section{Observables}
\label{sec:obs}

\subsection{Homogeneous nuclear matter}
\label{sec:NM}

Although not experimentally accessible, homogeneous nuclear matter is
an extremely useful system to characterize basic nuclear properties.
Nuclear matter properties (NMP) have been used over decades as key
parameters in macroscopic models \cite{Mol95a,Myers_1977,Hasse_1988}.
Extensive studies in this domain have developed reliable, although
model dependent, values for them. It is technically simple for every
mean-field model to compute NMP for homogeneous matter.  Moreover, NMP
can be given a physical interpretation.  Thus NMP are helpful
quantities to characterize a model and to compare different models.
We introduce here those NMP which will be used in the following.

Homogeneous nuclear matter is to be taken without Coulomb force
(it can be assumed to be neutralized by a homogeneous electron
background), pairing, and c.m. correction. It remains the energy per
particle as
\begin{equation}
  \frac{E}{A}(\rho_0,\rho_1,\tau_0,\tau_1)
  = 
  \frac{\mathcal{E}_\mathrm{kin}+\mathcal{E}_\mathrm{Sk}}{\rho_0}
  \quad.
\end{equation}
where we consider for a while $\rho$ and $\tau$ as independent
variables.  We keep the independence just to allow a simple computation
of effective masses.  Of course, a given system is characterized just
by the densities $\rho_T$ and the kinetic density depends on these
given densities as $\tau_T=\tau_T(\rho_0,\rho_1)$. Thus we have to
distinguish between partial derivatives $\partial/\partial_\tau$ which
take $\tau_T$ as independent and total derivatives $d/d\rho$ which
know only the $\rho_T$ dependence. The relation is
\begin{equation}
  \frac{d}{d\rho_T}
  =
  \frac{\partial}{\partial\rho_T}
  +
  \sum_{T'}\frac{\partial\tau_{T'}}{\partial\rho_T}
  \frac{\partial}{\partial\tau_{T'}}
  \quad.
\end{equation}
The standard NMP are defined at the equilibrium point
($\rho_0=\rho_\mathrm{eq}$, $\rho_1=0$) of symmetric nuclear matter.
\begin{table}[t]
\begin{center}
\begin{tabular}{lrcl}
\hline
\multicolumn{4}{c}{\rule{0pt}{12pt}isoscalar ground state properties}
\\[4pt]
\hline
 equilibrium density:\hspace*{-0.5em}
 &\rule{0pt}{16pt}
 $\rho_\mathrm{eq}$
 &$\leftrightarrow$&
 $\displaystyle\frac{d}{d\rho_0}\frac{E}{A}\Big|_\mathrm{eq}=0$
\\[12pt]
 equilibrium energy:\hspace*{-0.5em}
 &
 $\displaystyle\frac{E}{A}\Big|_\mathrm{eq}$
\\[12pt]
\hline
\multicolumn{4}{c}{\rule{0pt}{12pt}isoscalar response properties}
\\[4pt]
\hline
 incompressibility:\hspace*{-0.5em}
 &
  $  K_\infty$
  &=& 
  $\displaystyle
  9\,\rho_0^2 \, \frac{d^2}{d\rho_0^2} \,
       \frac{{E}}{A}\Big|_\mathrm{eq}
  $
\\[12pt]
  effective mass:\hspace*{-0.5em}
  &
  $\displaystyle
  \frac{\hbar^2}{2m*}
  $
  &=&
  $\displaystyle
   \frac{\hbar^2}{2m}
    + 
    \frac{\partial}{\partial\tau_0} \frac{{E}}{A}\bigg|_\mathrm{eq}
  $
\\[14pt]
\hline
\multicolumn{4}{c}{\rule{0pt}{12pt}isovector response properties}
\\[4pt]
\hline
  symmetry energy:\hspace*{-0.5em}
  &
  $J$
  &=&
  $\displaystyle\rule{0pt}{19pt}
 \frac{1}{2} \frac{d^2}{d\rho_1^2}
  \frac{{E}}{A} \bigg|_\mathrm{eq}
  $
\\[12pt]
  slope of $J$:\hspace*{-0.5em}
  &
  $L$
  &=&
  $\displaystyle
  \frac{3}{2}\rho_0 \frac{d}{d\rho_0}\frac{d^2}{d\rho_1^2}
  \frac{{E}}{A} \bigg|_\mathrm{eq}
  $
\\[12pt]
  TRK sum-rule enh.:\hspace*{-0.5em}
  &
  $\kappa_{\rm TRK}$
  &=& 
  $\displaystyle
  \frac{2m}{\hbar^2}
  \frac{\partial}{\partial\tau_1} 
  \frac{{E}}{A}\bigg|_\mathrm{eq}
  $
\\[14pt]
\hline
\multicolumn{4}{c}{\rule{0pt}{12pt}surface properties (semi-infinite matter)}
\\[4pt]
\hline
 surface energy:\hspace*{-0.5em}
 &
 $a_\mathrm{surf}$
 &&
 see \cite{Rei06a}
\\[4pt]
 surface symm. en.:\hspace*{-0.5em}
 &
 $a_\mathrm{surf,sym}$
 &&
 see \cite{Rei06a}
\\[12pt]
\hline
\end{tabular}
\end{center}
\caption{\label{tab:nucmatdef}
Definition of the nuclear matter properties (NMP).
All derivatives are to be  taken at the equilibrium point
of  symmetric nuclear matter.
}
\end{table}
They are summarized in table \ref{tab:nucmatdef}.
Most of the NMP are obvious.  

A few remarks on two more subtle features: The slope of symmetry
energy $L$ characterizes the density dependence of the symmetry energy
which allows to estimate the symmetry energy at half density, i.e. at
surface of finite nuclei.  The enhancement factor for the
Thomas-Reiche-Kuhn (TRK) sum rule \cite{Rin80aB} is a widely used way
to characterize the isovector effective mass which is obvious from the
given expression involving derivative with respect to $\tau_1$.

The NMP in table \ref{tab:nucmatdef} can be grouped into four classes:
first, the (isoscalar) ground state properties $\rho_\mathrm{eq}$ and
${E}/{A}\Big|_\mathrm{eq}$, second, isoscalar response properties $K$
and $m/m$, and third, isovector response properties $J$, $L$,
$\kappa_\mathrm{TRK}$. The response properties determine zero sound in
matter \cite{Tho61aB} and subsequently they are closely related to
giant resonance modes in finite nuclei as we will see later.  There is
a fourth category, the surface energies which go already beyond
homogeneous matter and are explored in the surface of semi-infinite
matter. Their computation in the context of quantum mechanical
mean-field theories is involved \cite{Rei06a}. But they are an important
ingredient in macroscopic models and thus should also be checked in
mean-field theories.

It is to be noted that the nine NMP in table \ref{tab:nucmatdef} taken
together are fully equivalent to the nine model parameters in the SHF
functional (\ref{eq:basfunct}), namely $C_T^\rho$,
$C_T^{\rho,\alpha}$, $\alpha$, $C_T^\tau$, and
$C_T^{\Delta\rho}$. There is a one-to-one correspondence between the
both sets. This allows to consider the NMP also as model
parameters. This is a more intuitive way to communicate the model
parameters and it allow direct comparison with other mean-field models
which are often also fully mappable to NMP, see for example the
comparison of symmetry energies in \cite{Nazarewicz_2013}.

Pure neutron matter is also an important system as there exists an
actual realization in neutron stars. Thus we will look
occasionally at the neutron equation-of-state (EoS)
$E/A\Big|_\mathrm{neut}(\rho)$ which is key input to the computation
of neutron star properties, see figure \ref{fig:NS-uncert}.  A way to
characterize the EoS by one relevant number is to look at the slope of
the EoS at a typical density. This is
$d/d\rho{E}/A\Big|_\mathrm{neut}$ taken at $\rho=0.1$ fm$^{-3}$.
This observable shows up amongst others in
figure \ref{fig:alignmatrix} which shows that this
observable (as most of the neutron EoS) is strongly related to static
isovector response.

\subsection{Finite nuclei}
\label{sec:finnuc}

The most prominent observables described by an energy-density
functional are, of course, energy and density. Total binding energy
$E_B$ is computed in straightforward manner with
eq. (\ref{eq:basfunct}). Although the local densities (\ref{eq:rtj})
are immediately available as ingredients of the mean-field
calculations, its relation to measurement is more involved and
deserves some explanation.

From the experimental side, only the charge density is easily
accessible through elastic electron scattering which allows a more or
less model free determination of the nuclear charge formfactor
$F_\mathrm{C}(\mathbf{k})$ \cite{Fri82a}.  From the theoretical side,
the charge formfactor is computed as \cite{Fri75a}
\begin{eqnarray}
  F_{\rm C} (k)
  \hspace*{-0.4em}&\!=\!&\hspace*{-0.8em}
  \sum_{q\in\{p,n\}} [ F_q G_{E,q} + F_{ls,q} G_M ]
  \exp{\left(\ffrac{\hbar^2 k^2}
             {8\langle \vec{\hat{P}}_{\rm cm}^2 \rangle}\right)}
  \,,
\\
  F_q (\vec{k})  
  \hspace*{-0.4em}&\!=\!&\hspace*{-0.8em}
  \int \! d^3r \; \exp{\I\vec{k}\!\cdot\!\vec{r}}\rho_q(\vec{r})
  \,,
\nonumber
\end{eqnarray}
where $F_{ls,q}$ is the form factor of $\nabla \cdot \vec{J}_q$
augmented by a factor $\mu_q/4m^2$ with $\mu_q$ being the magnetic
moment of the nucleon, $G_{E,q}$ is the electric form factor and $G_M$
the magnetic form factor of the nucleons (assumed to be equal for both
species). The overall exponential factor takes into account the
center-of-mass correction for the formfactor complementing the
corresponding energy correction (\ref{eq:cmcorr}).
It employs the same variance of the total momentum
$\langle\hat{\bf P}_{\rm cm}^2 \rangle$ and its physical
interpretation is an unfolding of the spurious vibrations of the
nuclear center-of-mass in harmonic approximation \cite{Sch91a}.  The
nucleon form factors $G_{E,q}$ and $G_M$ are taken from nucleon
scattering data \cite{Sim80aE,Wal86aPC}, for details see
\cite{Bender_2003}.

Exact DFT should, in principle, provide a reliable description of
density distributions, or formfactors respectively, in all details
\cite{Dre90}. However, actual functionals employ analytically simple
forms which are smooth functions of the densities as motivated by a
local-density approximation. It has been shown that this limits the
predictive value to the regime $k<2k_{\rm F}$ in the formfactor where
$k_{\rm F}$ is the Fermi momentum \cite{Rei92c}.  Fortunately, the
most crucial bulk properties of the nuclear density profile are
determined at low $k$. 
These are \cite{Fri82a}
\begin{equation}
  \begin{array}{rcl}
  \multicolumn{3}{l}{\mbox{charge r.m.s. radius:}}
  \\[4pt]
  \hspace*{0.9em}r_{\rm C}
  &=&\displaystyle
  \frac{3}{F_{\rm C}(0)} 
  \frac{d^2}{dk^2} F_{\rm C} (k) \bigg|_{k=0}
\\[12pt]
  \multicolumn{3}{l}{\mbox{charge diffraction radius:}}
  \\[4pt]
  R_{\rm C}
  &=& \displaystyle
  \frac{4.493}{k_0^{(1)}}
  \quad,\quad
  F_{\rm C} (k_0^{(1)}) = 0
\\[12pt]
  \multicolumn{3}{l}{\mbox{charge surface thickness:}}
  \\[4pt]
  \sigma_{\rm C}
  &=&\displaystyle
  \frac{2}{k_m} \log{\left(\frac{F_{\rm box}(k_m)}{F_{\rm C}(k_m)}\right)}
\\[12pt]
  &&\displaystyle
  F_{\rm box}(k) 
  = 
  3 \, \frac{j_1(kR_{\rm C})}{kR_{\rm C}} 
  \;,\;
  k_m = \frac{5.6}{R_{\rm C}}
  \quad,
\end{array}
\label{eq:formparams}
\end{equation}
where $k_0^{(1)}$ is the first zero of the formfactor $F_{\rm C}$.
The diffraction radius $R_{\rm C}$ parametrizes the overall
diffraction pattern which resembles those of a filled sphere of radius
$R_{\rm C}$ \cite{Fri82a}. The actual nuclear formfactor decreases
faster than the box formfactor $F_{\rm box}$ due to the finite surface
thickness $\sigma_C$ of nuclei which is thus determined by comparing
the height of the first maximum of the box equivalent formfactor and
of the mean-field result $F_{\rm C}$.

Binding energy $E_B$ and the three bulk parameters
(\ref{eq:formparams}) of the charge formfactor are the key observables
in the pool of fit data, see section \ref{seq:chi2}. Besides these
bulk properties, we also include in the fit pool odd-even staggering
of binding energies to fix the pairing strength.  It reads for
neutrons
\begin{equation}
 \Delta^{(3)}_n(Z,N)=\half(E_B(Z,N\!+\!1)-2E_B(Z,N)+E_B(Z,N\!-\!1))
\label{eq:oddeven}
\end{equation} 
and similarly for protons. 
It is interesting to note that the odd-even staggering
also refers to binding energies. But taking the difference of
them filters a much different information than the one seen in
binding energies as such.
Moreover we consider for the fits a few
selected spin-orbit splittings in the spectrum of s.p. energies
$\Delta\varepsilon_{nl}=\varepsilon_{nlj-\half}-\varepsilon_{nlj+\half}$
to fix the strength of the spin-orbit term.

Binding energy $E_B$ is also considered for nuclei outside the fit
pool as predicted (or extrapolated) observables, e.g., when estimating
the properties of exotic nuclei and super-heavy elements. Very useful
are also differences of binding energies as two-nucleon separation
energies or the energy expense for $\alpha$-decay
\begin{equation}
  Q_\alpha
  =
  E_B(Z,N)-E_B(N\!-\!2,Z\!-\!2)-E_B(2,2)
  \quad.
\label{eq:Qalpha}
\end{equation}
Differences act as amplifying glass. They subtract smooth bulk
properties and can reveal other aspects of a model as mentioned
already in connection with the $\Delta^{(3)}_{p/n}(Z,N)$. We will see
that also when discussing statistical correlations between
observables in section \ref{sec:correl}.

The charge distribution is mostly sensitive to the proton
distribution.  Unfortunately, experimental information on neutron or
mass density is plagued with model dependence \cite{Bat89aER}.
Promising measurements of neutron radii with high-energy particle
scattering are coming up \cite{Hor01b,Cla03}, but have yet to become
more precise. It is worth the effort because reliable information on
neutron radii will be of invaluable help to assess isovector
properties of the nuclear functional \cite{Naz10a}. Thus we will have
a look also at the neutron radius $r_n$ in the following. We do this
in terms of the neutron skin from r.m.s. radii 
\begin{equation}
  r_\mathrm{skin}
  =
  r_n-r_p
  \quad.
\label{eq:skin}
\end{equation}
Taking the difference of two similar bulk properties acts, again, as an
amplifying glass which filters particularly the isovector properties.

SHF is also adapted to deal with nuclear excitations. The most
prominent ones are the much celebrated giant resonances. They are
usually described by the random-phase approximation (RPA), a dynamical
self-consistent mean-field theory of small-amplitude oscillations see,
e.g., \cite{Rin80aB}. We treat it here with the efficient operator
techniques of \cite{Rei92a,Rei92b}. RPA delivers the full spectral
strength distribution. For the analysis in this paper, it is
preferable to have one number to characterize an excitation. Giant
resonances in heavy nuclei are the perfect candidates for that because
their spectrum is well concentrated in one resonance. Thus we will
consider in the following as a measure for typical resonance
excitation properties the peak energies of the isoscalar giant
monopole resonance (GMR), the isovector giant dipole resonance (GDR),
and the isoscalar giant quadrupole resonance (GQR), all in
$^{208}$Pb. The GMR and GQR are very well concentrated and thus can be
safely computed by the inexpensive fluid-dynamical approximation
\cite{Rei92b}. The GDR has a somewhat broader, though still peaked,
distribution and requires full RPA for correctly placing the resonance
peak. We take here the point of view that the giant resonances are
well described by time-dependent mean-field theory at the level of
RPA. We have to mention, however, that it is still a matter of debate
whether this description suffices. There are calculations including
complex configurations (coupling to phonons) which indicate that such
many-body correlations modify the giant resonances
\cite{Lyu15b}. Fortunately, they do that mainly for for the width of
the resonance. But some effects on the peak position cannot be
excluded. Another problem is that the trend of peak positions towards
lighter nuclei is not correctly reproduced by SHF-RPA
\cite{Erler_2010}. It is not yet clear whether this points to a
weakness of modeling isovector density dependence or whether it
becomes another argument in favour of many-body correlations in the
giant resonances \cite{Lyu12a}. Thus there are still open question in
the theoretical description of giant resonances. At present, we adopt
the view that RPA provides relevant peak positions, at least for heavy
nuclei. We will thus consider the resonances in $^{208}$Pb.

There is another crucial number which can be extracted from the dipole
strength distribution, the isovector dipole polarizability $\alpha_D$.
Just recently, there came up very careful experimental evaluations
which deliver $\alpha_D$ with high precision \cite{Tam11a,Has15a}. This is thus a welcome
data point to assess static isovector response in mean-field models.
The polarizability can be evaluated as integral over the dipole
strength with inverse energy weight. This is the standard way in
experimental evaluation and an option in theoretical
calculation. Mean-field theory allows an even more robust access as
the static response to an external dipole field. To that end, one
computes the nuclear ground state with a small additional dipole field
$V_\mathrm{ext}=\lambda\hat{d}$ where $\hat{d}$ is the isovector
dipole operator. The polarizability is then deduced from the dipole
response of the system
$\alpha_D=\partial/\partial_\lambda\,\langle\hat{d}\rangle$. In our
calculation, we use this robust and inexpensive option.

We will also consider fission barriers in super-heavy elements. The
fission path is computed by quadrupole constrained SHF as a series of
mean-field states $\Phi_{\alpha_{20}}$ with systematically increasing
quadrupole momentum $\alpha_{20}$.  This yields the collective
potential for fission as
$\mathcal{V}(\alpha_{20})=E_\mathrm{SHF}(\Phi_{\alpha_{20}})$.  It has
to include the full correlation energy (\ref{eq:Ecorrel}).
The fission barrier is then
\begin{equation}
  B_f
  =
  \mathcal{V}_\mathrm{max}-\mathcal{V}_\mathrm{min}+E_{\mathrm{coll},0}
\label{eq:fissbar}  
\end{equation}
where $\mathcal{V}_\mathrm{max}$ is the potential at the peak of the
barrier, $\mathcal{V}_\mathrm{min}$ at the minimum, and
$E_{\mathrm{coll},0}$ is the energy of the collective ground state
above $\mathcal{V}_\mathrm{min}$.  The $E_{\mathrm{coll},0}$ defines
the entry point for fission as it is explored in experiment.  Note
that we thus include for this observable some correlation effects
beyond SHF which means that $B_f$ is not a pure mean-field observable.
For details of this rather involved definition and calculations see
\cite{Sch09a,Erl12a}.  An extensive discussion of fission properties
in connection with statistical analysis is found in
\cite{Bar15a}. Here we spot only one example.

We will also have a look at low-lying quadrupole vibrations (lowest
$2^+$ states). These are related to large amplitude motion in
mid-shell nuclei and thus treated very similar to fission.
One generates a collective path with systematically changed
quadrupole deformation thus mapping the collective potential
$\mathcal{V}(\alpha_{20})$ for quadrupole motion. The potential
is augmented by a collective mass to be computed by 
self-consistent cranking mass (often called ATDHF mass) and
correction for the spurious zero-point energy. This together
defines a collective Schr\"odinger equation whose solution
then yields the collective excitation energies and associated B(E2)
values. For details of this rather involved formal framework see,
e.g., \cite{Kluepfel_2008,Rei87aR}.

\section{Calibration of the model and analysis of predictive power}
\label{sec:calibration}

\subsection{Least-squares fit}
\label{seq:chi2}

We have used in section \ref{sec:motivateSHF} some formal reasoning to
motivate the form of the SHF functional (\ref{eq:basfunct}).  The
actual values of the parameters $C_T^\mathrm{(type)}$ remain open. Up
to now, there exists no quantitatively successful derivation of the
parameters from ab-initio calculations and this will remain so for a
while because the case of nuclear many-body theory is not yet
unambiguously settled. It is standard practice since decades to
calibrate the SHF functional (and competing mean-field models) to
empirical data. Various strategies have been used for that in the
past. The most systematic approach is probably a least-squares
($\chi^2$) fit \cite{Bev69aB} which was used in an SHF context first
in \cite{Fri86a} and has meanwhile become the standard method for
calibration of self-consistent mean-field models. There remains a
great variety in the bias set by the choice of empirical data for a
fit. To name a few examples out of many: some concentrate on spherical
nuclei with negligible correlation effects \cite{Kluepfel_2009}, others are
particularly concerned with deformed nuclei
\cite{Nik04a,Nik08a,Kor12a}, still others try to adjust also spectra of
s.p. energies \cite{Bro98a,Kor14a}. The properties of the resulting
functionals have much in common, although some differences may be
found in details. This indicates some robustness of the modeling.
We will discuss in section \ref{sec:groups} the impact of variations
of fit data. The standard pool of fit data for this publication will
be explained at the end of this section. Before that, we briefly
summarize the $\chi^2$ technique.

Center piece of $\chi^2$ fits is  a global quality measure by
summing the squared deviations from the data as
\begin{eqnarray}
  \chi^2
  &=&
  \sum_\mathrm{obs}
  \chi_\mathrm{obs}^2
  \;,
\nonumber\\
  \chi_\mathrm{obs}^2
  &=&
  \sum_\mathrm{nucl}
  \frac{\mathcal{O}^\mathrm{(th)}_\mathrm{obs,nucl}
        -
        \mathcal{O}^\mathrm{(exp)}_\mathrm{obs,nucl}}
       {\Delta\mathcal{O}_\mathrm{obs,nucl}}
  \;,
\label{eq:chi2}
\end{eqnarray}
where $\mathcal{O}$ stands for an observable, ``obs'' for a type of
observables ($E$, $r_\mathrm{rms}$, ...), ``nucl'' for a nucleus
(defined by $Z$,$N$), the upper index ``th'' for a calculated value,
and ``exp'' for the experimental value. The denominator
$\Delta\mathcal{O}_\mathrm{obs,nucl}$ quantifies the adopted error of
that observable. It renders each contribution dimensionless and
regulates the relative weights of the various terms.  It is to be
noted that $\Delta\mathcal{O}_\mathrm{obs,nucl}$ cannot be identified
with the experimental error on the given observable which is usually
much smaller.  The limiting factor are limitations at the side of the
model, i.e., the quality we can expect from a mean-field description.
A self-regulating choice is to tune all
$\Delta\mathcal{O}_\mathrm{obs,nucl}$ in one group ``obs'' the same
way such that $\chi_\mathrm{obs}^2$ yields in the average a
contribution of about one for each nucleus \cite{Bev69aB,Bra97aB}, for
details in the nuclear context see \cite{Dob14a}.

The total quality measure is a function of all model parameters, i.e.
$\chi^2=\chi^2(\vec{p})$ where $\vec{p}=(p_1,...,p_F)$. The optimal
parameters $\vec{p}_0$ are those which minimize $\chi^2$, i.e.
\begin{equation}
  \vec{p}_0\,:\quad
  \chi^2(\vec{p}_0)=\chi^2_0=\mbox{min}
  \,.
\end{equation}
The minimium condition looks straightforward. But finding the absolute
minimum is a non-trivial task because the $\chi^2$ landscape has
several minima and often discontinuities in between \cite{Ben07a}. We
employ the minimization technique from Bevington \cite{Bev69aB} which
works well in the vicinity of minima and combine that with Monte-Carlo
steps to explore the rough $\chi^2$ landscape in a broader range. For
large scale searches, there are more elaborate methods around,
e.g. more robust iteration to a minimum \cite{Kortelainen_2010} or
genetic algorithms for particularly obstinate cases \cite{Bue04b}.

Finally, we comment briefly on the choice for the pool of fit data
used later on. Some more details are provided in appendix
\ref{sec:pool}. We are using exactly the same data as in the
survey of \cite{Kluepfel_2009}. In that paper, the experimental values
and the adopted errors are explained in great detail. Thus we need not
to repeat it here. The set of fit data includes the bulk
properties $E_B$, $r_\mathrm{rms}$, $R_\mathrm{diffr}$, and $\sigma$
plus pairing gaps and some spin-orbit splittings (see section
\ref{sec:finnuc}). The nuclei are restricted to be spherical and
carefully selected to have small ground-state correlations
\cite{Kluepfel_2008}. This guarantees that we deal with nuclei which
can reliably well be described by a mean-field model. Another group
which has small correlation effects would be strongly deformed
nuclei. We omit them for practical reasons because they are costly to
compute. All transitional nuclei require corrections from vibrational
correlations and must be discarded from any fit pool.

\subsection{Statistical error analysis}
\label{sec:errors}

Not only the optimal parametrization $\vec{p}_0$, but also the
parameters in the vicinity of $\vec{p}_0$ deliver a reasonable
reproduction of data.  This is systematically quantified in
statistical analysis \cite{Bev69aB,Bra97aB}.  Although we will see
later that the residual errors
$\mathcal{O}^\mathrm{(th)}_\mathrm{obs,nucl}-\mathcal{O}^\mathrm{(exp)}_\mathrm{obs,nucl}$
(also called simply ``residuals'') in $\chi^2$ fits of nuclear
mean-field models are not really statistically distributed, we will
employ the well developed schemes of statistical analysis for
estimating extrapolation errors and correlations between
observables. It remains in any case a powerful tool to explore the
$\chi^2$ landscape and the information it contains about the interplay
of model and fit data.

Assuming a statistical distribution of residuals, one postulates a
probability distribution of reasonable model parameters as
\cite{Bra97aB,(Tar05)}
\begin{equation}
  W(\vec{p})
  \propto
  \exp(-\chi^2(\vec{p}))
  \quad.
\label{eq:probab}
\end{equation}
Their domain is characterized by
$
  \chi^2(\vec{p})
  \leq
  \chi^2_0+1
$
(see Sec.~9.8 of Ref.~\cite{Bra97aB}). The range of $\vec{p}$
fulfilling this condition is usually small and we can
perform a Taylor expansion
\begin{eqnarray}\label{chi2a}
  \chi^2(\vec{p})\hspace*{-0.5em}
  &\approx&
  \hspace*{-0.5em}\chi^2_\mathrm{0}
  \!+
  \!\!\!\sum_{\alpha,\beta=1}^{N_p} (p_\alpha\!-\!p_{0,\alpha})(\mathcal{C}^{-1}_{\mbox{}})_{\alpha\beta}(p_\beta\!-\!p_{0,\beta}),
\\
  (\mathcal{C}^{-1}_{\mbox{}})_{\alpha\beta}\hspace*{-0.5em}
  &=&
  \hspace*{-0.5em}{\textstyle\frac{1}{2}}\partial_{p_\alpha}\partial_{p_\beta}\chi^2\Big|_{\,\vec{p}_0}
  {\simeq}
  \sum_i{J}_{i\alpha}{J}_{i\beta}
  \;,
\\
  {J}_{i\alpha}\hspace*{-0.5em}
  &=&
  \hspace*{-0.5em}\frac{\partial_{p_\alpha}\mathcal{O}_i\Big|_{\,\vec{p}_0}}{\Delta\mathcal{O}_i}
  \;,
\end{eqnarray}
where $\hat{J}$ is the rescaled Jacobian matrix and $\mathcal{C}$ the
covariance matrix. The latter plays the key role in covariance
analysis. The domain of reasonable parameters is thus given by
\begin{equation}
  \chi^2(\vec{p})-\chi^2_0
  \approx
  \vec{p}\cdot\hat{\mathcal{C}}^{-1}\cdot\vec{p}
  \leq
  1
\label{eq:reasonable}
\end{equation}
which defines a confidence ellipsoid in the space of model
parameters. It is related to the Taylor expanded probability
distribution (\ref{eq:probab}) which becomes
\begin{equation}
  W(\vec{p})
  =
  \left[(2\pi)^F\mbox{det}\{\hat{\mathcal{C}}\}\right]^{-1/2}\,
  \exp(-\frac{1}{2}\vec{p}\cdot\hat{\mathcal{C}}^{-1}\cdot\vec{p})
\label{eq:probab2}
\end{equation}
where the fore-factor guarantees proper normalization $\int d^Fp\,W=1$.

Any observable $A$ is a function of model parameters
$A=A(\vec{p})$. The value of $A$ thus varies within the confidence
ellipsoid, and this results in some uncertainty $\Delta A$. Usually,
one can assume that $A$ varies weakly such that one can linearize it
\begin{subequations}
\begin{eqnarray}
\label{eq:linear}
  A(\vec{p})
  &\simeq&
  A(\vec{p}_0)+\vec{G}^A\cdot(\vec{p}- \vec{p}_0)
  \quad,
\\
  \vec{G}^A
  &=&
  \bm{\partial}_{\bm{p}}A\Big|_{\,\vec{p}_0}
  \quad.
\label{eq:GA}
\end{eqnarray}
\end{subequations}
This together with the Gaussian probability distribution
(\ref{eq:probab2}) allows to compute analytically averages and variances.
The average becomes
\begin{subequations}
\begin{equation}
  \overline{A}
  =
  \int d^Fp\,W(\vec{P})\,A(\vec{p})
  =
  A(\vec{p}_0)
\end{equation}
as one would have expected. The variance quantifies the fluctuation
of $A$ around $A(\vec{p}_0)$ and is
\begin{eqnarray}
  \overline{\Delta A}
  &=&
  \sqrt{\overline{\Delta^2A}}
  \;,
  \label{eq:variance}
\\
  \overline{\Delta^2A}
  &=&
  \int d^Fp\,W(\vec{P}),(A(\vec{p})-\overline{A})^2
\nonumber\\
  &=&
  \sum_{\alpha\beta}G^A_\alpha{\mathcal{C}}_{\alpha\beta}G^A_\beta  
  \;,
\nonumber
\end{eqnarray}
with $G^A_\alpha$ as given in eq. (\ref{eq:GA}). 
As it is derived from a statistical interpretation of the
$\chi^2$ landscape, it is also coined statistical error.
Along the same lines, one can also define
a cross-variance between two different observables. It reads
\begin{eqnarray}
  \overline{\Delta A\,\Delta B}
  &=&
  \int d^Fp\,W(\vec{P}),(A(\vec{p})-\overline{A})(B(\vec{p})-\overline{B})
\nonumber\\
  &=&
  \sum_{\alpha\beta}G^A_\alpha{\mathcal{C}}_{\alpha\beta}G^B_\beta  
\label{eq:crossvar}
\end{eqnarray}
This then allows to defined the covariance, or correlation
coefficient, 
\begin{equation}
  {c}_{AB}
  =
  \frac{|\,\overline{\Delta A\,\Delta B}\,|}
       {\overline{\Delta A}\,\overline{\Delta B}}
\quad.
\label{eq:correlator}
\end{equation}
\end{subequations}
It quantifies the statistical correlations between two observables $A$
and $B$. A value ${c}_{AB}=1$ means fully correlated where knowledge
of $A(\vec{p})$ fully determines $B(\vec{p})$.  A value ${c}_{AB}=0$
means uncorrelated, i.e. $A(\vec{p})$ and $B(\vec{p})$ are
statistically independent where knowledge of $A(\vec{p})$ carries no
information whatsoever on $B(\vec{p})$, for examples see
\cite{Reinhard_2013,Reinhard_2013a}.

\section{Strategies for estimating errors}
\label{sec:estim}

Although very powerful, a $\chi^2$ fit is a black box and seduces the
user to use it without much understanding of the physics beyond.  One
plugs in a model, chooses a couple of relevant fit data, and grinds
the mill until one is convinced to have found the absolute minimum
$\chi^2_0$ together with the optimal parameters $\vec{p}_0$. Thus far
one may find a satisfying description of the fit data but may live
with little understanding of the relevance and reliability of the
model. The challenging task remains to understand the model thus
achieved, in particular its reliability in extrapolations to other
observables. This is the quest for error estimates, see \cite{Dob14a}
for a basic discussion of error estimates in the context of nuclear
models. The problem does not have a simple and unique
answer. Statistical errors as explained in section \ref{sec:errors}
are well under control. The hard part is to estimate the systematic
errors coming from prejudices in the modeling or choice of data. The
bad news is that there is no systematic way to assess systematic
errors. From a critical point of view, every physical theory holds
only preliminarily and is always threatened by hidden insufficencies
which may be revealed some days by new observations.  As a
consequence, we need to handle our theories with a great deal of
restless error awareness. The best way to enhance confidence in a
model is to scrutinize it employing a great variety of different
checks and so piece-wise put together a view of the sources of
uncertainty. In the following, we summarize a couple of possible checks:
\begin{enumerate}

 \item\label{it:theo} {\bf Exploring a model from the theoretical side}
 \begin{enumerate}

 \item Estimates of beyond mean-field effects:\\
   The first step at all is to check whether the model under
   consideration is appropriate to describe the wanted data. This means for
   SHF to check many-body correlations beyond mean field.  Density
   matrix expansion (see \cite{Neg72a,Neg75a} and
   section\ref{sec:motivateSHF}) indicates that short-range
   correlations can be mapped into an energy functional. However,
   correlations from low lying $2^+$ states fluctuate strongly and
   cannot be mapped into a smooth functional. Nuclei which acquire
   large correlation corrections should be excluded from the pool of
   fit data \cite{Kluepfel_2009,Kluepfel_2008}. The search for
   possible correlation effects has not yet come to an end, for an
   example see section \ref{sec:abinit}.

 \item Exploring the stability of the model:\\ Models which rely on
   expansions are likely to have regions of instability. There is also
   the problem that $\chi^2(\vec{p})$ can have non-analyical
   points. It is crucial to explore those dangerous regimes and set
   rules to avoid them, or to cure the problem.  These considerations
   have accompanied the development of SHF all along, for recent
   critical studies, particularly of the tensor term, see
   \cite{Ben07a,Ben09c,Dav09a,Hel13a}.

 \item\label{it:formconnet}
   Formal analysis of inter-dependencies:\\
   Correlation analysis (see point \ref{it:correl}) can reveal
   dependencies between observables. To make sure that these are more
   than statistical correlations, one ideally establishes the
   connection at a purely formal level. Literature on such studies is
   unoverseeable. We mention here only two examples, one for the
   relation between effective mass $m^*/m$ and level density
   \cite{Mah85aR} and another one for the mapping of SHF parameters to
   NMP via the LDM \cite{Roc13a}.

 \item\label{it:varyfom} Variations of the functional:\\
    A test for a model from within is to vary the basic ansatz.  This
    means, e.g., for the SHF model to omit or to add terms or to vary
    the density dependence \cite{Erl10c}.  Another way is to compare
    and/or accumulate the results from different models as done by
    combination with RMF in \cite{Reinhard_2013a,Piekarewicz_2012}.
    We will exemplify in section \ref{sec:hierarchy} adding and
    omitting terms.

 \end{enumerate}

 \item\label{it:statist} {\bf Straightforward statistical analysis}
 \begin{enumerate}
 \item\label{it:uncert} Extrapolation uncertainties:
 \\
   This employs eq. (\ref{eq:variance}) to estimate
   the statistical error on extrapolations. It serves as useful
   indicator for safe and unsafe regions of the model see, e.g.,
   section \ref{sec:extrap}.

 \item\label{it:correl} 
   Correlations between observables:\\
   The covariances (\ref{eq:correlator}) reveal dependencies between
   observables. This helps to estimate the information content of
   a new measurement, for examples see \cite{Reinhard_2013}
   and section \ref{sec:correl}. It may be combined with a variation
   of fit data (see point \ref{it:varydata}) to distinguish between
   model properties and data dependence. It can also serve to
   motivate searches for formal connections (see point
   \ref{it:formconnet}). 

 \item Sensitivity of model parameters $\vec{p}$:\\ 
   The Jacobian matrix $\hat{J}$ together with the covariance matrix
   $\hat{\mathcal{C}}$ allows to explore the impact of each single
   model parameter $p_\alpha$ on a given fit observable
   $\hat{\mathcal{O}}_i$. Examples are found in
   \cite{Kor12a,Dob14a,Kortelainen_2010}.

 \end{enumerate}

 \item\label{it:dedicvar} {\bf Exploring the model by dedicated variations}
 \begin{enumerate}

 \item\label{it:residuals} Unresolved trends of residual errors:\\
    A perfect model should produce a purely statistical (Gaussian)
    distribution of residuals
    $\mathcal{O}^\mathrm{(th)}_\mathrm{obs,nucl}-\mathcal{O}^\mathrm{(exp)}_\mathrm{obs,nucl}$.
    Well visible
    trends of residuals of one sort of observable over the series of
    nuclei indicate deficiencies of the model. All nuclear mean-field
    models produce still strong unresolved trends see, e.g.,
    \cite{Kluepfel_2009,Erler_2010,Kortelainen_2010} 
    and section \ref{sec:bulkprop}.

 \item\label{it:varobs} Dedicated variations of parameter or observable:\\
    As a complement to correlation analysis (see point
    \ref{it:correl}), one obtains a direct view of dependencies by
    dedicated variation of a parameter $p_\alpha$ or observable
    $A(\vec{p})$. To that end, one fits a parametrization with
    constraint on a fixed value of $p_\alpha$ or $A$ and
    systematically varies the constrained value
    \cite{Kluepfel_2009}. Then one looks at the trend of other
    observables $B(p_\alpha)$, or $B(A)$ respectively.  This test is
    particularly instructive when using variations of NMP, see
    \cite{Kluepfel_2009} and section \ref{sec:varyNMP}.

 \item\label{it:varydata} Variations of fit data:\\
    The fit observables $\hat{\mathcal{O}}_\mathrm{obs,nucl}$ stem
    from different groups of observables ``obs'' as, e.g., energy or
    radius.  One can omit a group from the pool of fit
    data. Comparison with the full fit allows to explore the impact of
    the omitted group, see \cite{Erl14b} and section \ref{sec:groups}.

 \item\label{it:predict} Test of predicted observables:\\ A natural
   test is to compare a prediction (extrapolation) with experimental
   data wherever available.  It is a strong indicator for a systematic
   error if the deviation exceeds significantly the extrapolation
   error (\ref{eq:variance}).  An example for such a critical case is
   the energy of super-heavy nuclei where systematic deviations point
   to a problem of SHF, see \cite{Erler_2010} and section
   \ref{sec:bulkprop}.

 \end{enumerate}
 
\end{enumerate}
The above list of testing strategies shows that establishing a model
requires a broad palette of tools and counter-checks. The present
paper can not exemplify all of them. We will address the groups
\ref{it:statist} and \ref{it:dedicvar} as well as point
\ref{it:varyfom} from group \ref{it:theo}.

\section{Results}
\label{sec:results}


\subsubsection{Bulk properties}
\label{sec:bulkprop}

\begin{figure}[tbp]
\centerline{\includegraphics[width=\linewidth]{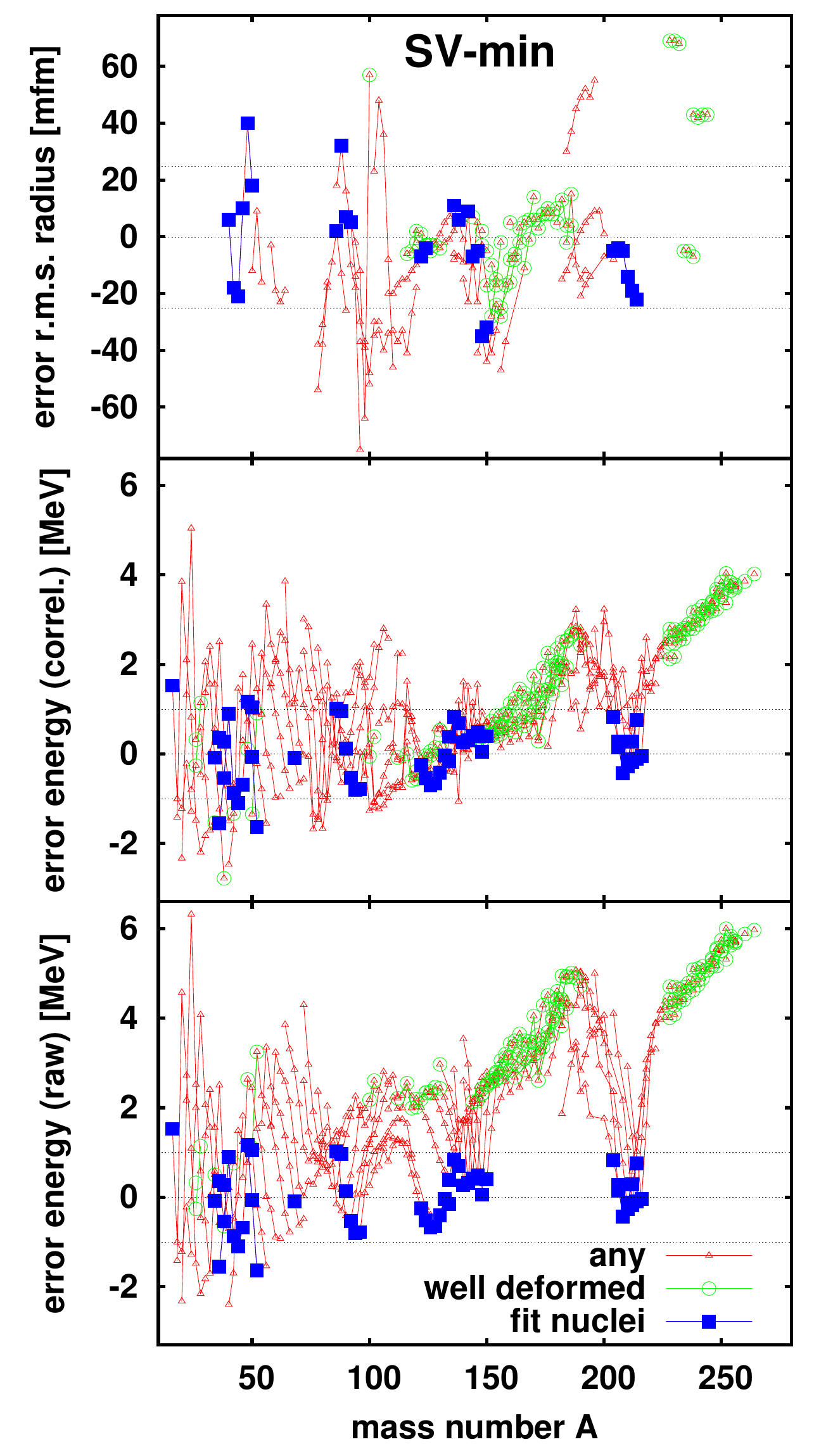}}
  \caption{Deviations of SHF calculations from experimental data for
    all nuclei for which data were available.  Fit nuclei are shown
    with filled blue squares, well deformed nuclei with green circles,
    and all other with red triangles. Isotopic chains are connected by
    a line. Lower panel: for binding energies computed with (deformed)
    SHF including only the c.m. correction.  Middle panel: for binding
    energies with SHF plus corrections from collective correlations
    (vibration and rotation).  Upper panel: For charge r.m.s. radius
    computed with (deformed) SHF.  }
\label{fig:trends-SVmin}
\end{figure}

Modern SHF parametrization have reached a high quality of reproduction
of nuclear ground state properties. As example, we quote here the
r.m.s. errors for SV-min which was fitted to ground state properties
of semi-magic spherical nuclei \cite{Kluepfel_2009}. They are 0.6 MeV
for binding energies $E_B$, 0.029\,fm for charge diffraction
radii $R_\mathrm{diffr}$, 0.022\,fm for charge surface
thicknesses, and 0.017\,fm for charge r.m.s. radii. Similar
quality is found for all modern parametrizations which appeared in the
last decade. 

Although small, the question is how the errors are distributed over
the nuclei. We do that here for SV-min \cite{Kluepfel_2009}.  Figure
\ref{fig:trends-SVmin} shows the residuals
$\mathcal{O}^\mathrm{(th)}_\mathrm{obs,nucl}-\mathcal{O}^\mathrm{(exp)}_\mathrm{obs,nucl}$
of r.m.s. radii (upper panel) and of binding energies (middle and
lower panel) for all nuclei for which data were
available. Different symbols (and colors) distinguish nuclei from the
fit pool, well deformed nuclei, and everything else.  The lower panel
shows binding energies $E_B$ from SHF calculations with
c.m. correction, but without any other correlation correction
(\ref{eq:Ecorrel}).  The fit nuclei gather nicely within a small error
band as given by the above mentioned r.m.s. errors. That is no
surprise because they were selected to have negligible correlations
\cite{Kluepfel_2009,Kluepfel_2008}. The other nuclei deviate
significantly. The middle panel shows the results with all corrections
from collective correlations, see eq. (\ref{eq:Ecorrel}). This brings
all results closer to a narrow error band. The deformed nuclei in the
medium heavy region ($A<200$) perform already very well while
transitional nuclei still overshoot.  

The errors are obviously not statistically distributed but show strong
trends which are not resolved by the given SHF model.  Even with
correlation corrections (middle panel) errors are smallest towards
shell closures and largest in between, particularly for the
transitional nuclei (red triangles). This is clearly a shell
fluctuation not fully accounted for by the present estimate of
correlations (only from the lowest $2^+$ state). More than that, there
is a strong trend to underbinding with increasing system size $A$ for
deformed nuclei (green circles). This becomes obvious in the regime of
actinides and super-heavy elements (A$>$220). Even attempts with
variations of density dependence \cite{Erl10c} and including deformed
nuclei in the fit \cite{Erler_2010} could not resolve the problem. It
seems to be inherent in the way the present SHF functional
(\ref{eq:basfunct}) models density dependence.  For example,
traditional RMF functionals which model density dependence quite
differently by non-linear self-coupling of the $\sigma$ meson 
have the opposite problem: they tend to
overbind super-heavy nuclei \cite{Bue98a}.  The unwanted
trends in figure \ref{fig:trends-SVmin} suggest that there is still
a need for improvement in nuclear self-consistent mean-field models.

The upper panel shows deviations for charge r.m.s. radii. Correlation
effects are smaller for this observable and not shown. There are also
less data which leads to less points on the plot. The deviations gather 
better within the error band from the fit nuclei. Nonetheless, they
are also not statistically distributed. We observe again the shell
fluctuations between shell closure and mid-shell nuclei. The
pronounced trend for deformed, super-heavy nuclei which was seen for
binding energies is absent for r.m.s. radii.

\subsubsection{Far extrapolations}
\label{sec:extrap}

An important quantity delivered by the $\chi^2$ technique is the
variance (\ref{eq:variance}) of an observable. It provides an estimate
for the extrapolation uncertainty which is extremely useful to get an
impression how reliable an extrapolation may be. We exemplify that
here for an extrapolation to super-heavy nuclei and an extremely far
extrapolation to neutron star properties. The case of super-heavy
nuclei continues the discussion from the previous section \ref
{sec:bulkprop} of an unresolved trend for heavy nuclei.

\begin{figure}
\centerline{\includegraphics[width=0.99\linewidth]{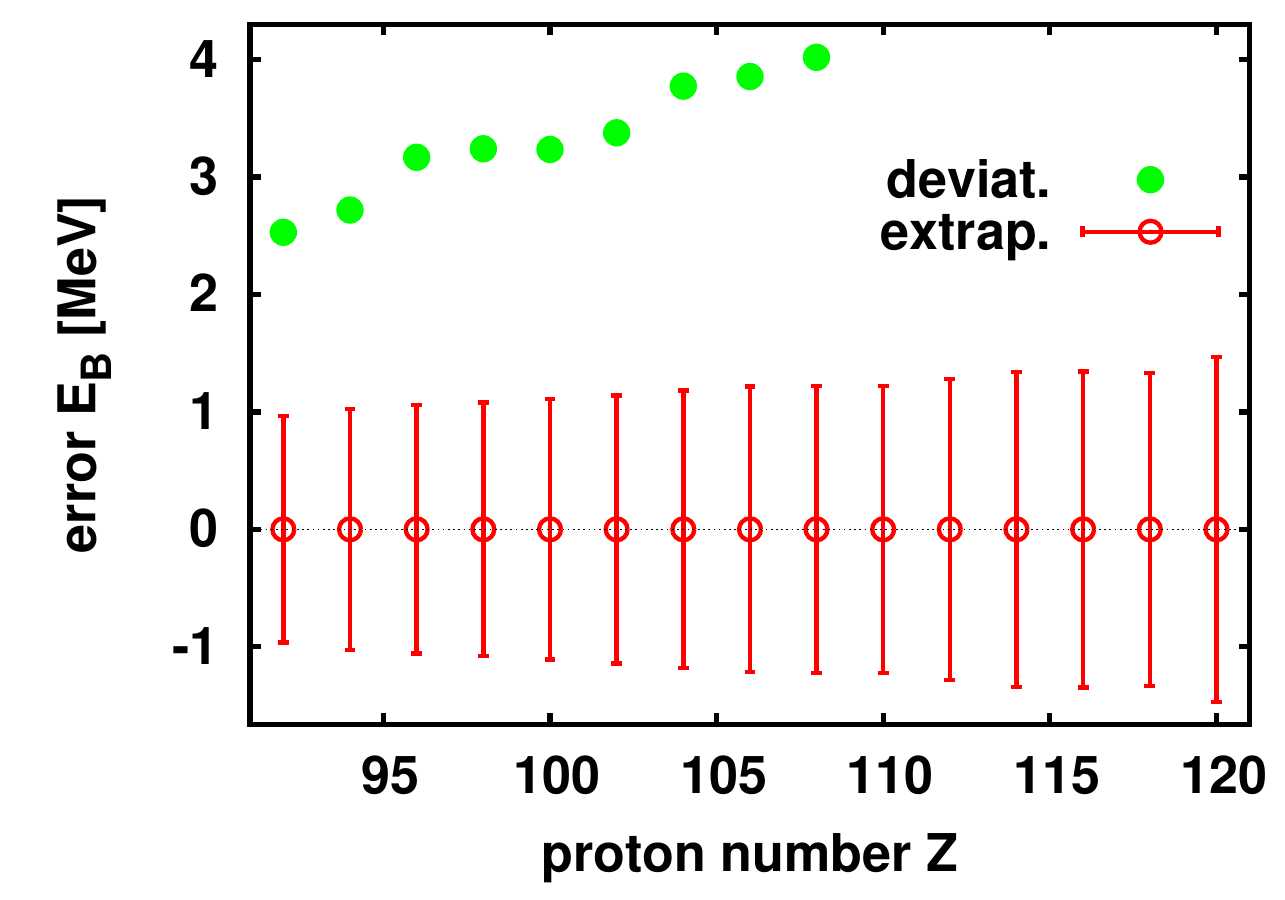}}
\caption{Evolution of errors for a chain of super-heavy elements with
  increasing proton number $Z$ in steps of 2. The neutron number has
  been increased accordingly in steps of 4, occasionally steps of 2,
  to cover the landscape of super-heavy elements ending up near the
  magic neutron number $N=184$. Shown are extrapolations errors
  (denoted ``extrap.'')  and deviations from data
  $E_\mathrm{SHF}-E_\mathrm{exp}$ (denoted ``deviat.'') where
  $E_\mathrm{SHF}$ includes correlation corrections as explained in
  connection with figure \ref{fig:trends-SVmin}. The parametrization
  SV-min was used.  }
\label{fig:extrap-SHE}
\end{figure}
Figure \ref{fig:extrap-SHE} shows extrapolation uncertainties computed
with eq.  (\ref{eq:variance}) and actual deviations from data (where
available) computed with SV-min for a chain of super-heavy nuclei from
$Z=90$ to $Z=120$. The sequence of neutron numbers has been chosen to
cut a path through the landscape spanning from Z/N=90/126 to 120/182.
It was: $N=$126, 130, 134, 138, 142, 146, 148, 150, 152, 154, 158,
162, 166, 170, 174, 178, and 182 The extrapolation uncertainties
(shown as error bars) grow steadily when moving away from the region
of fit nuclei. This is an expected behavior: the farther away the
extrapolation the larger the uncertainty. The chosen sequence still
allows to compare with data for more than half of the sample. The
actual deviations from available experimental values (shown as filled
circles) are significantly larger than the estimates of uncertainty
from statistical analysis. The figure demonstrates nicely the possible
devastating effect of a systematic error. In this case we are in the
lucky situation that we could already identify the problem because
there are sufficient data for checking, see figure
\ref{fig:trends-SVmin} and the discussion thereof. The case becomes
uncontrollable for totally new observables for which one does not
dispose of benchmark data. Nonetheless, with the statistical
extrapolation error we have an indispensable tool to get an idea about
the uncertainties to expect and a safe lower estimate for them. Mind
the the extrapolation uncertainty in figure \ref{fig:extrap-SHE} has
at least the correct order of magnitude.

\begin{figure}
\centerline{\includegraphics[width=0.99\linewidth]{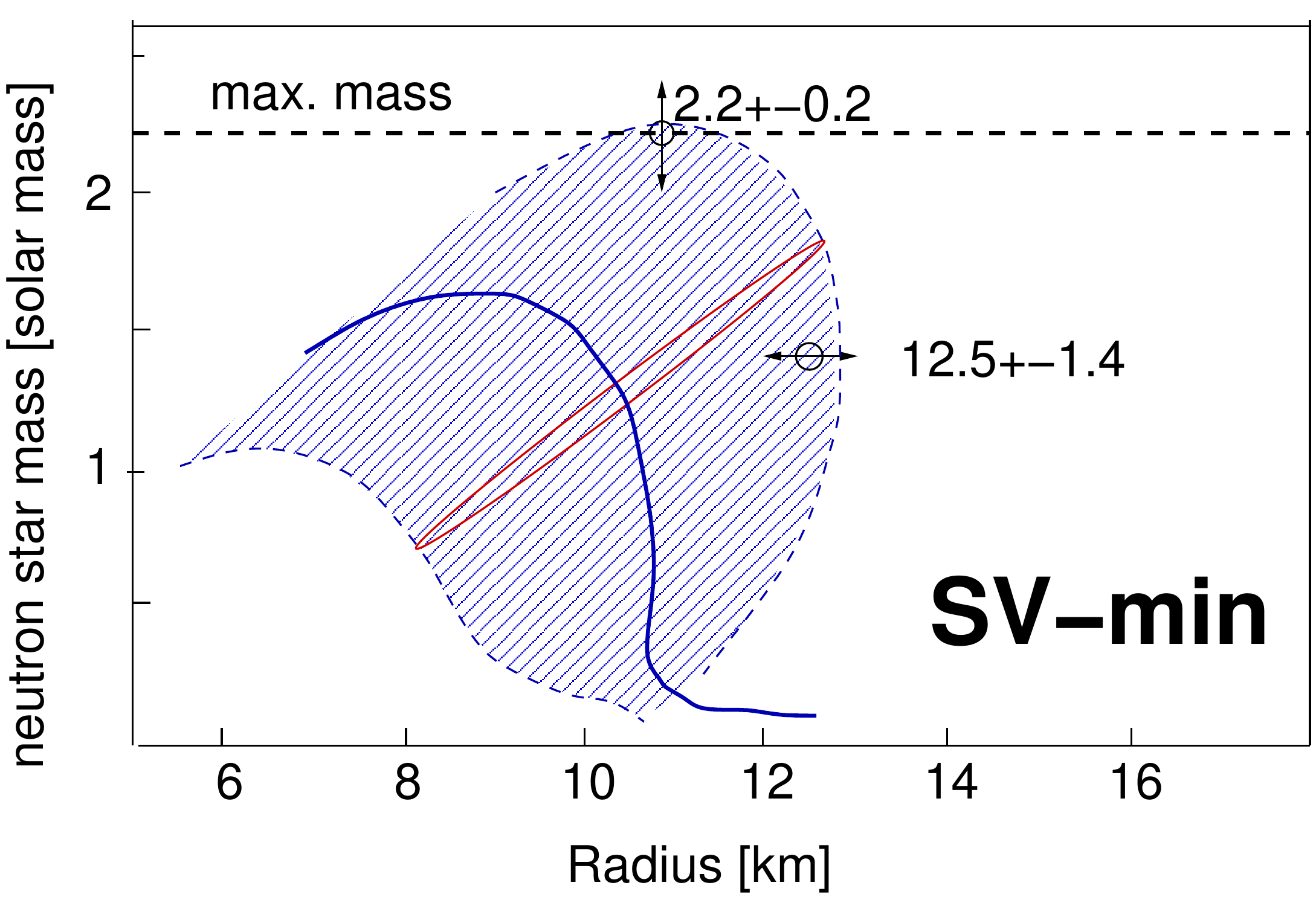}}
\caption{Mass-radius relation of a neutron star computed for SV-min
(heavy blue line).
The shaded area indicates the range of the extrapolation errors. 
The mass-radius correlation ellipsoid is shown by a read line.
The dashed horizontal line indicates the mass of the 
heaviest identified neutron star. The black dots with errors
stand for the maximum mass and maximum radius from SV-min.
Data taken from \cite{Erl13a}.}
\label{fig:NS-uncert}
\end{figure}
Figure \ref{fig:NS-uncert} addresses as next example the extrapolation
to neutron star mass and radii. The heavy line shows the results of
the Tolman-Oppenheimer-Volkov equation for the mass-radius relation
using as input the neutron equation-of-state from SV-min
\cite{Erl13a}. The faint horizontal line indicates the maximum mass of
a neutron star found so far \cite{Dem10a}. This sets a lower limit for
the maximal mass because one cannot exclude that an even heavier star
may be found some day. The prediction from SV-min stops far below the
wanted maximum mass. However, the outlook improves when taking into
account the extrapolation errors, marked by the shaded area. The
uncertainties are huge which is understandable because we deal here
with a really far extrapolation, from finite nuclei to comparatively
infinite star matter. The good news is that the error band scratches
the wanted maximum mass. This indicates that inclusion of wanted
neutron star properties can be accommodated by a $\chi^2$ fit without
sacrifices.  This was done with success in \cite{Erl13a} yielding the
parametrization TOV which indeed hits correctly the mass line and has
also a much smaller uncertainty band due to the inclusion of star
data. The figure \ref{fig:NS-uncert} contains one more detail. This is
the error ellipsoid plotted inside the error band.  It indicates the
region in which the probability distribution (\ref{eq:probab2}) is
larger than 1/2, this is the region of predictions from the range of
``reasonable parametrizations''. In this example, the ellipsoid looks
more like a needle going diagonal through the error band. This means
that we encounter here a highly correlated scenario where a given
value of radius fixes inevitably a corresponding value for mass. It is
a prototype example for a case with large covariance
(\ref{eq:correlator}) close to 1.

\subsection{Hierarchy of terms in the functional}
\label{sec:hierarchy}

The SHF functional (\ref{eq:basfunct}) was motivated by a Taylor
expansion of the a supposed microscopic effective interaction, see
section \ref{sec:motivateSHF}. This implies a hierarchy of importance
of the terms in the functional with the zeroth order of expansion (the
purely density dependent parts) representing the leading terms.  A
further ordering principle comes from the fit data. The landscape of
known nuclei extends widely over size $A$ but only over a narrow band
of isotopic chains. This means that isoscalar properties are well
determined by the long trends with $A$ while knowledge of isovector
properties is limited by the small extension in isovector direction
$N-Z$ together with the fact that only one sort of density is well
known empirically, namely the proton density measured by electron
scattering \cite{Fri82a}. This combined sorting principles led to the
assignment of a ``minimal set of terms'' as indicated by yellow (gray)
shading in eq.  (\ref{eq:basfunct}). In this section, we are going to
explore the importance of terms in detail by starting from the most
sparse model using only density dependent terms and adding step by
step more terms up to the full model (\ref{eq:basfunct}).  At each
stage, the model consisting out of the chosen terms is optimized by
minimization of $\chi^2$ always using the same data pool from
\cite{Kluepfel_2009}, see end of section \ref{seq:chi2} and apendix
\ref{sec:pool}.  This tries to make the best out of the given
functional. The strategy differs from a similar investigation of
stages in \cite{Erler_2011}.  There, we started from ab-initio
calculations in bulk matter to define the background of LDA similar as
in electron systems \cite{Dre90} and added stepwise derivative terms
with information on finite nuclei. Here we optimize each stage as far
as possible.

\begin{figure}[p]
\centerline{\includegraphics[width=0.9\linewidth]{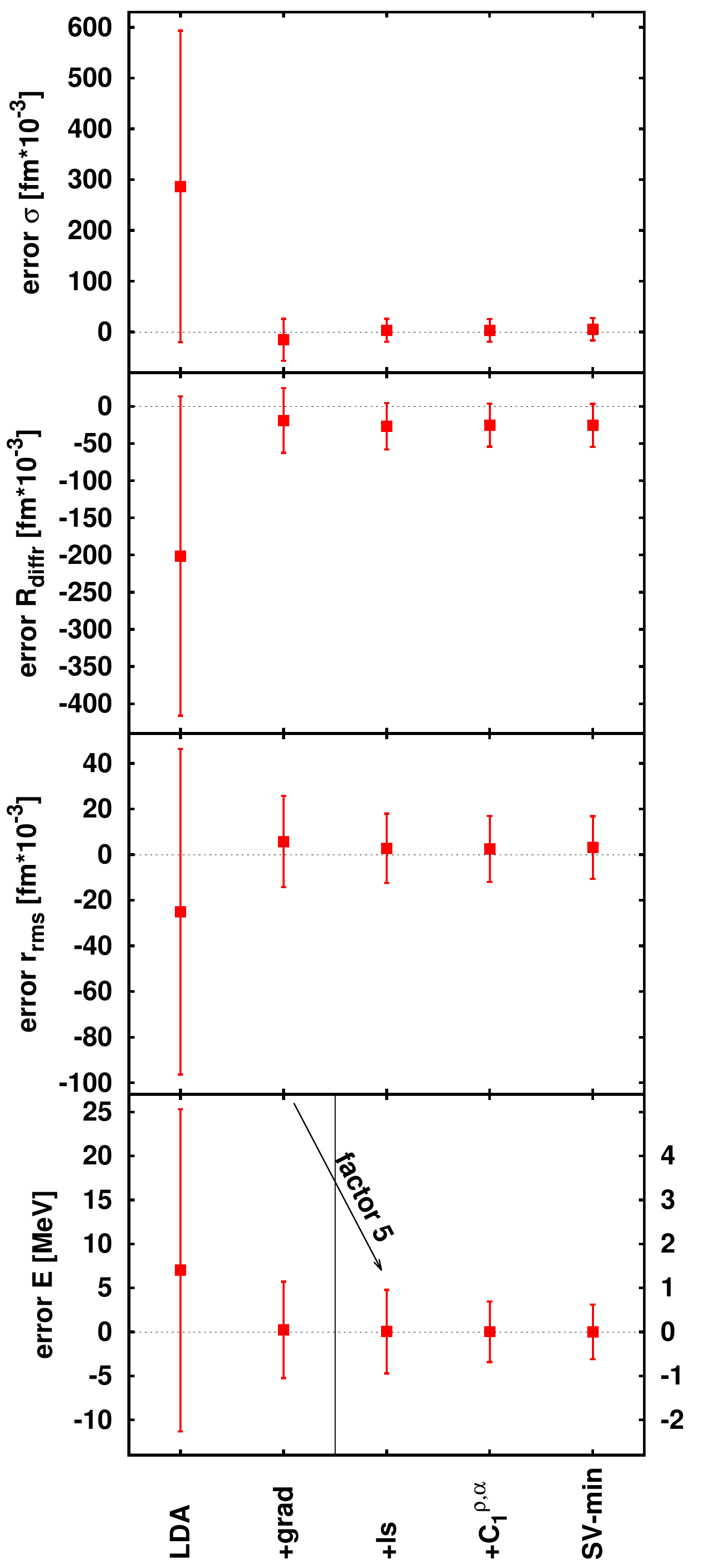}}
\caption{Average and r.m.s. errors on basic bulk properties, binding energy
    $E_B$, charge r.m.s. radius $r_\mathrm{rms}$, charge diffraction
    radius $R_\mathrm{diffr}$, and charge surface thickness
    $\sigma_\mathrm{surf}$. Averages are taken over the pool of fit
    nuclei. The average deviation from data is shown by filled boxes
    and the r.m.s. deviation by the error bars around the boxes.
    Results are shown for the series of fits with increasing number
    of terms in the functional (\ref{eq:basfunct}), see text.
    The lower panel for errors on binding energy is split into two
    energy scales. The first two entries (``LDA'' and ``+grad'') are
    plotted with respect to the scale on the left y-axis, the other
    three with respect to the right y-axis which is stretched by
    factor 5 with respect to the left scale.
 }
\label{fig:abs-quality}
\end{figure}

We consider the following stages of the model:
\begin{description}
  \item{LDA:}\\ This first stage invokes the terms with the parameters
    $C_0^{\rho}$, $C_1^{\rho}$, $C_0^{\rho,\alpha}$, and $\alpha$.  It
    covers only the purely density dependent terms which are active in
    homogeneous matter. We call it LDA because that choice produces a
    purely density dependent functional (no derivative terms
    wherever). But we optimize it empirically. It is not LDA in
    its strict sense where exact solution for bulk matter are used to
    deduce a functional \cite{Dre90}.
  \item{+grad:}\\ To the previous stage, the term
    $\propto{C}_0^{\Delta\rho}$ is added. This gives the model the
    flexibility to account for surface properties.
  \item{+ls:}\\ To the previous stage, the isoscalar spin-orbit
    $C_0^{\nabla{J}}$ term is added. This allows to put shell closures
    at the right place.
  \item{+$C_1^{\rho,\alpha}$:}\\ To the previous stage, the density
    dependent isovector term $\propto{C}_1^{\rho,\alpha}$ is added.
    This allows to resolve more isovector trends in the ground state
    properties, if there are still any.
  \item{SV-min:}\\
    This is result from invoking the full SHF functional
    (\ref{eq:basfunct}). It was published in
    \cite{Kluepfel_2009} an discussed in great detail there.
\end{description}
In all cases, we include, of course, the pairing functional (\ref{eq:ep}).
The first two stages (``LDA'' and ``+grad'') are too rough to deserve
a detailed pairing adjustment. Here we use volume pairing,
i.e. $\rho_{0,\mathrm{pair}}\rightarrow\infty$, with fixed
$V_\mathrm{pair,p}=V_\mathrm{pair,n}=300$ MeV\,fm$^3$. For the further
three stages which provide a detailed modeling, we optimize all three
pairing parameters together with the free  parameters of the SHF
model. 

Figure \ref{fig:abs-quality} shows for the above listed sequence of
stages of the model average and r.m.s. deviations for the four fitted
bulk properties where averages are taken over the pool of fit data.
The errors for the simplest stage, LDA, are large, particularly for
the binding energy where the error band reaches $\pm 20$ MeV.  But
this is, in fact, a satisfying result. It shows that already these few
leading terms are capable of accommodating roughly the energies
over all $A$, and even better for radii and surface thickness. Of
course, details are bound to be wrong because we are yet missing the
spin-orbit term and thus put shell closures at the wrong nucleon
numbers.

Adding the gradient term allows to tune the nuclear surface energy.
This improves the average quality of energy by half an order of
magnitude. With $\pm 4$ MeV it is as far as on can come without
appropriate shell effects. The description is even better for radii
and surface thickness because shell fluctuations play a minor role in
these observables. Here one has reached almost the final quality. The
great improvement due to the gradient term shows that the surface
energy is a crucial contribution to a nuclear energy functional. This
complies with the same findings in the purely macroscopic droplet
model of the nucleus \cite{Myers_1977,Hasse_1988}.

The next substantial improvement is achieved with adding the
spin-orbit force, stage ``+ls''. This yields a mean-field model where
shell closures emerge at the right nucleon number and also all other
shell effects acquire the correct place and magnitude. Consequently, we
see another dramatic reduction of the energy error. The model has now
reached a high quality. All further terms deliver only gradual
improvement.  The stage ``+ls'' with its 6 free SHF parameters thus
represents a minimal model which produces already an excellent
description of the fit observables (bulk properties of the nuclear
ground states). 

In the step to stage ``+$C_1^{\rho,\alpha}$'', we explore the gain by
adding more isovector flexibility. There is minimal improvement in the
quality of radii and surface thickness and there is a 25\% reduction
in the energy error, not dramatic but non-negligible either. There is
thus sufficient isovector information in the fit pool to fix this
second isovector parameter. 

Finally with ``SV-min'', we have activated all terms in the functional
(\ref{eq:basfunct}). The gain as compared to the previous stage is
extremely small. The additional terms in that step are all of
isovector type and the pool of fit data does not contain enough
isovector information to fix them. Thus ``SV-min'' leaves large
uncertainties in these isovector parameters.  Further data are
required to determine the isovector model parameters more
precisely. Isovector response properties (GDR, dipole polarizability, and
NMP, see section \ref{sec:varyNMP}) and the neutron skin
(\ref{eq:skin}) add valuable information as we will see later.

\begin{figure}
\centerline{\includegraphics[width=\linewidth]{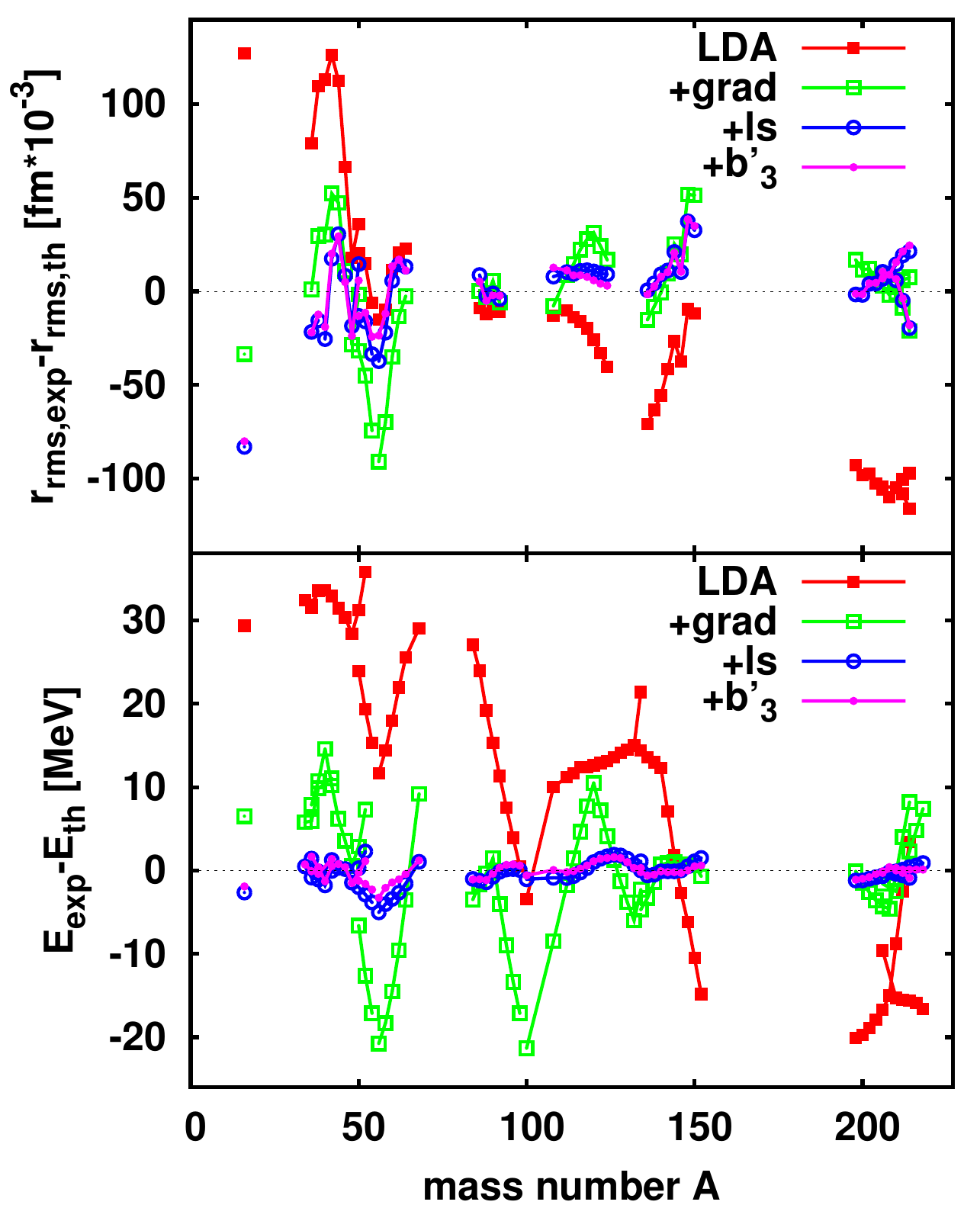}}
  \caption{Trends of deviations from data for the series of fits with
    increasing number of terms in the functional, see section
    \ref{sec:functional}.  The lower panel shows deviations for the
    binding energy $E_B$ and the upper panel for the charge
    r.m.s. radius $r_\mathrm{rms}$.  Results along isotopic or
    isotonic chains are connected by a line.  }
\label{fig:trends-quality}
\end{figure}

Figure \ref{fig:trends-quality} shows the detailed map of deviations
from data for the above introduced stages of the SHF functional. For
simplicity, the plot is limited to errors in energy and
r.m.s. radius. The diffraction radius and surface thickness behave
similar. The errors for LDA show a strong trend over mass number $A$.
The fit for LDA has found a global compromise between light and heavy
nuclei. But the model at LDA level is not capable to deal properly
with system size. The step to ``+grad'' resolves the global trend at
once. A straight line through the errors is identical with the zero
line. But there remain the strong shell fluctuations which are due to
the fact that shell closure lie at the wrong nucleon numbers. These
fluctuations are much reduced with the step to ``+ls'' whose energy
errors (lower panel) then gather closely around the zero line.  The
errors on radii (upper panel) seem to be more resistive.  But this
impression is to a large extend a matter of scale.  LDA delivered
already quite good values for the radii. In turn, the discrepancy between
LDA and further stages is not as dramatic for the radii as it is for
energy (see also figure \ref{fig:abs-quality}). Thus we see the
deviations on radii with higher resolution and that reveals more
clearly the irresistible unresolved trends.

The further steps deliver only gradual improvement, hardly
distinguishable at the scales of this plot. The remaining errors,
however, are still not statistically distributed. There remain some
unresolved trends seemingly related to shell structure (shell
closures versus mid shell). This calls for further refinement of the
model. But it is yet open whether this can be achieved within the SHF
and pairing functional or whether we are forced to invoke correlation
effects.

\begin{figure}
\centerline{\includegraphics[width=0.8\linewidth]{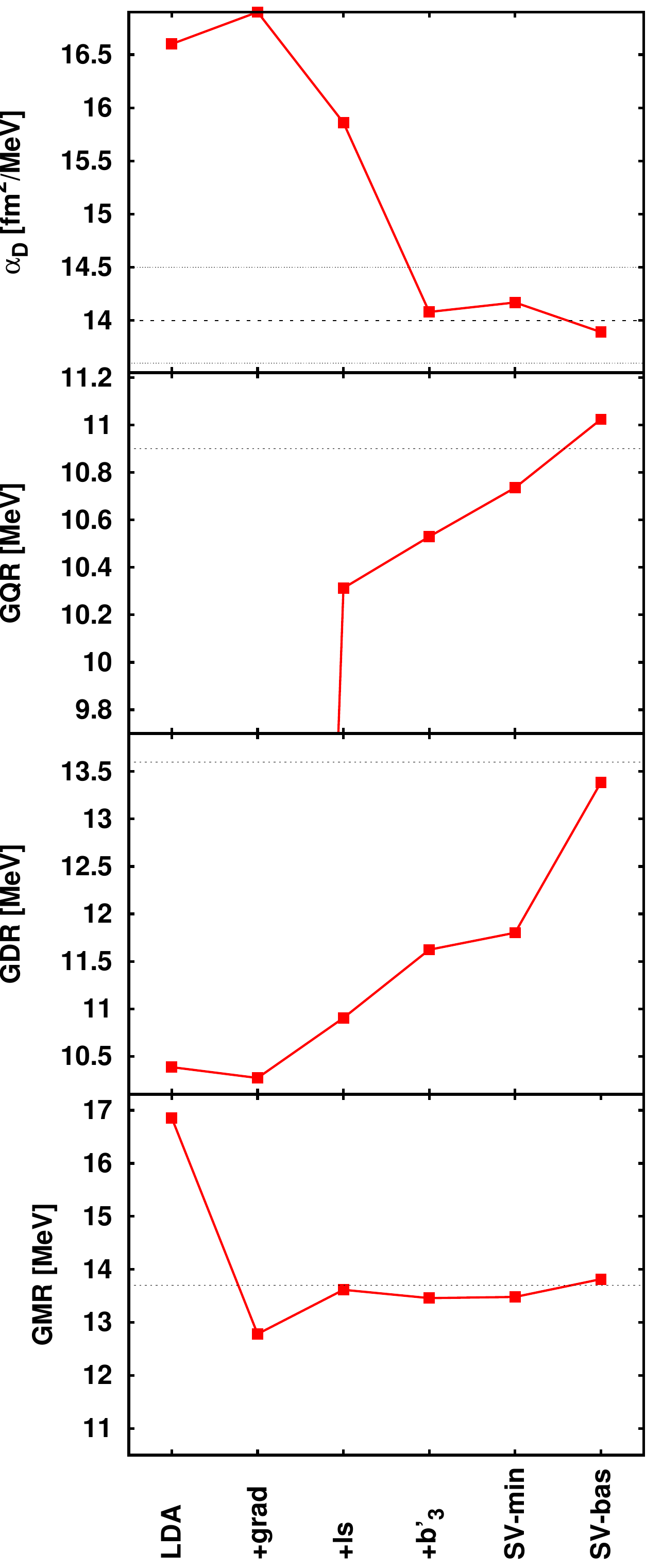}}
\caption{ Predictions for giant resonances and dipole polarizability
  in $^{208}$Pb compared with experimental data. Results are shown for
  the series of fits with increasing number of terms in the
  functional. A new entry as compared to figure \ref{fig:abs-quality}
  and \ref{fig:trends-quality} is the parametrization SV-bas. It is
  fitted the same way as SV-min but includes constraint an NMP in
  order to allow also a good reproduction of giant resonance energies,
  see \cite{Kluepfel_2009}. }
\label{fig:GDR-quality}
\end{figure}
Figure \ref{fig:GDR-quality} shows the peak energies for giant
resonances and the dipole polarizability for the stages of the SHF
functional, all for the nucleus $^{208}$Pb. The lower panel shows the
GMR. It improves with each stage and reaches already at the minimal
model, stage ``+ls'', good agreement with the experimental value.
This is not surprising. The GMR is closely tied to the
incompressibility $K$ which characterizes the response to changes in
density \cite{Bla76a,Bla80aR}. The incompressibility, in turn, depends
on the parametrization of density dependence in the model. And this
seems to be well determined by the fit data which span a broad range
of mass numbers (isoscalar information). We learn that the isoscalar
aspects are already well converged with the stage ``+ls''.
 
The second panel from below shows results for the GDR. They are all
far below the experimental point even.  The reason is clear for all
stages before SV-min because these do no invoke the effective mass
terms $C_T^\tau$ and it is known that the GDR depends sensitively on
TRK sum-rule enhancement $\kappa_\mathrm{TRK}$ which is closely
related to $C_1^\tau$ \cite{Rei99c,Kluepfel_2009}.  But even the step
to for SV-min, which includes all terms of the functional does not
improve the situation.  The fit data are not strict enough to push
$\kappa_\mathrm{TRK}$ to the right value.  On the other hand, they do
not fix $\kappa_\mathrm{TRK}$ so much.  The uncertainty in the GDR
energy is 1.4 MeV for SV-min. This leaves sufficient leeway to tune
$\kappa_\mathrm{TRK}$. This is done with the parametrization SV-bas
which optimizes some NMP to tune the giant resonances in
$^{208}$Pb\cite{Kluepfel_2009}.  This leads obviously to a good
reproduction of the GDR.

A similar situation is encountered for the GQR (second panel from
above). All stages up to including SV-min underestimate its resonance
energy. This is no surprise for the stages LDA through
``+$C_1^{\rho,\alpha}$'' because they do not invoke the term
$\propto C_0^\tau$ and thus have a mean field with effective mass
$m^*/m=1$. But the GQR is known to depend sensitively on $m^*/m=1$
and require a lower value 0.8--0.9 \cite{Bra85aR}. This option is
set free with SV-min. Similar as for the GDR, SV-min does not
put $m^*/m$, and with it the GQR, at the right place, but leaves
sufficient leeway to tune it properly. This is again achieved
by SV-bas.

The upper panel in figure \ref{fig:GDR-quality} shows the dipole
polarizability $\alpha_D$ for the various stages of the functional. It
starts for LDA with a rather wrong value, improves with each step, and
reaches the data already at the stage ``+$C_1^{\rho,\alpha}$'' where
the functional was given sufficient isovector freedom to take
advantage of the (static) isovector information in the fit data. 

It is interesting to note that two data points, GMR (related to
incompressibility $K$) and $\alpha_D$, are well described already from
the fits to ground state data while the two other, GDR and GQR, need
extra measures. Mind that $K$ and $\alpha_D$ are static response
properties. No surprise then that they are well determined by ground
state data. The GDR and GQR, on the other hand, are related to the
dynamic response properties $m^*/m$ and $\kappa_\mathrm{TRK}$. It is
plausible that dynamical properties require dynamic data for
calibration. Thus we need some information from excited states to
fully fix the functional (\ref{eq:basfunct}).

\subsection{Impact of groups of data}
\label{sec:groups}

In this section, we take up point \ref{it:varydata} of the analyzing
strategies outlined in section \ref{sec:estim}, the variation of fit
data. An analysis following this strategy was already given in
\cite{Erl14b}. We repeat part of that material for a reduced set of
variations and, on the other hand, add new variations.
Basis is the pool of fit data as developed in \cite{Kluepfel_2009},
see also end of section \ref{seq:chi2} and appendix \ref{sec:pool}. An
unconstrained fit to the full set yields the parametrization
SV-min. Now we are interested in fits with deliberate omission of
groups of fit data.
\begin{table}[bht]
\begin{center}
\begin{tabular}{|l|c|ccc|}
\hline
  & $E_B$  &  $r_{\mathrm{rms,C}}$ & $R_{\mathrm{diff,C}}$ & $\sigma_\mathrm{C}$ \\
\hline
 ``half data'' &   x/2 & x/2 & x/2 & x/2 \\
 SV-min &   x & x & x & x \\
 ``$E$+$r_\mathrm{rms}$ only'' &   x & x & - & - \\
 ``$E$ only'' &  x & - & - & - \\
 ``$r_\mathrm{rms}$ only'' & - & x & - & -  \\
\hline
\end{tabular}
\end{center}
\caption{\label{tab:names} Included data sets from the standard pool
  of fit data from finite nuclei \cite{Kluepfel_2009}.  A ``x'' means
  included, ``-'' stands for excluded,and ``x/2'' means that only half
  of the data were included. Pairing gaps $\Delta_\mathrm{pair}$ and
  spin-orbit splitting $\varepsilon_\mathrm{ls}$ are included in all
  data sets and not listed explicitly above.  The observables were
  explained in section \ref{sec:finnuc}.  }
\end{table}
%
This yields the parametrizations as listed in table \ref{tab:names}.
We have also studied omission of $\varepsilon_\mathrm{ls}$ or
$\Delta_\mathrm{pair}$. This showed only minor effects and will be
not reported here.

\begin{figure}
\centerline{\includegraphics[width=0.99\linewidth]{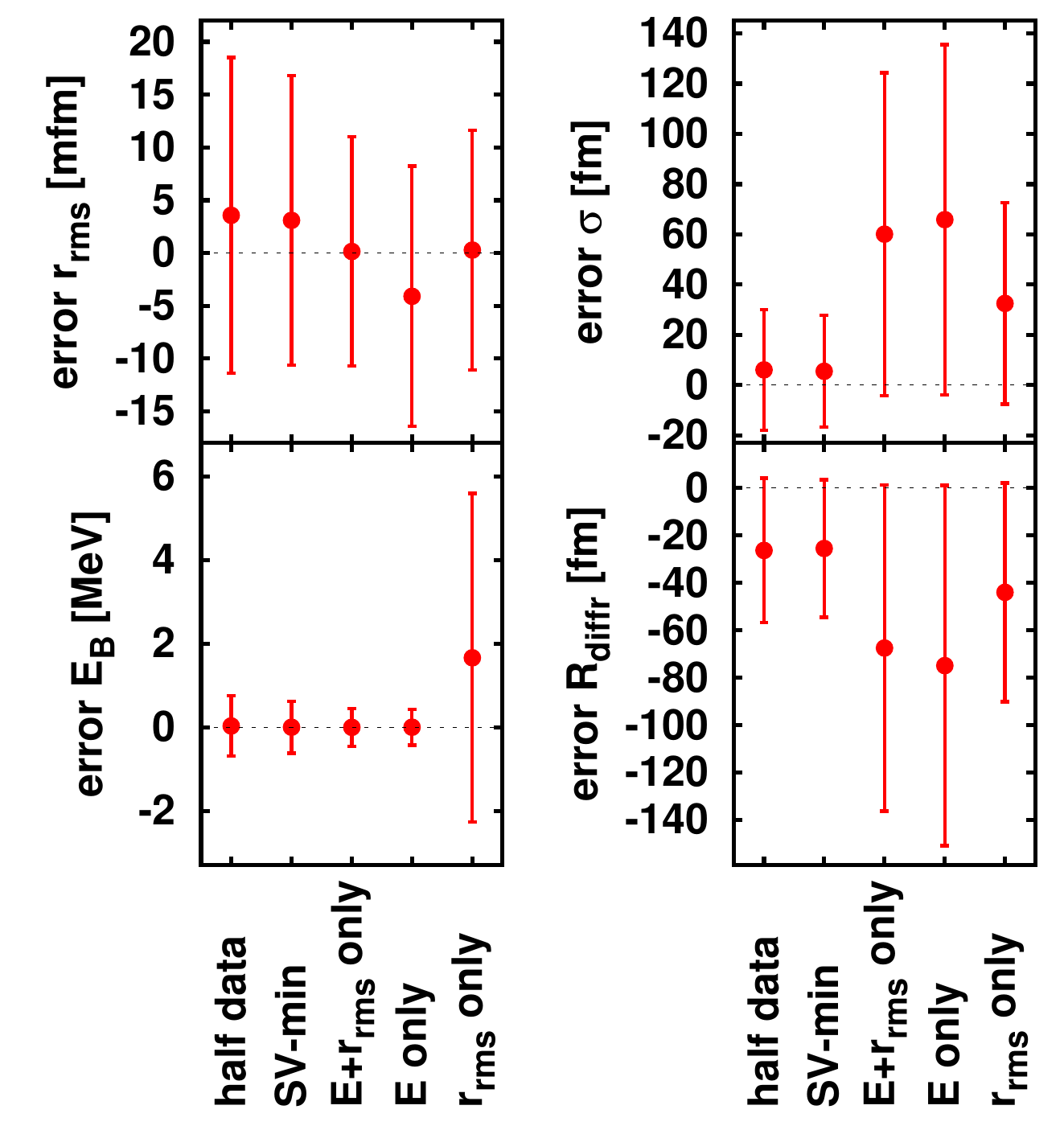}}
\caption{Average and r.m.s. errors on basic bulk properties, binding
  energy $E_B$, charge r.m.s. radius $r_\mathrm{rms}$, charge
  diffraction radius $R_\mathrm{diffr}$, and charge surface thickness
  $\sigma_\mathrm{surf}$. Averages are taken over the pool of fit
  nuclei. The average deviation from data is shown by filled boxes and
  the r.m.s. deviation by the error bars around the boxes.  Results
  are shown for the series of fits with different selections of groups
  of data as listed in table \ref{tab:names}.
  }
\label{fig:collect-errorblocks-reduced}
\end{figure}

Figure \ref{fig:collect-errorblocks-reduced} demonstrates the effect
of varied fit data on the average and r.m.s. residuals taken
separately over a group of data. Mean values deviating significantly
from zero within the scale set by the uncertainties indicate some
basic incompatibility of the observable with the model. Changes on the
r.m.s. error indicate the sensitivity to a group of observables. 

First, we should realize that the scales in this figure are generally
much smaller than those of the various stages of the functional in
figure \ref{fig:abs-quality}. All fits with reduced data sets produce
very similar models. Even when taking ``$E$ only'' or even
``$r_\mathrm{rms}$ only'' we obtain an agreeable model. This gives us
some confidence that the model has predictive power and allows
extrapolations to other observables or other regions of nuclei.

However, looking closer at the results in figure
\ref{fig:collect-errorblocks-reduced} we see at some places
differences which carry some information on tensions within the
model and at other places continued similarity which indicates 
the strongholds of the model.

Let us first compare the set ``half data'' with SV-min. Both sets
produce exactly the same averages. In fact, looking at detailed
residuals (not shown here) we find that both fits produce very similar
results. This can be read off also globally from the fact that the
r.m.s. errors (taken over all data) are similar.  This proves that the
SHF functional interpolates well.

The step from SV-min to ``$E$+$r_\mathrm{rms}$ only'' has dramatic
consequences for $R_\mathrm{diffr,C}$ and $\sigma_\mathrm{C}$.  The
average error in these two observables makes a large jump upward and
similarly the r.m.s. error.  This indicates that there is some
incompatibility between $r_\mathrm{rms,C}$ and $R_\mathrm{diffr,C}$ in
the present model. That could already be spotted in SV-min. Already
here, we find a large average error for $R_\mathrm{diffr,C}$. In fact,
the r.m.s. error is almost exhausted by the average error.  The same
behavior persists in the step to ``$E$ only''. But here, we also find
the same increase in the average error for $r_\mathrm{rms,C}$ which
indicates a slight incompatibility of $E_B$ and $r_\mathrm{rms,C}$. It
is likely that both incompatibilities, for $R_\mathrm{diffr,C}$ as
well as for $r_\mathrm{rms,C}$ are related to insufficencies in the
surface profile and thus to the model of density dependence.

The average error of energy $E_B$ is near zero for all fits which
include $E_B$. This indicates that the SHF functional is well suited
to adjust the energy over all ranges of nuclei. The model
``$r_\mathrm{rms}$ only'' produces a slight mismatch of the average.
However, these stay well within the huge error bars in this case.  The
large size of the error bars can be viewed as an ``automatic'' warning
from the $\chi^2$ analysis. It signals that the data set
``$r_\mathrm{rms}$ only'' does not contain sufficient information to
fix the energies. It is remarkable, though, that the other direction,
determining $r_\mathrm{rms}$ by a fit ``$E$ only'', works much better.
Energies are the leading input to the pool of fit data.

\begin{figure*}
\centerline{\includegraphics[width=0.99\linewidth]{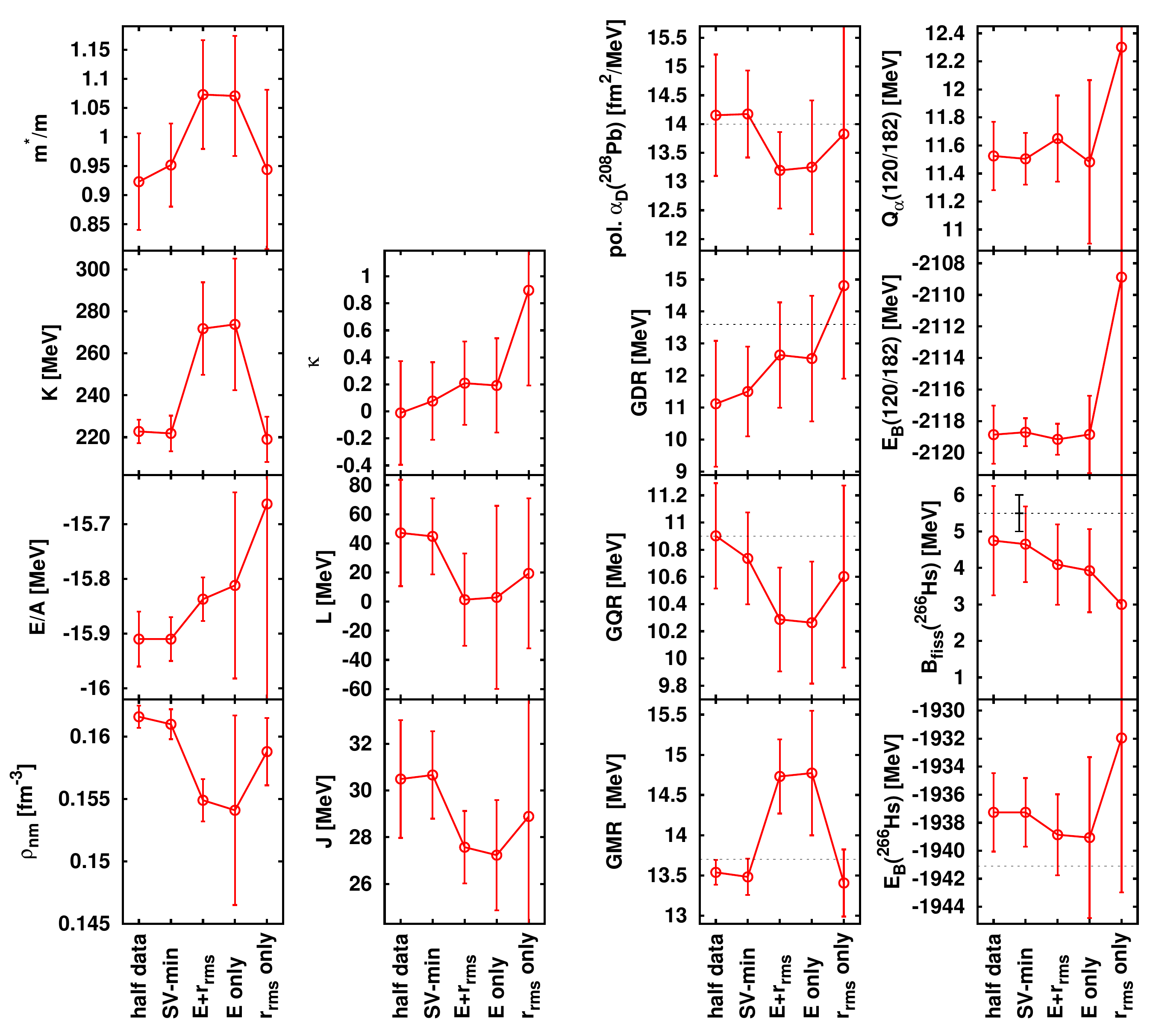}}
\caption{Results on NMP (left two columns) and
specific observables in finite nuclei (right two columns) for the
three fits to different data sets: ``SV-min'' = fit to the full standard
data pool of \cite{Kluepfel_2009}, ``$E+r_\mathrm{rms}$ only'' =
energies and r.m.s. radii in the pool plus gaps and l*s splittings,
``$E$ only'' = energies  in the pool plus gaps and l*s splittings.
Left column: isoscalar NMP. Second from left column: isovector NMP.
Third column: Giant resonance energies and polarizability in
$^{208}$Pb.
Right column: binding energy and fission barrier in $^{266}$Hs
and binding energy and $\alpha$-decay energy in the the hypothetical
super-heavy nucleus Z=120/N=182.
}
\label{fig:collect-NMPfin-reduced}
\end{figure*}
Figure \ref{fig:collect-NMPfin-reduced} shows the effect of varied fit
data on predicted/extrapolated observables, nuclear matter properties
(NMP) and key observables of the three nuclei $^{208}$Pb, $^{266}$Hs
and Z=120/N=182.  For Z=120/N=182, we consider binding energy $E_B$
and $\alpha$-decay energy $Q_\alpha$ according to
eq. (\ref{eq:Qalpha}). No data are available yet for this nucleus. For
$^{266}$Hs, we consider binding energy $E_B$ \cite{Wan12} and fission
barrier $B_f$ \cite{Pet04}.  The experimental value for $B_f$ is given
with error bar as it is associated still with large uncertainty.  The
binding energy includes the rotational zero-point energy
(\ref{eq:zperot}) which is obligatory for deformed nuclei.  In
$^{208}$Pb, we consider response properties, the peak energies of GMR,
GDR,and GQR and the dipole polarizability $\alpha_D$ (see section
\ref{sec:finnuc}).
The NMP considered are: binding energy $E/A$, density $\rho_{nm}$,
incompressibility $K$, isoscalar effective mass $m^*/m$, symmetry
energy $a_\mathrm{sym}$, and Thomas-Reiche-Kuhn (TRK) sum rule
enhancement factor $\kappa_\mathrm{TRK}$, all taken at the equilibrium
point of symmetric matter.  Figure~\ref{fig:collect-NMPfin-reduced}
shows the effect of variation of fit data on the observables and their
uncertainties.  Changes in uncertainty indicate the importance of the
omitted data group on the observable.  A shift of the average shows
what data are pulling in which direction.

The effects on NMP (left two columns) are generally large.  A most
pronounced shift is produced by omitting the formfactor information
of $R_\mathrm{diffr,C}$ and $\sigma_\mathrm{C}$ in ``$E+r_\mathrm{rms}$
only'' and ``$E$ only''. This leads to a large jump in bulk
equilibrium density $\rho_\mathrm{nm}$ and incompressibility
$K$. There is also a jump in the isovector response $a_\mathrm{sym}$
in addition to the generally strong changes.  It is also interesting
to note that formfactor information keeps the effective mass $m^*/m$ down
to values below 1 while fits without radii let $m^*/m$ grow visibly
above one.  The reason is probably that $m^*/m$ has an impact on the
surface profile, thus on $r_\mathrm{rms,C}$ and $\sigma_\mathrm{C}$,
and, in turn, also on $R_\mathrm{diffr,C}$.  This suggests that the
formfactor is closely related to the surface profile of the
nucleus. This, in turn, seems to be connected with modeling the
density dependence as one can read off from the strong impact on $K$
(second derivative with respect to density).  The step to
``$r_\mathrm{rms}$ only'' restores approximately the values of
$\rho_\mathrm{nm}$ and $K$ from the full fit SV-min.  This indicates
that there is also some competition between
$r_\mathrm{rms}$ and $E_B$.  These two quantities pull in
different direction which indicates that they would prefer different
surface profiles.  Mind, however, that the deviations in the average
errors are fully compatible within the error bars (characterizing the
r.m.s. errors).  This means that all choices of subsets of data are
compatible when taking into account the message about uncertainties of
a fit which is quite appropriately delivered by statistical analysis.
Nonetheless, the observed variations of average errors indicate that
the SHF functional may deserve further fine tuning to reduce these
conflicts between observables. The most likely aspect to work on is
the density dependence of the functional.

The variances usually grow when omitting data which is plausible
because less data mean less determination.  Particularly large
errors emerge for the sets ``E only'' and ``$r_\mathrm{rms}$ only''.
This shows that
combined information with  the charge formfactor is extremely useful to confine
the parametrization.  But mind that the error
bars for ``E only'', although being generally larger, are still in
acceptable ranges. This shows that data from energy alone can already
provide a reasonable parametrization. One may argue similarly with the
errors for ``$r_\mathrm{rms}$ only''. But here the error on $E/A$ is
rather large which was already foreseeable from the also large error
on energies of finite nuclei seen in
\ref{fig:collect-errorblocks-reduced}. Besides $E/A$, the
parametrization ``$r_\mathrm{rms}$ only'' yields also a very large
error on the symmetry energy $J$. This happens because the selection
of nuclei with radius data is smaller along isotopic chains thus
carrying even less isovector information.

Column 3 of figure \ref{fig:collect-NMPfin-reduced} shows the effect
of variation of fit data on response observables in $^{208}$Pb.  The
three giant resonances and the polarizability $\alpha_D$ are known to
have a one-to-one correspondence with each one NMP
(see sections \ref{sec:varyNMP}, \ref{sec:correl}, and ref.
\cite{Kluepfel_2009}): the GMR with $K$, the GQR with $m^*/m$, the GDR
with $\kappa_\mathrm{TRK}$, and $\alpha_D$ with
$a_\mathrm{sym}$. These pairs of highly correlated observables show
exactly the same trends in the figure which confirms the
correspondence also from this point of view. Unlike NMP, observables
in finite nuclei allow a comparison with experimental data.  GMR and
$\alpha_D$ are related to the static response properties $K$ and
$J$. Comparison with data shows clearly that inclusion of formfactor
information drives into the right direction while a fit without
$R_\mathrm{diffr,C}$ and $\sigma_\mathrm{C}$ yields less favorable
values. This indicates that information from the formfactor carries
correct physics about the nuclear response although these data were
found in the previous figure to stay slightly in conflict with energy
information. The situation is mixed for the dynamical response
properties. The GDR improves with omitting $R_\mathrm{diffr,C}$ and
$\sigma_\mathrm{C}$ while GQR deteriorates this way. This is very likely
pure chance because dynamic response is not embodied in ground state
data. An explicit fit of dynamic response, as done e.g. in SV-bas
\cite{Kluepfel_2009}, is anyway recommended.

The situation is mixed for the super-heavy elements (SHE) shown in
column 4 of figure \ref{fig:collect-NMPfin-reduced}.  As several times
before, all predictions agree with each other within their error
bars. There remains to comment a few trends of average values.
Skipping $R_\mathrm{diffr,C}$ and $\sigma_\mathrm{C}$ (parametrization
``$E$+$r_\mathrm{rms}$ only'') improves $E_B(^{266}$Hs$)$ but spoils
$B_f(^{266}$Hs$)$.  Skipping, further more $r_\mathrm{rms,C}$ to
parametrization ``$E$ only'' does not make much of further
changes. However, the parametrization ``$r_\mathrm{rms}$ only'' spoils
both $E_B(^{266}$Hs$)$ and $B_f(^{266}$Hs$)$. There are even less
trends in the averages for the SHE Z=120/N=182, except for
``$r_\mathrm{rms}$ only'' which is obviously totally inappropriate for
SHE. This is also signaled by huge error bars which
``$r_\mathrm{rms}$ only'' produces in all four observables of SHE. The
data basis of r.m.s. radii is obviously too weak to allow for far
extrapolations. The examples demonstrate that statistical analysis
combined with variation of model or conditions delivers very useful
insight into the model.

\subsection{Systematic variation of nuclear matter properties (NMP)}
\label{sec:varyNMP}

In this section, we exemplify the strategy \ref{it:varobs} (see
section \ref{sec:estim}), namely systematic variation of a model
parameter or property. We have argued in section \ref{sec:NM} that the
NMP as defined in table \ref{tab:nucmatdef} are fully equivalent to
the parameters of the SHF functional and have the additional advantage
that they carry an intuitive physical meaning. We follow here the
strategy of \cite{Kluepfel_2009} and define a base point with the
following four NMP fixed: $K=234$ MeV, $m^*/m=0.9$, $J=30$ MeV, and
$\kappa_\mathrm{TRK}=0.4$. These values were chosen because they lead
to a good reproduction of the four basic response properties in
$^{208}$Pb: GMR, GDR, GQR, and $\alpha_D$. A fit with these four NMP
fixed delivers the parametrization SV-bas. Starting from this base
point, we have produced four chains of parametrizations where one of
the NMP was varied systematically while the other three were kept
fixed. A detailed description of this set of parametrizations is
given in  \cite{Kluepfel_2009}. In the following we use this set for
the examples.

\begin{figure}
\centerline{\includegraphics[width=0.99\linewidth]{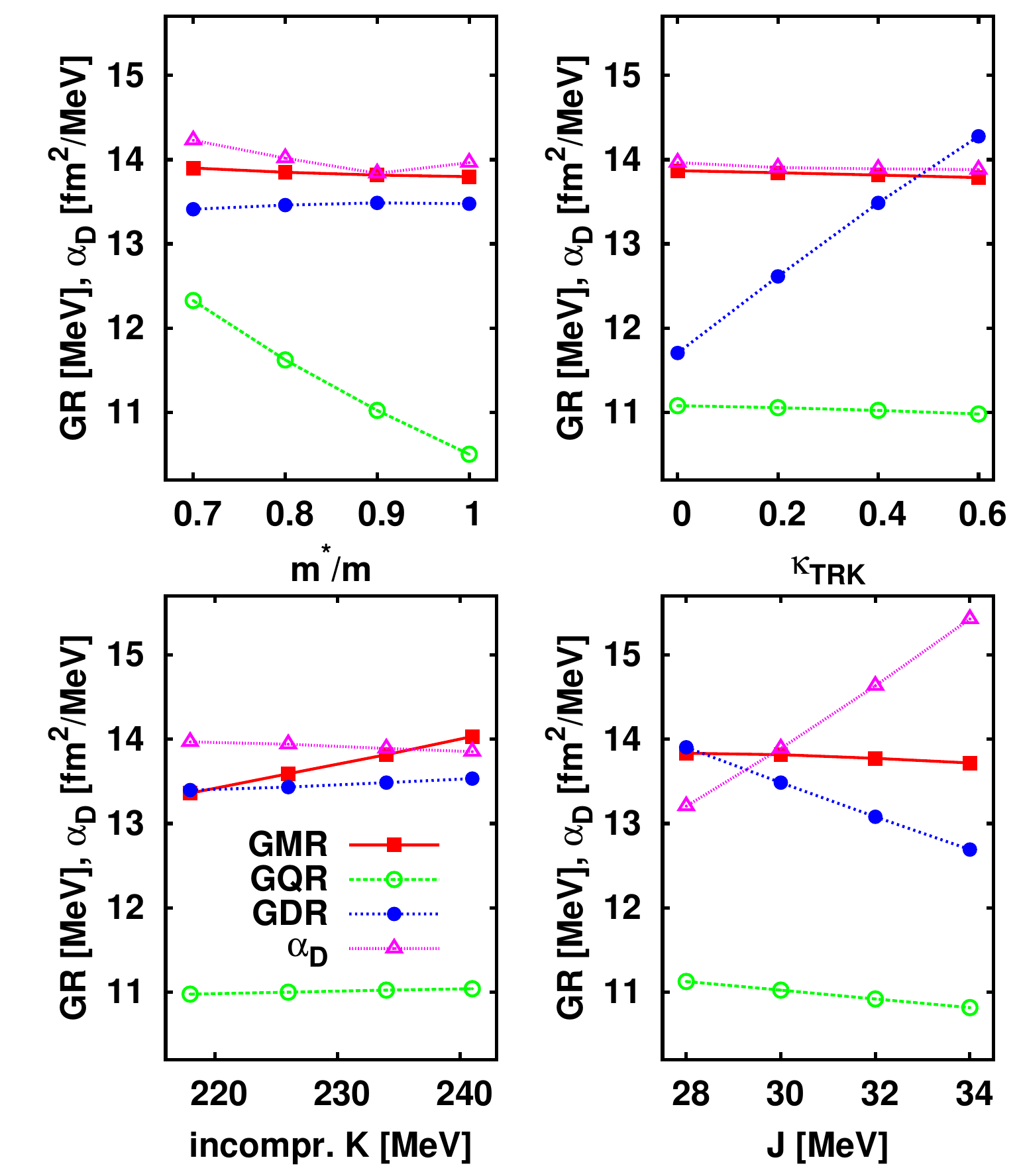}}
\caption{Peak energies of giant resonances (GMR, GDR, GQR) and dipole
  polarizability $\alpha_D$ in $^{208}$Pb for sets of parametrizations
  with systematically varied NMP from \cite{Kluepfel_2009} as
  indicated on the $x$-axes.  }
\label{fig:collect-GR}
\end{figure}
Figure \ref{fig:collect-GR} shows the trends of the four crucial
response properties in $^{208}$Pb with the four NMP. The lower left
panel shows the variation with $K$. Only the GMR moves with $K$ while
the other three response properties react only weakly. This makes it
obvious that the GMR is uniquely related to $K$. The upper left panel
demonstrates in the same way the close relation between $GQR$ and
$m^*/m$. And similarly, we see the unique relation between GDR and
$\kappa_\mathrm{TRK}$ in the upper right panel. The lower right panel
shows the trends with $J$. Here it is the dipole polarizability
$\alpha_D(^{208}\mathrm{Pb})$ which shows the strongest dependence.
There is some trend for the GDR, but this remains smaller than the
trend for $\alpha_D$ versus $J$ or for the GDR's dependence on
$\kappa_\mathrm{TRK}$. Thus we conclude from these four variations
that there is a one-to-one relation between the four response
properties and the four varied NMP.  These findings are confirmed even
more clearly by correlation analysis in section \ref{sec:correl} and
figure \ref{fig:alignmatrix} therein.

One may wonder why a variation of $L$, the slope of symmetry energy,
is not considered. The point is that $J$ and $L$ are highly correlated
with each other (covariance near 1) for the present SHF functional and
pool of fit data. Thus a plot versus $L$ shows the same trends as a
plot versus $J$. This feature may change for extended SHF functionals
which deliver more flexibility in the isovector part of density
dependence.  An example in that direction is the extension
(\ref{Esk-dens2}) which indeed reduces the correlation between $J$ and
$L$ as will be seen later in the middle panel of figure
\ref{fig:alignmatrix}.

\begin{figure*}
\centerline{\includegraphics[width=0.8\linewidth]{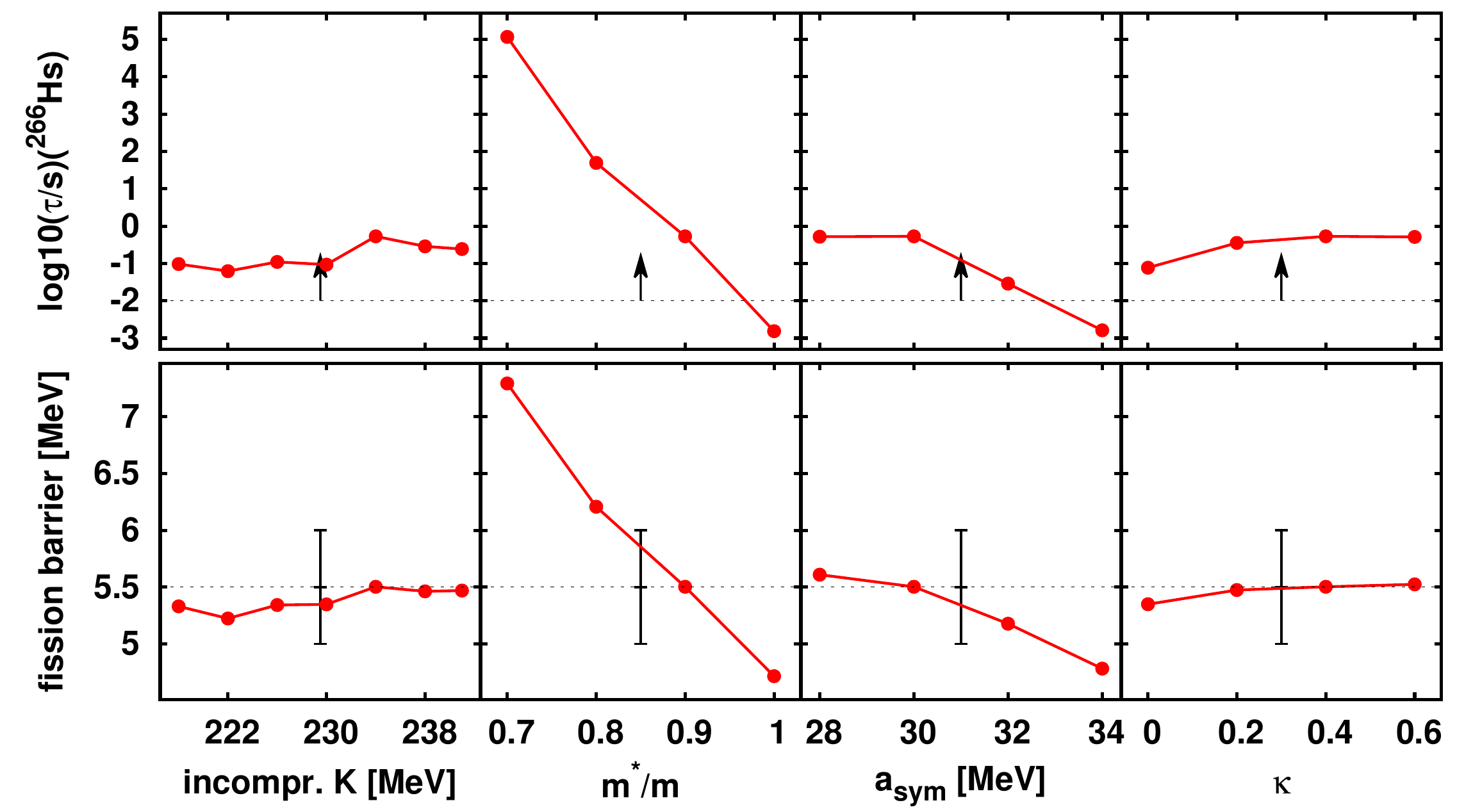}}
\caption{Fission barrier $B_f$ in $^{266}$Hs (lower panels) and
fission halflife $\tau_f$ (upper panels) for sets of
parametrizations with systematically varied NMP as indicated
on the $x$-axes. 
}
\label{fig:collect-fiss}
\end{figure*}
Figure \ref{fig:collect-fiss} shows the trends of fission barrier
$B_f$ and fission lifetime $\tau_f$ of $^{266}$Hs (for its definition
see section \ref{sec:finnuc}) for varied NMP. The dominant influence
comes here from $m^*/m$. This is not surprising because fission is
driven by shell effects and these are highly sensitive to $m^*/m$.
The figure is, however, to some extend misleading. It gives the
impression that fission is exclusively related to $m^*/m$. This view
is not confirmed by correlation analysis \cite{Bar15a,Erl14b} where
the covariance of $B_f$ with $m^*/m$ is about 0.2 which is some
correlation but not a dominant one. The reason that there are
several other agents which all have some impact on the fission path.  The
next larger influences come from the surface energy, spin-orbit term
(also acting through shell effects), and pairing. The example
demonstrates that systematic variation of force parameters is very
instructive to reveal influences, but should be augmented by
correlation analysis (section \ref{sec:correl} and point
\ref{it:correl} in section \ref{sec:estim}) to avoid premature
conclusions.

\begin{figure}
\centerline{\includegraphics[width=0.99\linewidth]{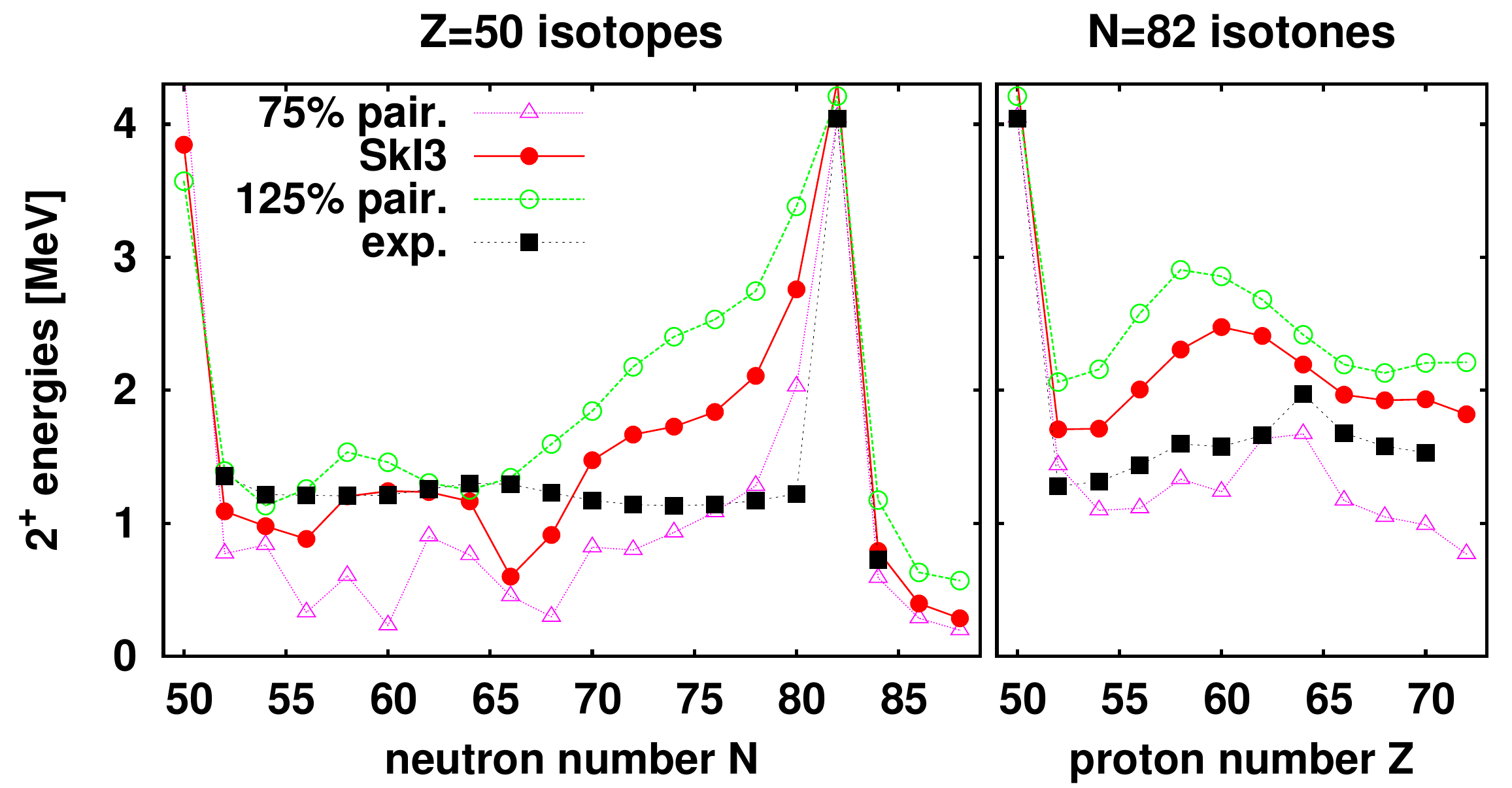}}
\caption{Energy of the low-lying $2^+$ state in the isotopic chain of
  Sn (Z=50) and the isotonic chain with N=82. Shown are results for
  SkI3 \cite{Rei95d} and a variation of pairing strength while keeping the SHF
  parameters of SkI3 fixed. 
  Data taken from \cite{Kluepfel_2008}.
}
\label{fig:Equad_ski3_DI_pairing}
\end{figure}
Figure \ref{fig:Equad_ski3_DI_pairing} shows the $E_{2^+}$ energies of
low-lying quadrupole states in the chain of even Sn isotopes and even
N=82 isotones.  We show here the effect of a variation of pairing
strength on the results.  Similar as for the fission path, the
collective path for large amplitude quadrupole vibrations is most
strongly influenced by pairing and grabs a piece of influence from
many other ingredients. Thus we show here only the strongest effect
and refer the reader to \cite{Kluepfel_2008} for the many more subtle
variations. First, we point out that SkI3 with
standard pairing provides satisfying results, particularly for Sn in
the mid shell region. We see again that the SHF functional is capable
to describe at once also large amplitude motion without having been
particularly fitted for that. Larger deviations occur near shell
closures that is the region where collectivity is not so large such
that the mapping of the collective path by a mere quadrupole
constraint becomes insufficient \cite{Kluepfel_2008}. One should
employ here variationally optimized paths as delivered by adiabatic
TDHF, see e.g. \cite{Rei87aR}. The effect of varied pairing is
dramatic showing that low-lying $2^+$ states are a sensitive probe for
the pairing functional. This is a feature which has not yet been fully
exploited in calibration strategies. The effect is particularly dramatic for
the case of reduced pairing (purple line with filled triangles).
The excitation energies shrinks occasionally to near zero. These are
cases where pairing breaks down and it shows that the standard values
of pairing strength are rather weak (i.e. close to the break-down
regime).

\subsection{Impact of ab-initio data}
\label{sec:abinit}

The nuclear many-body problem is much more involved than the
electronic one. The microscopic nucleon-nucleon interaction has a huge
repulsive core at short distances and yet one cannot ignore long-range
correlations from zero-sound modes. Moreover, it is not even correct
to speak of a nucleon-nucleon interaction because nucleons are
composite particles from quarks and gluons, and these constitute a
highly non-linear field theory. In spite of all these enormous
complications, much progress has been made with nuclear ab-initio
calculations during the last two decades
\cite{Epe09aR,Mac11aR,Ham13aR}. Time is coming close that we can refer
to ab-initio calculations in bulk matter, perhaps also in finite
nuclei \cite{Eks15a}.  Earlier attempts to use that for calibration of
mean-field models were promising, but not yet fully satisfying, see
e.g. the study in \cite{Erler_2011} on the basis of relativistic
Brueckner-Hartree-Fock data of \cite{Dal07a}. Ab-initio calculations
have still further improved since then. In particular, there is one
point which seems to be well settled. This is neutron matter at very
low densities. This stage is called a correlated Fermi gas (CFG).  In
the limit of low densities $\rho$, or Fermi momenta $k_F$
respectively, the CFG energy can be parametrized in terms of the
neutron kinetic energy as
\begin{equation}
  \frac{E}{N}_\mathrm{neut}(k_f)
  =
  \xi\frac{E}{N}_\mathrm{kin,neut}(k_f)
\label{eq:CFG}
\end{equation}
where $\xi$ is, in principle, function of $k_F$. We are interested
here in the limit $k_F\longrightarrow 0$ for which $\xi$ becomes
just one number. Many-body theory predicts a value
$\xi(0)=0.38--0.44$ for the CFG \cite{Cha03a,Ast04a}.

It is straightforward to compute $\frac{E}{N}_\mathrm{neut}$
for the SHF functional. With the constituents of the
total energy (\ref{eq:Etot}), we can deduce
$
\frac{E}{N}_\mathrm{neut}=
\left(\mathcal{E}_\mathrm{kin}+\mathcal{E}_\mathrm{Sk}\right)/\rho_0
$
and thus
\begin{equation}
  \xi
  =
  1+\frac{\mathcal{E}_\mathrm{Sk}}{\mathcal{E}_\mathrm{kin}}
  \quad.
\end{equation}
To estimate $\xi$ in the low-density limit, we recall
$k_F\propto\rho^{1/3}$ and discuss the limit in terms of $\rho$
because the SHF functional (\ref{eq:basfunct}) is given that way.  The
kinetic density in homogeneous matter becomes
$\mathcal{E}_\mathrm{kin}\propto\tau\propto\rho^{5/3}$ \cite{Mar10aB}.
The leading term in the SHF functional is the two-body contact
interaction $\propto\rho^2$. The density dependent term behaves as
$\propto\rho^{2+\alpha}$. Successful parametrizations require
$\alpha>1$. This term, therefore, vanishes faster than $\rho^2$ in the
limit $\rho\rightarrow 0$ and can be neglected. The kinetic
interaction term behaves as $\rho\tau\propto\rho^{8/3}$ which, again,
can be neglected. It remains
$\xi\propto 1\!+\!\rho^2/\rho^{5/3}\longrightarrow{1}$ in the limit
$\rho\longrightarrow 0$. Thus we see in purely
analytical manner that the standard SHF-functional (\ref{eq:basfunct})
cannot meet at all the requirement $\xi(0)\approx{0.42}$.

In order to allow for a description of $\xi<1$, we extend the SHF
functional by adding to $\mathcal{E}_\mathrm{Sk}$ another density
dependent term
\begin{eqnarray}
  \mathcal{E}_\mathrm{Sk,dens3}
  &=&
   \left[
     C_0^{\rho,-1/3}  \rho_0^{2}
     +C_1^{\rho,-1/3} \rho_1^2
    \right]\rho_0^{-1/3}
  \;.
\label{Esk-dens2}
\end{eqnarray}
It has the form of the density dependent term in the functional
(\ref{eq:basfunct}) with $\alpha=-1/3$. Handled as additional term, it
allows to maintain the quality of the standard ansatz while opening
the chance to accommodate the CFG limit.

\begin{figure}
\centerline{\includegraphics[width=0.99\linewidth]{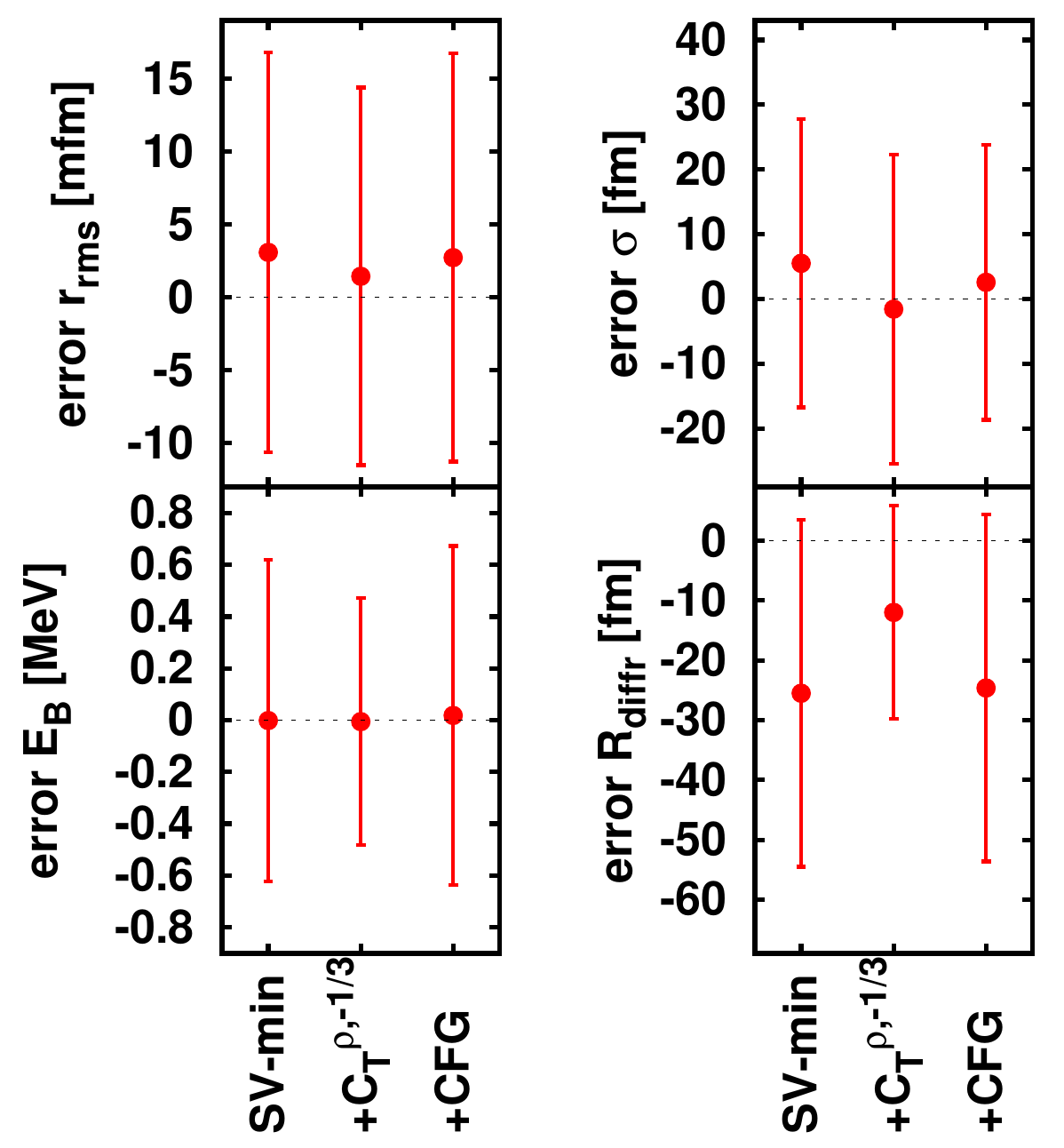}}
\caption{Average and r.m.s. errors on basic bulk properties, binding
  energy $E_B$, charge r.m.s. radius $r_\mathrm{rms}$, charge
  diffraction radius $R_\mathrm{diffr}$, and charge surface thickness
  $\sigma_\mathrm{surf}$ for the series of fits with the additional density dependent
  term (\ref{Esk-dens2}) and four NMP fixed.
  Averages are taken over the pool of fit
  nuclei. The average deviation from data is shown by filled boxes and
  the r.m.s. deviation by the error bars around the boxes.
  The series is: ``SV-min'' = fit of standard functional to full data pool from
  \cite{Kluepfel_2009}; ``+$C_T^{\rho,-1/3}$'' = fit as SV-min but
  including the new term (\ref{Esk-dens2}) in the functional;
  ``+CFG'' = fit as ``+$C_T^{\rho,-1/3}$'' but including
  the CFG limit as data point.
 }
\label{fig:collectLD-errorblocks}
\end{figure}
We have produced two new parametrizations with the extended
functional, now including also the term (\ref{Esk-dens2}).  Starting
point is here, again, SV-min, a fit to the standard pool of fit data.
The first new set is ``+$C_T^{\rho,-1/3}$'' and uses the same data as
SV-min. The second new set ``+CFG'' adds to these data the CFG condition
$\xi(0)={0.42}$ as further data point.

Figure \ref{fig:collectLD-errorblocks} shows the resulting average and
r.m.s. errors for the basic blocks of observables. The averages stay
at zero level throughout for $E_B$ and vary somewhat for radii and
surface thickness. The fit ``+$C_T^{\rho,-1/3}$'' with more freedom in
density dependence, yet without enforcing the CFG point, leads always
to lower average and r.m.s. errors. This confirms the impression
gained in previous section that the slight conflict between energy and
formfactor calls for improved density dependence. However, the step to
``+CFG'' deteriorates again radii and surface thickness in the average
approximately back to the status of SV-min. The freedom coming along
with the new term has been fully exploited by the CFG point.

\begin{figure}
\centerline{\includegraphics[width=0.99\linewidth]{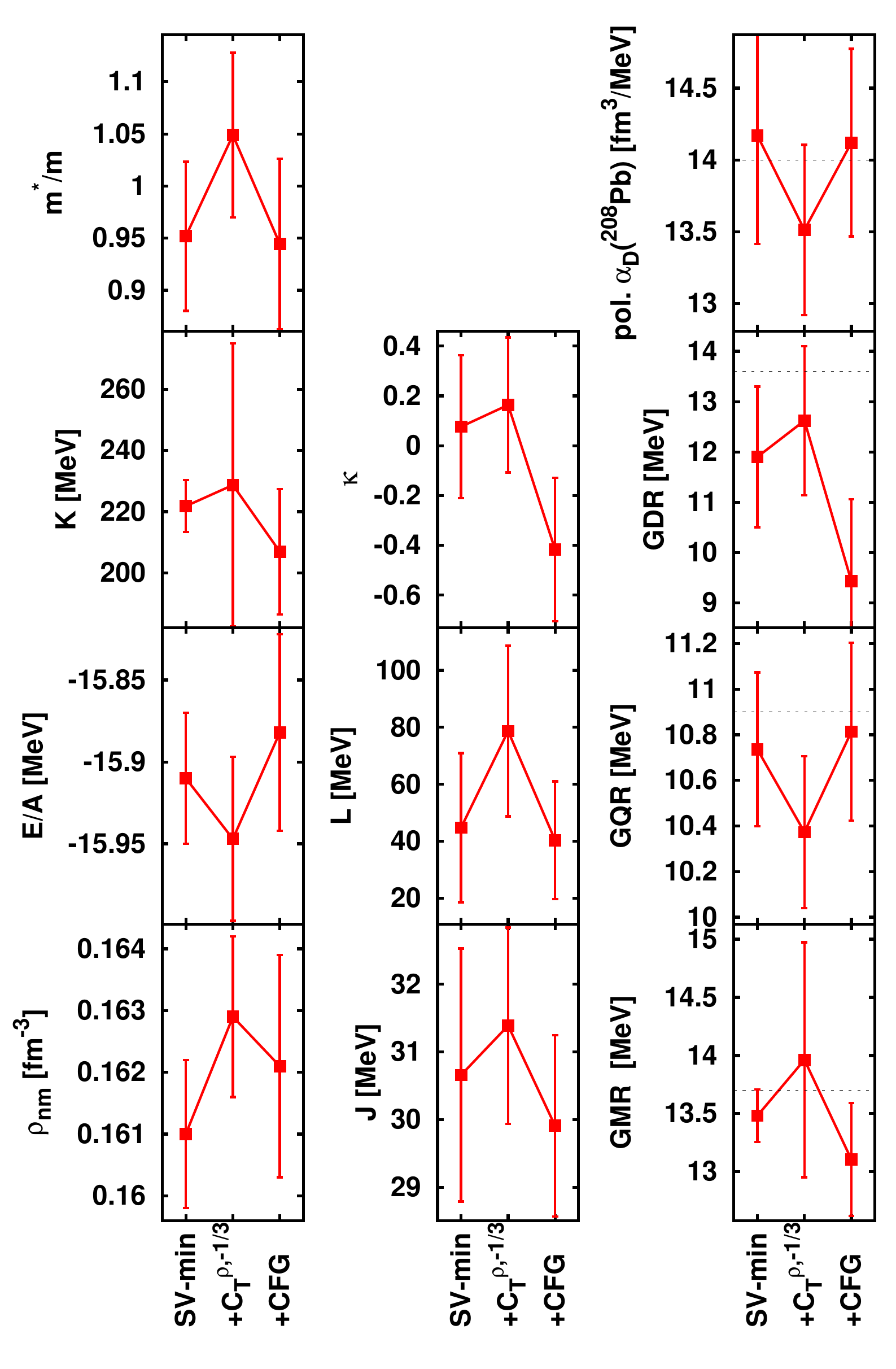}}
\caption{Results on NMP (left two columns) and
response properties of $^{208}$Pb (right column) for the
series of three forces as in figure \ref{fig:collectLD-errorblocks}.
}
\label{fig:collectLD-NMPfin}
\end{figure}
Figure \ref{fig:collectLD-NMPfin} shows NMP and response properties in
$^{208}$Pb for the parametrizations introduced in this section.  There
are marked changes in the results from the three parametrizations. But
they are all compatible within the error bars.  Nonetheless, it may be
instructive to discuss the changes in the center values. Similar to
the previous figure, we find that the results of SV-min and ``+CFG''
are closer together while ``+$C_T^{\rho,-1/3}$'' makes generally the
largest changes. Exceptions are the pair $\kappa_\mathrm{GDR}$ with
GDR and $K$ with GMR where ``+CFG'' produces too low values. The strong impact
of the CFG point is plausible for $K$ and GMR because these
observables have direct relation to density dependence. The fact that
the CFG point pulls the GMR away from the data indicates that the
extended model is not yet ideal, although the data may be corrected by
fitting with constraint on $K$ in the manner as was done for SV-bas
\cite{Kluepfel_2009}. It is also noteworthy that the error bars for
the free fit of the extended density dependence in
``+$C_T^{\rho,-1/3}$'' produces particularly large uncertainties for
$K$ and GMR, the two observables sensitive to density dependence.
This large uncertainty is much reduced by the step to ``+CFG'' which
indicates once more that the CFG point has a strong impact on the
density dependence of the model.  

The example is also quite revealing
as it shows that some long-range correlations (at least those at
extremely low density) are not incorporated into the SHF
functional. This is annoying because these corralations are smooth
with mass number and thus should be contained in a smooth
functional. Again, we see that we need to improve on the density
dependence of the SHF functional.

\subsection{Correlation analysis}
\label{sec:correl}

In section \ref{sec:errors}, correlations, or covariances
respectively, between observables had been defined with
eq. (\ref{eq:correlator}). In this section we will give a few examples
for a small, but relevant, selection of observables.  The selection
covers: the four NMP $K$, $m^*/m$, $J$, $\kappa_\mathrm{TRK}$ together
with the four related response properties in $^{208}$Pb, namely GMR,
GDR, GQR, and $\alpha_D$; for static isovector response we look also
at $L$, neutron skin $r_\mathrm{skin}(^{208}\mathrm{Pb})$, and slope
of the neutron equation of state $\partial_\rho
E/A_\mathrm{neut}\Big|_{\rho=0.1}$; finally we check extrapolations to
exotic nuclei, $E_B$ for the very neutron rich $^{140}$Sn and $E_B$ as
well as $Q_\alpha$ for the super-heavy element Z=120/N=182.
Three parametrizations will be considered: SV-min as the standard, ``E
only''from section \ref{sec:groups} to exemplify the impact of the
choice of fit data, and ``+$C_T^{\rho,-1/3}$'' from section
\ref{sec:abinit} to explore the impact of an extended functional.

\begin{figure}
\centerline{\includegraphics[width=0.93\linewidth]{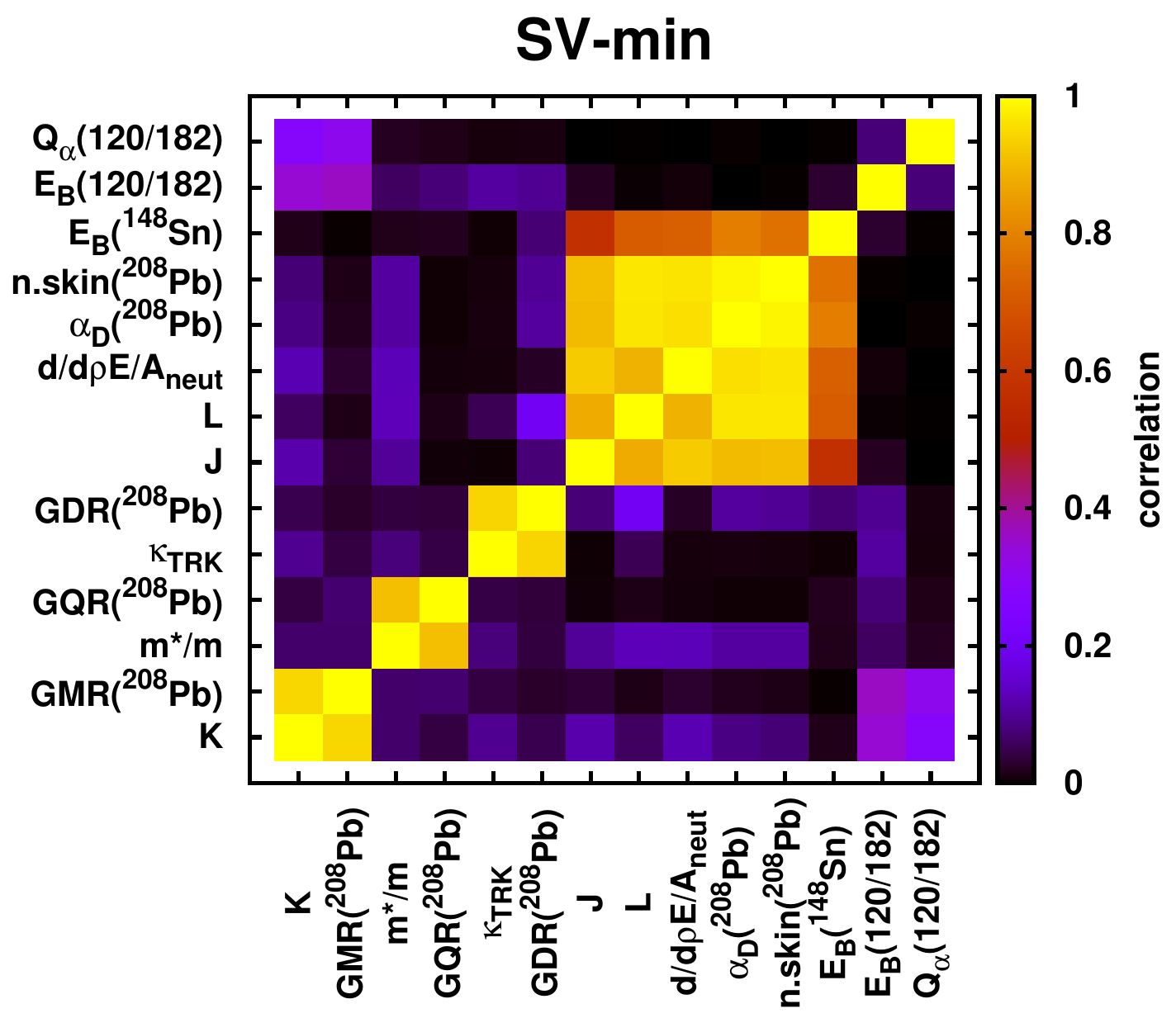}}
\centerline{\includegraphics[width=0.93\linewidth]{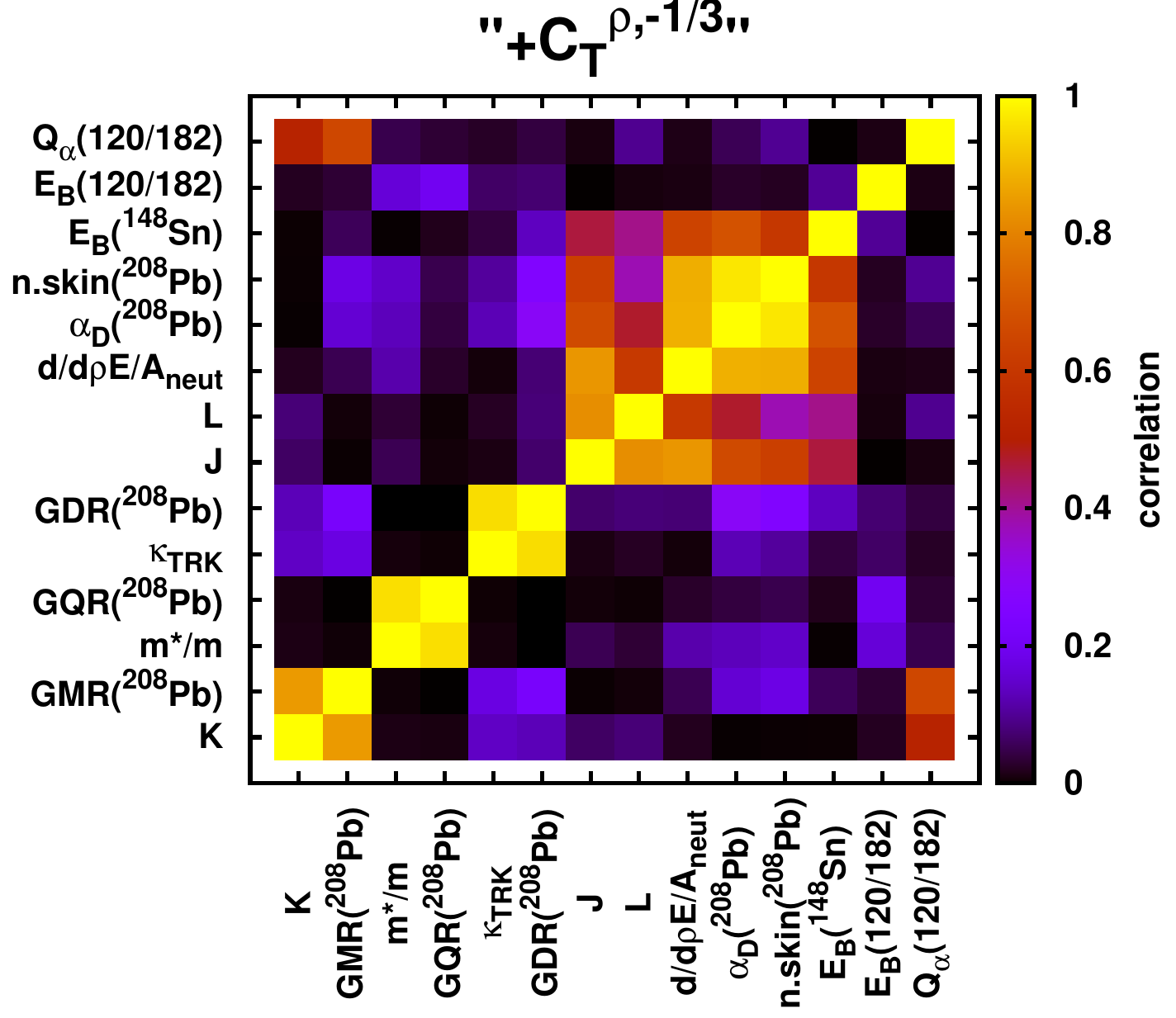}}
\centerline{\includegraphics[width=0.93\linewidth]{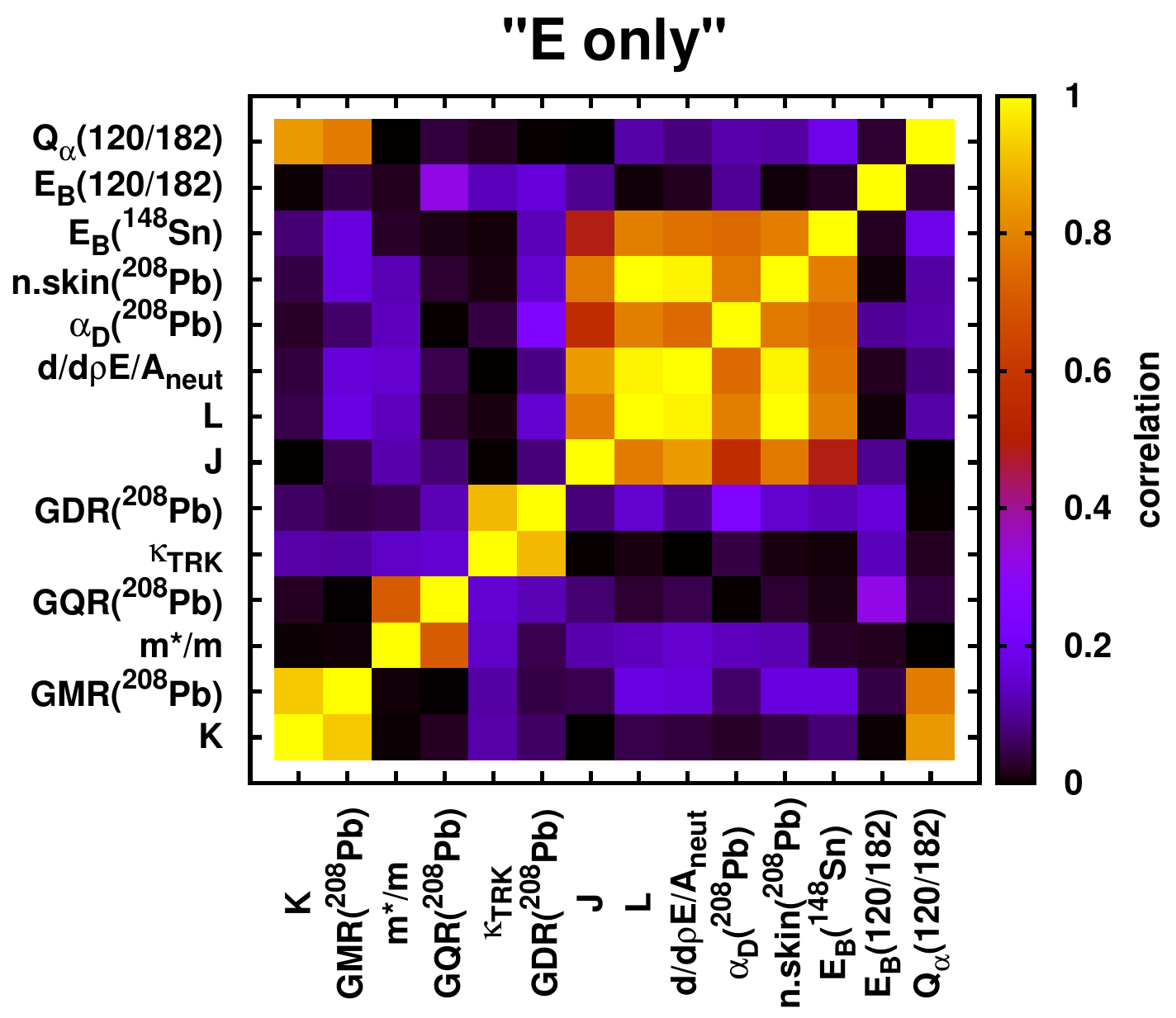}}
\caption{Correlation matrix for a selection of observables as
  indicated.  Three cases are shown: SV-min = fit of standard SHF
  functional to the standard pool of data from \cite{Kluepfel_2009},
  ``+$C_T^{\rho,-1/3}$'' = unconstrained fit as SV-min but with the
  functional extended by the density dependent term (\ref{Esk-dens2}),
  ``$E$ only = fit of standard functional to standard data pool
  without radii and surface thickness (see section \ref{sec:groups}).
}
\label{fig:alignmatrix}
\end{figure}
Figure \ref{fig:alignmatrix} shows the correlations between the
selection of observables as indicated on the axes. As every observable
can be related to any other, we show it as correlation matrix. Let us
start with the parametrization SV-min, the straightforward fit to the
standard pool of data \cite{Kluepfel_2009}.  We see clearly the strong
correlation between $K$ and GMR (static isoscalar response), $m^*/m$
and GQR (dynamic isoscalar response), $\kappa_\mathrm{TRK}$ and GDR
(dynamic isovector response), and for the whole group of static
isovector response covering here the NMP $J$, $L$,
$d/d\rho{E}/A\Big|_\mathrm{neut}$ (see section \ref{sec:NM}) together
with the finite nuclei observables $\alpha_D$ and $r_\mathrm{skin}$
(labeled ``n.skin'' in the figure).  For the exotic nuclei, there is
some correlation of $E_B(^{140}\mathrm{Sn})$ to isovector static
response which is plausible for a extremely neutron rich nucleus with
very large isospin.  The two data in the super-heavy element
Z=120/N=182 are nearly decoupled from the other groups of data. One
may admit some correlation with $K$\&GMR. This super-heavy region is
dominated by shell effects and not so much by the bulk properties
included in this correlation matrix. It is also interesting to remark
that $E_B$ and $Q_\alpha$ are almost uncorrelated. Differences of
energies often filter quite different aspects of a model than the
energy as such.

We now look at the correlations produced by the next two
parametrizations, with less data in ``$E$ only'' and with more terms
in the model in ``+$C_T^{\rho,-1/3}$''. In both cases, we see
generally reduced correlations what we could have expected. It is,
however, remarkable how small the reduction is leaving the gross
structure of the correlations fully alive. There remain clearly the
blocks for the four different responses. Larger changes are seen for
isovector static block. And here, the changes are very different for
the two different variations of the fit. For example, while ``$E$
only'' degrades the correlations for $J$ thus making $L$ the leading
bulk parameter for isovector static response, the other parametrization
``+$C_T^{\rho,-1/3}$'' degrades just $L$ and leaves $J$ more
intact. In such details, we see indeed a dependence on model and data
pool. But recall that the general structure of correlations was
robust. 

There is another interesting detail concerning the super-heavy element
Z=120/N=182. The independence of $E_B$ and $Q_\alpha$ persists. But
there is now more correlation between $Q_\alpha$ and $K$\&GMR than was
seen for SV-min. This shows that correlations are not necessarily
reduced by less data or a large model. It can also happen that more
freedom allows to explore connections which were hindered before by a
too rigid model.


\subsection{Spin sensitive applications: spin modes, odd nuclei, TDHF}
\label{sec:spin}

This section is devoted to the spin properties of SHF. Spin properties
are crucial for a broad range of applications. They play a role
already for ground state properties, namely in case of odd nuclei, see
e.g. \cite{Pos85a,Ben87b,Dug02a,Pot10a}. They are key
players in spin excitation modes, see
e.g. \cite{Ben02c,Cao09a,Ves09a}, they play a role in rotating nuclei,
see e.g.  \cite{Fle79a,Dob95c}, and they can contribute to energetic
heavy-ion collision described by time-dependent Hartree-Fock (TDHF)
with the SHF functional, for an example where the impact of a spin
excitation is discussed explicitly see \cite{Mar06a}. Unfortunately,
the spin terms are presently the least well controlled part of the SHF
functional. Their proper calibration is yet an open problem and they
are most prone to introduce instabilities into the SHF equations, for
a critical analysis see \cite{Ben07a,Ben09c,Dav09a,Hel13a}.  The
development of the spin branch in SHF is still in its first stages,
this means in the language of section \ref{sec:estim} at the
preparatory level of point \ref{it:theo}, although one can find also
some steps of the error analysis according to point \ref{it:dedicvar}
and sub-points \ref{it:residuals} together with \ref{it:predict}
therein. This latter aspect will shine up briefly in the subsequent
example of odd nuclei and of TDHF. But generally, the aim of the
present section is mainly to demonstrate the capabilities of SHF
to deal also with spin properties.

Before going on, we have to look at the spin aspects of the basic SHF
functional (\ref{eq:basfunct}), so far not discussed. Some spin terms,
namely the time odd spin-orbit terms $\propto C^{\nabla J}_T$ in 
the time-odd part of the functional (\ref{eq:ESkodd}), are fixed
by the requirement of Galilean invariance. Yet open remain the
parameters of the genuine spin terms $C^{\sigma}_T$ and the tensor
spin-orbit terms $\propto C^J_T$.
\begin{table}
\begin{center}
\begin{tabular}{|c|}
\hline
  $
  \begin{array}{rcl}
  C_0^{s}
  &=&
  -\ffrac{1}{3}\left(3C_1^\rho+2C_0^\rho\right)
  \;,
\\[4pt]
  C_1^{s}
  &=&
  -\ffrac{1}{3}C_0^\rho
  \;,
\\[12pt]
  C_0^{\Delta s}
  &=&\displaystyle
  -\frac{C_0^{\Delta\rho}\!-\!C_1^{\Delta\rho}}{2}
  +\frac{C_0^\tau\!-\!C_1^\tau}{8}
  \;,
\\[4pt]
  C_1^{\Delta s}
  &=&\displaystyle
    \frac{C_0^\tau\!-\!C_1^\tau}{8}
    -\frac{C_0^{\Delta\rho}\!+\!C_1^{\Delta\rho}}{6}
  \;,
\\[16pt]
  C_0^J
  &=&
  \ffrac{1}{12}\left(C_0^\tau-3C_1^\tau
                     -16C_1^{\Delta\rho}\right)
  \;,
\\
  C_1^J
  &=&
  -\ffrac{1}{12}\left(C_0^\tau-3C_1^\tau\right)
  \;,
  \\[12pt]
  C_1^{\nabla J}
  &=&
  C_0^{\nabla J}
  \;.
  \end{array}
  $
\\
\hline
\end{tabular}
\end{center}
\caption{\label{tab:restrict}
Restrictions on the parameters of the general SHF functional
(\ref{eq:basfunct}) in case that the functional is derived from the
Skyrme force ansatz (\ref{eq:SHFforce}).
}
\end{table}
From a strictly density functional perspective, one can consider these
all as free parameters of the model.  This leaves still a large number
of 7 more free parameters whose independent calibration would require
access to a great amount of reliable, spin-sensitive
data. Alternatively, one can approach the problem from a formal side.
The density-matrix expansion sketched in section \ref{sec:motivateSHF}
yields not directly an energy functional, but rather a zero-range
effective interaction, the much celebrated Skyrme force. For its
detailed form and a critical discussion of the notion of an effective
interaction see appendix \ref{app:force}.  Deriving the SHF functional
from the Skyrme force yields a well defined connection between the
spin terms and the leading terms, see table \ref{tab:restrict}.
These restrictions on the spin-orbit parameters from the force
definition of the SHF functional are often discarded. Many
parametrizations simply ignore the tensor spin-orbit terms, thus
supposing $C_0^J=0$ and $C_1^J=0$.  There are also reasons to override
the restriction on the spin-orbit parameter 
$C_1^{\nabla J}=C_0^{\nabla J}$ \cite{Rei95d}. All functionals used in this paper
ignore the tensor spin-orbit terms and treat $C_1^{\nabla J}$ as
independent parameter.  The spin gradient terms $\propto{C}_T^{\Delta
  s}$ when taken as given in table \ref{tab:restrict} lead for most
functionals to dramatic instabilities. They must be discarded. What
remains then from the force concept is the parameter connection from
Galilean invariance and the determination of the pure spin terms
$\propto{C}_T^\sigma$. Galilean invariance is the unavoidable minimum
requirement and will always be used. Above that we have the freedom to
employ the spin and/or tensor spin-orbit terms.  Although one cannot
take the full ``force'' terms consistently into account, it is
convenient to exploit it for the spin terms and the tensor spin-orbit
part yielding the connections as given in table \ref{tab:restrict}. As
this is an ad-hoc decision, justification can come only from
experience.

\begin{figure}
\centerline{\includegraphics[width=0.99\linewidth]{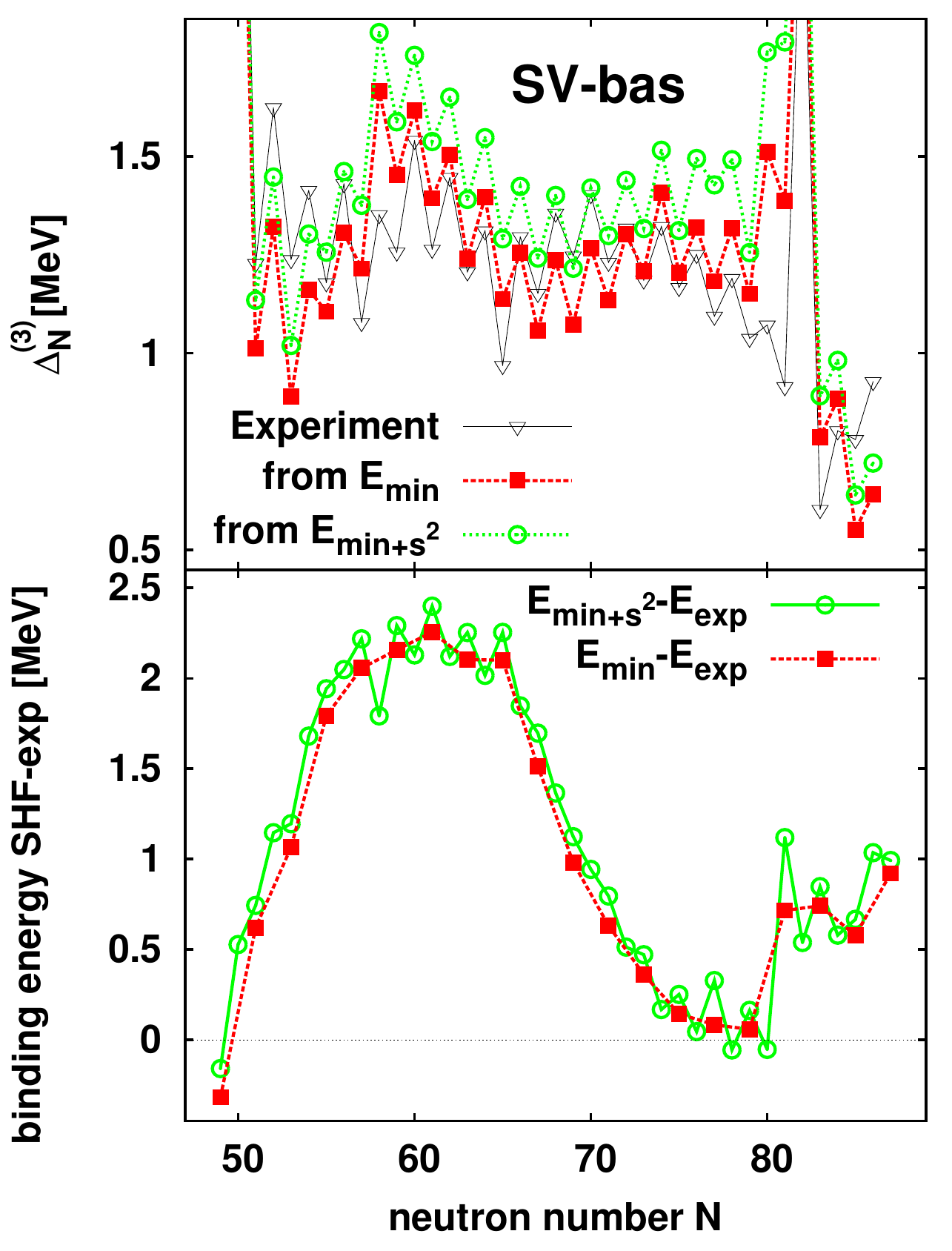}}
\caption{ Difference of theoretical and experimental binding energies
  (lower panel) and neutron odd-even staggering $\Delta^{(3)}_N$
  (upper panel) along the isotopic chain of Sn nuclei computed with
  the parametrization SV-bas \cite{Kluepfel_2009}.  The
  $E_\mathrm{min}$ stands for the minimal Galileo invariant functional
  and $E_{\mathrm{min},\sigma^2}$ for the the functional including
  additionally the $\sigma^2$ terms according to the force definition,
  see table \ref{tab:restrict}.  Data extracted from \cite{Pot10a}.  }
\label{fig:odd-Sn}
\end{figure}
As a first example, we address odd nuclei picking a key result from
\cite{Pot10a}.  Figure \ref{fig:odd-Sn} shows the deviation of binding
energies from experimental values and the odd-even staggering
(\ref{eq:oddeven}) along the chain of Sn isotopes. Two stages of the
functional are considered, the minimal Galilean invariant form
$E_\mathrm{min}$ corresponding to $C_T^\sigma=0$ and the inclusion is
the $\vec{\sigma}_T^2$ terms with parameters as given in table
\ref{tab:restrict}. The trend of the binding (lower panel) is
generally very similar for odd and even isotopes. There is a good
reproduction in the region $N=70-82$.  The larger deviations mid shell
indicate that these nuclei require collective correlations (not
included here) for a proper description. Odd nuclei make no difference
in that respect.  Of course, both variants yield the same values for
even isotopes because the time-odd terms are inactive there. It
remains to look at differences for odd isotopes.  The effect of the
spin terms is small, of the order of 0.2 MeV and in the majority of
cases enhancing slightly the deviations.  Thus the results for binding
energies give a slight preference for the $E_\mathrm{min}$ model.

The upper panel of figure \ref{fig:odd-Sn} shows the odd-even
staggering.  It is very satisfying that the parametrization SV-bas
provides at once a satisfying reproduction of the data. At a more
detailed level, we see differences between the two choices for the
spin terms. There are cases where $E_{\mathrm{min},\sigma^2}$ performs
better (particularly at the side of small $N$). But, again, we find
that the majority of isotopes looks better with the minimal Galilean
invariant choice $E_\mathrm{min}$.  This slight preference for
$E_\mathrm{min}$ was also found for other examples and other
parametrizations \cite{Pot10a}. But the differences are probably too
small to be decisive and more studies are still necessary to settle
the case. Recall that the most important result of this brief
demonstration is that the SHF functional yields immediately an
acceptable description also of odd nuclei.

\begin{figure}
\centerline{\includegraphics[width=0.7\linewidth]{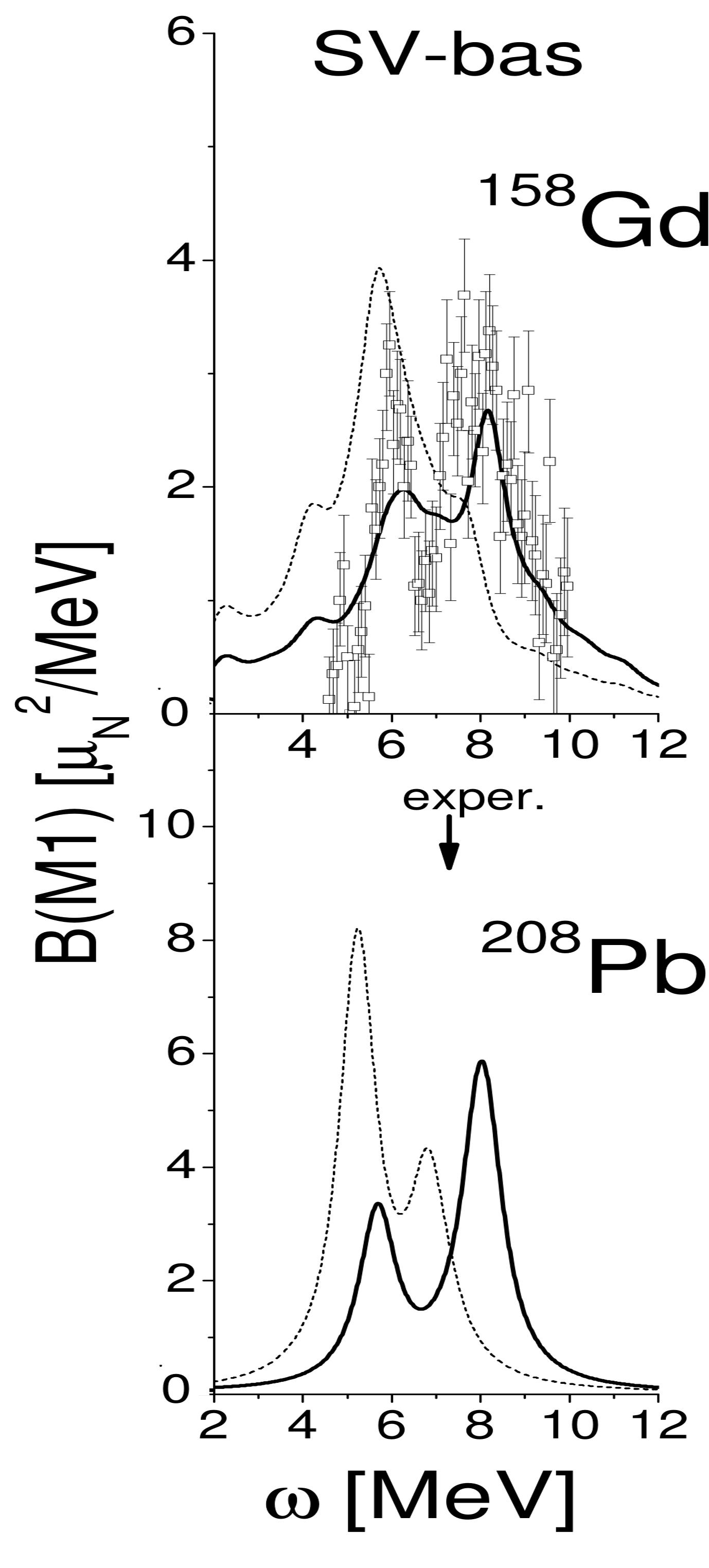}}
\caption{Isoscalar M1 strength distribution for the spin-flip mode in
  deformed $^{158}$Gd and the spherical $^{208}$Pb computed with
  separable RPA using the parametrization SV-bas.  The RPA results are
  shown by heavy line and the strength distribution from pure
  two-quasiparticle excitations (from static mean field only) by faint
  lines. The experimental distribution is shown for $^{158}$Gd. The
  remarked experimental resonance peak is indicated by an arrow for
  $^{208}$Pb.  Adapted from \cite{Nes10b}.  }
\label{fig:spinflip}
\end{figure}
Figure \ref{fig:spinflip} shows as next example a result for spin
excitations, here the M1 strength distribution of the spin-flip mode,
computed including the spin terms of te SHF functionals.
Two rather different nuclei are considered, the well deformed
mid-shell nucleus $^{158}$Gd and the spherical, doubly magic
$^{208}$Pb. The RPA result is shown together with the mere
two-quasiparticle excitations (i.e. without the dynamical effects from
the RPA residual interaction \cite{Mar10aB}). One sees a marked
upshift of strength by RPA (also called collective shift) due to the
repulsive residual interaction in the isoscalar spin channel.  This
upshift brings indeed the strengths into the right energy range.  Thus
far we see once more that a proper SHF functional is also appropriate
for spin modes, at least what gross features is concerned. Looking
closer at the case, we see a problem. The experimental distribution
differs between deformed and spherical nuclei. It has two pronounced
peaks for the deformed $^{158}$Gd but only one for the spherical
$^{208}$Pb. The theoretical result shows a two-peak structure for both
cases. This has been studied extensively in \cite{Ves09a,Nes10b} for a
broad selection of SHF parametrizations. There are some
parametrizations which produce one single peak for $^{208}$Pb, but
then one peak only also for $^{158}$Gd. With presently available
functionals it seems impossible to describe both nuclei equally
well. The source of the problem lies most probably in an incomplete
inclusion of spin-orbit terms. A satisfying solution has yet to be
worked out.

Finally we demonstrate the enormous capabilities of self-consistent
mean field theories by examples from time-dependent Hartree-Fock
(TDHF). TDHF allows to simulate a wide range of nuclear processes
covering resonance dynamics, non-linear excitations, and particularly
large amplitude motion with substantial rearrangement of structure in
the course of time as it occurs, e.g., in heavy-ion reactions.  First
nuclear TDHF calculations came up already shortly after the appearance
of static SHF calculations \cite{Bon76a}.  Two decades ago, TDHF has
been revived because computing capabilities had evolved sufficiently
far to allow realistic large-scale calculations, for examples of a
couple of very different applications see
\cite{Kim97a,Sim03a,Nak05a,Mar06b}. A combination of TDHF with
density-constrained Hartree-Fock even allows to compute sub-barrier
fusion cross sections \cite{UO09b,Obe10a}. Stepping forward to still
larger scales, TDHF is now being applied to simulate matter under
astro-physical conditions in neutron stars and super-nova explosions
\cite{Uma14a,Sch14a}. For recent reviews on TDHF with SHF functionals
see \cite{Sim12aR,Mar14aR}. In all these applications, one found that
SHF functionals provide at once a fair description of the processes
studied. There are open ends though because spin terms can play a role
in energetic collisions and we have seen above that just these terms
are not yet well settled. We will exemplify in the following both, the
capabilities and the uncertainties of TDHF simulations.

\begin{figure}
\centerline{\includegraphics[width=0.8\linewidth]{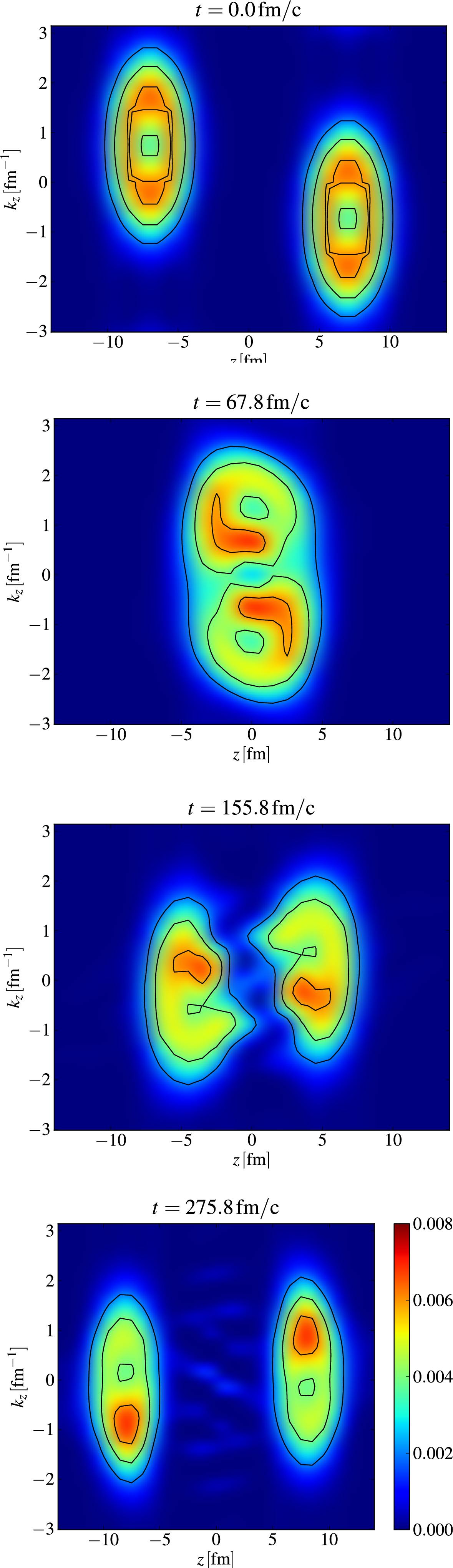}}
\caption{Snapshots of the Wigner distribution in the $z$-$k_z$-plane
 from a TDHF simulation of a central $^{16}$O+$^{16}$O collision using
 the parametrization SkI3 \cite{Rei95d}. Adapted from \cite{Loe11a}.
 }
\label{fig:HI-coll}
\end{figure}
Figure \ref{fig:HI-coll} shows snapshots of the Wigner
function from a TDHF simulation of a central collision $^{16}$O+$^{16}$O
\cite{Loe11a}. The simulation was performed on a Cartesian 3D mesh,
for details of the physics and numerics of TDHF and the 3D code see
\cite{Mar14aR}.  The Wigner function is constructed from the one-body
density matrix $\varrho(\vec{r},\vec{r}')$ by Fourier transforming
with respect to $\vec{y}=\vec{r}-\vec{r}'$, i.e.
$
f(\overline{\vec{r}},\vec{k})=
\int d^3y e^{\mathrm{i}\vec{k}\cdot\vec{y}}
\varrho(\overline{\vec{r}}+\vec{y}/2,\overline{\vec{r}}-\vec{y}/2).
$
Originally introduced in \cite{Wigner}, it is the attempt to provide a
phase-space picture of a quantum state, for details see
\cite{Bra97bB}.  The Wigner function has the weakness that it is not
positive semi-definite, thus preventing a strict probabilistic
interpretation. Nonetheless, it can serve as a useful illustration of
a dynamical process and the availability of all detailed wavefunctions
in a TDHF calculation allows us to produce this insight.  The Wigner
function in a 3D calculation is six-dimensional and thus rather
difficult to handle.  We thus look at a 2D cut through the 6D
function, namely along the reaction axis $z$ and associated wavenumber
$k_z$.  The snapshots are sorted from above to below in order of
increasing time. The initial stage shows nicely the two Fermi spheres
of the $^{16}$O nuclei, unperturbed except for the c.m. boost to
positive $k_z$ for the left nucleus and to negative $k_z$ for the
right one.  The second panel shows the situation at the point of
closest contact.  Although the corresponding local density represents
a compact compound system, the phase space picture shows that the
compound stage has not lost its memory at the initial state. There are
two distinct fragments in phase space. This feature persists to the
next plot the fragments separating. They are strongly perturbed, but
kept their identity from the initial nuclei. And in the final stage
(lowest panel), two fragments are flying apart, internally excited and
with reduced c.m. motion, but still keeping the two-fragment structure
of the initial state. This indicates that pure mean-field motion does
not allow full melting into one thermal compound state. This becomes
possible only if two-body collisions are added to the description.
This is achieved in the semi-classical framework by the
Vlasov-Uehling-Uhlenbeck scheme \cite{Ber88}. Adding a collision term
to quantum mechanical TDHF turns out to be much more involved,
although just recently promising attempts have come up \cite{Sur14a}.
This example indicates that TDHF dynamics is likely to underestimate
dissipation processes. It still provides relevant guidelines for the
gross structure of a process and can be even quantitatively correct in the
entrance state of a heavy-ion reaction where nucleon-nucleon
collisions did not yet have a chance to play a role. For example, one
can describe very well fusion cross sections with TDHF, see
e.g. \cite{UO09b}.

\begin{figure}
\centerline{\includegraphics[width=0.9\linewidth]{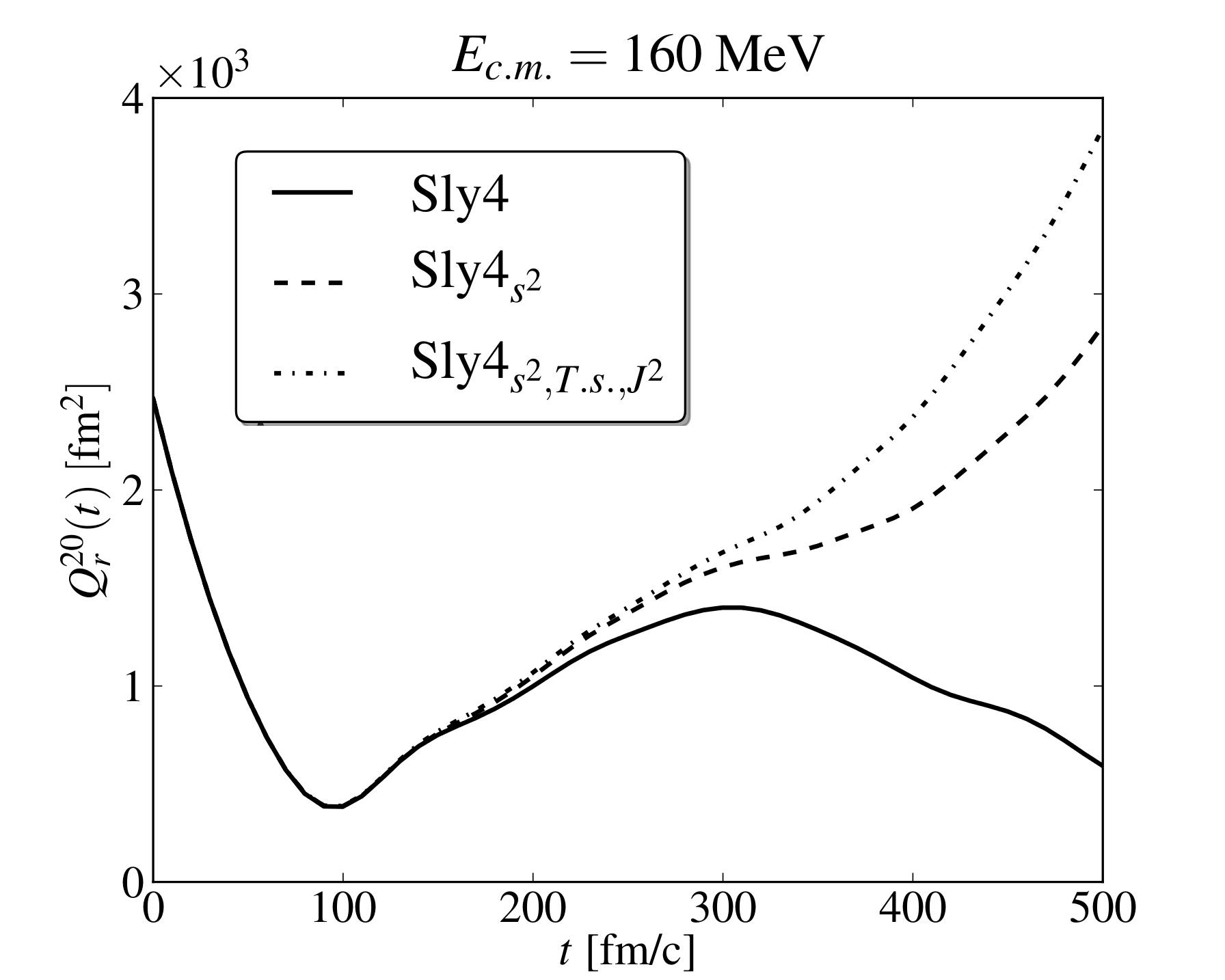}}
\caption{
Time evolution of the quadrupole momentum $Q_{20}$ for a central
collision $^{40}$Ca+$^{40}$Ca computed for the parametrization SLy4
\cite{Cha98a}
with three different options for the spin terms as indicated.
 Adapted from \cite{Loe12a}.
 }
\label{fig:TDHF-Q20}
\end{figure}
Figure \ref{fig:TDHF-Q20} shows the result from a TDHF calculation for
a central collision of $^{40}$Ca+$^{40}$Ca at relative center of mass
energy 240 MeV \cite{Loe12a}. Shown is the quadrupole momentum along
the reaction axis as function of time. This quantifies the relative
distance of the fragments and illustrates the type of reaction as well
as damping of c.m. motion. In this case, three different choices for
the SHF functional are compared: The parametrization SLy4 just with
the minimal Galilean invariant terms, SLy4 with the $\vec{\sigma}^2$
terms included, and SLy4 with $\vec{\sigma}^2$ plus tensor spin-orbit
terms included. Unlike the case of ground states of odd nuclei, we see
in this highly dynamical case a great sensitivity to the spin and
spin-orbit terms. Their inclusion turns a fusion dynamics (heavy line
for SLy4 without spin terms) to inelastic scattering which means that
they reduce dissipation. It ought to be mentioned, however, that the
leading spin orbit term $\propto{C}_T^{\nabla J}$ acts in different
direction, significantly enhancing dissipation
\cite{Lee87,Rei88d}. Thus there is a very subtle interplay of the
various terms in the SHF functional.  With the new generation of
computers and codes, one will now be in a position to systematically
investigate the intrinsic structure of heavy-ion reactions. And, as in
all previous example of this section, we emphasize that the SHF
functional delivers in general a relevant picture of structure and
dynamics also for cases where spin plays a role. This is a safe basis
for then going into details.

\section{Concluding remarks}

The Skyrme-Hartree-Fock (SHF) model for a self-consistent mean-field
description of nuclear structure and dynamics has been inspected from
different perspectives: formal background, range of applicability,
performance, estimates of extrapolation uncertainties, and correlation
(covariance) between different observables.  SHF is based on an
effective energy-density functional which is formally motivated by a
low-momentum expansion of a most general effective interaction and its
free parameters are calibrated by a fit to empirical data, mostly from
nuclear ground states.  This mixed origin calls for particularly
careful exploration of the capability, reliability, and limits of the
model. One key task is to estimate the uncertainties in
extrapolations. Related to that is the other task to find out the
dependencies between observables (and model parameters) which helps to
identify, e.g., bottlenecks in the model and to cure them. A great
manifold of strategies to attack these task was outlined in section
\ref{sec:estim}. A large part of them was exemplified in the
subsequent result section. The same pool of fit data was used in all
examples to render the tests consistent with each other and so
maintain comparability. These data were bulk properties of nuclear
ground states in chains of semi-magic nuclei selected to be good
mean-field nuclei (only very small effects from collective
ground-state correlations).  We summarize the basic findings of that
survey in key sentences:
\begin{enumerate}
\item
  There is no single ideal strategy to estimate the uncertainties in
  predictions from a model. A bundle of different methods has to be
  invoked to develop a sound intuition for the limits and capabilities
  of a model. Statistical analysis in connection with $\chi^2$ fits is
  one important part. Harder is an estimate of the systematic
  error. It has to be approached from different sides, formal analysis
  as well as variations of the model, and yet will always leave open
  ends. 
\item Concerning the basic performance, modern SHF functionals have
  reached a high quality of description of ground state properties in the
  whole known nuclear landscape, from light nuclei, as $^{16}$O, up to
  super-heavy elements. For example, energies are reproduced with a
  r.m.s error of about 0.6 MeV.
\item\label{it:othapp}
  Although not explicitly fitted for that, modern SHF functionals can
  be used successfully for many other observables and processes as
  resonance excitations, astro-physical nuclear matter, nuclear
  rotation at high spin, and large-amplitude collective motion
  (low-energy quadrupole vibrations, fission, fusion, heavy-ion
  scattering). The latter example goes, in fact, often beyond a mere
  mean-field treatment.
\item Although small, the residual errors are not statistically
  distributed, but show still unresolved trends as typical
  fluctuations between doubly-magic and mid-shell nuclei and, more
  puzzling, a small but systematic increase of under-binding towards
  heavy and super-heavy nuclei. The latter trend is probably due to
  yet insufficient modeling of density dependence of the SHF
  functional.
\item
  Extrapolation errors from statistical analysis provide a lower
  estimate (yet without contributions from systematic errors) which is
  already extremely useful to distinguish safe and risky
  extrapolations. Not surprisingly, errors increase with increasing
  distance to the fit data. For example, extrapolations to super-heavy
  elements are still found to be reliable while error bars grow huge
  when predicting properties of neutron stars.  On the other hand,
  interpolations in the vicinity of the fit data are safe and yield
  deviations as small as the average errors in the fit pool.
\item
  There is a hierarchy of importance within the terms of the SHF
  functional. From a formal side, importance shrinks with increasing
  order of derivatives involved. From the empirical side, isoscalar
  properties are better determined than isovector ones due to the
  limited extension of known nuclei along isotopic chains. Taking the
  most important terms delivers a minimal model with only 6 free
  parameters which provides already the high quality of ground state
  properties as the final model employing all 11 free parameters of
  the SHF functional. The further terms above the minimal model serve
  to optimize further features as, e.g., giant resonance excitations.
\item\label{it:drift} The impact of the fit data was checked by
  performing fits with successively omitting a group of data.  From
  the four groups of fit data (binding energy $E_B$, charge
  r.m.s. radius $r_\mathrm{rms}$, charge diffraction radius
  $R_\mathrm{diffr}$, charge surface thickness $\sigma_C$), $E_B$ is
  found to be the most decisive one. Fits with $E_B$ alone yield
  already an acceptable model. Adding $r_\mathrm{rms}$ to the pool of
  $E_B$ comes another step closer to the full fit while considering
  $r_\mathrm{rms}$ alone is insufficient. The small drifts of average
  errors and of predicted observables when adding $R_\mathrm{diffr}$
  and $\sigma_C$ shows that each group pulls the fit into a slightly
  different direction. This conflict between observables points to a
  rigidity in modeling the nuclear surface profile which, again, is
  probably related to the modeling of density dependence.
\item\label{it:varyNMP} A particularly instructive analyzing tool is
  the systematic variation of model features, here exemplified by the
  variation of nuclear matter parameters (NMP). It illustrates the
  impact of such a parameter on an observable of interest. This
  allowed to demonstrate the near one-to-one correspondence of NMP and
  giant resonances in $^{208}$Pb, namely incompressibility $K$ with
  the giant monopole resonance, isoscalar effective mass $m^*/m$ with
  the giant quadrupole resonance, symmetry energy $J$ with the dipole
  polarizability $\alpha_D$, and the TRK sum-rule enhancement
  $\kappa_\mathrm{TRK}$ (alias isovector effective mass) with the
  giant dipole resonance.  A much different behaviour is seen for
  fission barrier and lifetime in the super-heavy nucleus $^{266}$Hs.
  These observables gather influences from many different parameters
  and have no prevailing correlation.
\item
  Ab-initio data are not yet regularly included in the calibration of
  the SHF functional. As an example for ab-initio input, the neutron
  gas at very low density was considered. This data point can
  meanwhile be computed very well with the theory of the correlated
  Fermi gas (CFG) and it is exclusively determined by long range
  correlations from zero-sound modes of the gas.  It turns out that
  the standard SHF functional is not able to accommodate this data
  point from CFG. A new density dependent term must be added to allow
  an adjustment which then works acceptably well. The extended model 
  allows also to test the effect of an additional density dependence
  when fitting to the standard pool of data (excluding the CFG point).
  This fit resolves the conflict between observables discussed
  under point \ref{it:drift} thus confirming that modeling of
  density dependence is still an issue.
\item
  Correlation (or covariance) analysis was exemplified for a small
  selection of relevant observables covering the four essential groups
  of response properties, isoscalar static, isoscalar dynamic,
  isovector static, and isovector dynamic. This confirmed very clearly
  the one-to-one correspondence between NMP and giant resonance,
  already worked out under point \ref{it:varyNMP}. Noteworthy is a
  large block of isovector static observables covering, beside $J$ and
  $\alpha_D$, the neutron skin, pure neutron matter, and the slope of
  symmetry energy $L$. All these relations are rather robust 
  under changing conditions of modeling and selection of fit data.
\item
  Spin terms in the SHF functional leave still open questions.  A part
  of them is determined by the requirement of Galilean invariance.
  Other terms can be fixed by the concept of a Skyrme ``force''.  But
  this setting often leads to principle instabilities in the spin
  channel. Particularly dangerous are the spin-gradient terms which
  thus are discarded in all applications. There remains, again, a
  decision based on empirical data. The case is not yet settled and
  requires further research.
\item
  The pairing part of the functional has not been discussed in this
  paper. Similar as the spin terms, it is also not yet fully
  explored. The present form is optimized to reproduce the odd-even
  staggering in isotopic and isotonic chains of semi-magic nuclei and
  serves to produce well defined ground states for open shell nuclei. 
  Pairing is found to have a large impact on low-lying quadrupole
  excitations and fission. This may require still a further fine tuning 
  for such processes.
\item
  As mentioned in point \ref{it:othapp}, the SHF functional is often
  successful even in applications beyond pure mean field.  Such
  methods require the energy overlap between different mean field
  states.  These can be evaluated with the SHF functional as long as
  the different states remain close (Gaussian overlap approximation).
  However, projection onto good particle number or angular momentum
  require also overlaps between remote states and the SHF functional
  in its standard form become inapplicable for this task. New modeling
  based strictly on a force concept is needed, and underway to resolve
  this problem.
\end{enumerate}
Altogether, the SHF functional has proven to deliver an extremely
useful and reliable reproduction of static and dynamic features of
nuclei as far as they are accessible to a mean-field description.  One
could be highly satisfied with what has been achieved. However, the
success shows the capability of SHF and motivates to ask for more.
Such more is also highly desirable in view of the the remarkable
progress in producing exotic nuclei and of the far extrapolations
implied in astro-physical applications (extremely neutron rich
isotopes in the $r$-process, neutron stars, nuclear matter under
supernova conditions). Work on further scrutinizing and improving the
SHF functional along the lines summarized above is in progress in many
research groups around the world.

\bigskip

\noindent
Acknowledgment: 
This work has been supported by a grant from the
German ministry of Science and Technology, grant number 05P12RFFTG,
and from the Deutsche Forschungsgemeinschaft, grant number RE
322-14/1.

\appendix

\section{The Skyrme force}
\label{app:force}

The SHF functional was motivated in section \ref{sec:motivateSHF}
through the density-matrix expansion \cite{Neg72a,Neg75a,Rei94aR}.
This expansion acts on a supposed microscopic interaction, the $T$
matrix from some ab-initio model, and delivers the ``Skyrme force'' 
in the form (tensor terms are ignored here)
\begin{eqnarray}
  \hat{V}_\mathrm{Sk}
  &=&
  t_0(1\!+\!x_0 \hat{ P}_\sigma)\delta(\mathbf{r}_{12})
\nonumber\\
  &&
  \!+
   \frac{t_3}{6}(1\!+\!x_3\hat P_\sigma)
  \rho^\alpha\left(\mathbf{r}_1\right)
  \delta(\mathbf{r}_{12})
\nonumber\\
  &&
  \!+ \frac{t_1}{2}(1\!+\!x_1\hat{P}_\sigma)
  \left(
   \delta(\mathbf{r}_{12})\hat{\boldsymbol k}^2
   +
   {\hat{\boldsymbol{k}}}'^{2}\delta(\mathbf{r}_{12})
  \right)
\nonumber\\
  &&
  \!+ t_2(1\!+\!x_2\hat P_\sigma)\hat{\boldsymbol k}'
  \delta(\mathbf{r}_{12})\hat{\boldsymbol k}
\nonumber\\
  &&
  \!+\I t_4(\hat{\boldsymbol{\sigma}} _1 + \hat{\boldsymbol{\sigma}}_2)
  \cdot \hat{\boldsymbol k}' \times \delta(\mathbf{r}_{12})
  {\mathbf{k}}
  \quad,
\label{eq:SHFforce}
\\
  \mathbf{r}_{12}
  &=&
  \textbf r_1-\textbf r_2
  \quad,
\nonumber
\\
  \hat P_\sigma
  &=&
  \frac{1}{2}(1+\hat{\boldsymbol \sigma}_1\hat{\boldsymbol{\sigma}}_2)
  \quad,\quad
\nonumber\\
  \hat{\boldsymbol k} 
  &\!=\!&
  -\frac{i}{2 }\left(\stackrel{\rightarrow}{\boldsymbol\nabla}_1
           -\stackrel{\rightarrow}{\boldsymbol\nabla}_2\right)
  \,,\,
  \hat{\boldsymbol k}'
  =
  \frac{i}{2 }\left(\stackrel{\leftarrow}{\boldsymbol\nabla}_1
      -\stackrel{\leftarrow}{\boldsymbol\nabla}_2\right)
  \,.
\nonumber
\end{eqnarray}
where $\hat{\boldsymbol k}$ acts to the right and
$\hat{\boldsymbol{k}}'$ to the left.  We put the notion ``force'' in
quotation mark because this object depends on the density which is
produced by the wave function on which this force acts. This is not a
standard two-body interaction, but an effective force designed for the
only purpose to derive an energy functional (for a detailed discussion
of the principle problems see \cite{Erler_2010}).  The evaluation of
the expectation value of $\hat{V}_\mathrm{Sk}$ with a BCS state yields
indeed a functional of the form (\ref{eq:basfunct}). In fact, it
yields more than that. The 10 parameters of the ``force''
(\ref{eq:SHFforce}) are distributed over the 23 parameters of the SHF
functional (when counting each term with a separate free
parameter). This establishes a couple of relations between the
parameters of the SHF functional. They embrace, of course, the
restriction from Galilean invariance already embodied in the time-odd
part (\ref{eq:ESkodd}) and add seven more relations which fix all spin
terms as shown in table \ref{tab:restrict}.

It is interesting to note that the Skyrme force is particularly
restrictive with respect to the spin-orbit terms. The tensor
spin-orbit terms $\propto\mathbb{J}^2$ are fully determined by the
kinetic terms as $C_T^J$ is given by $C_T^\tau$ and $C_T^{\Delta\rho}$
and the $C_1^{\nabla J}=C_0^{\nabla J}$ ties the isovector to the
isoscalar spin-orbit term. The first restriction is neither harmful
nor beneficial as this tensor spin-orbit term can easily be
compensated by the normal spin-orbit term.  The second restriction is
more serious and to some extend questioned by phenomenology
\cite{Rei95d}. Most traditional parametrizations employ this
restriction. Recent fits, particularly those following
\cite{Kluepfel_2009} discard the restriction to maintain independence
of the isovector spin-orbit parameter.

\section{Pool of fit data}
\label{sec:pool}

\begin{table}
\begin{center}
\begin{tabular}{|l|l|}
\hline
$E$: &\rule[4pt]{0pt}{8pt} $^{36-52}$Ca, $^{68}$Ni, $^{100,126-134}$Sn,  $^{204-214}$Pb,
\\[-1pt]&
       $^{34}$Si, $^{36}$S, $^{38}$Ar, $^{50}$Ti,
       $^{86}$Kr, $^{88}$Sr, $^{90}$Zr,
\\
    &
        $^{92}$Mo, $^{94}$Ru, $^{96}$Pd, $^{98}$Cd, 
       $^{134}$Te, $^{136}$Xe,
\\&
      $^{138}$Ba, $^{140}$Ce, $^{142}$Nd, $^{144}$Sm, $^{146}$Gd,
\\
    & $^{148}$Dy,
       $^{150}$Er, $^{152}$Yb, 
      $^{206}$Hg, $^{210}$Po,
\\&
      $^{212}$Rn, $^{214}$Ra,  $^{216}$Th, $^{218}$U
\\
$R_\mathrm{diff}$:
     & $^{16}$O, $^{40-44,48}$Ca, $^{58-64}$Ni, $^{118-124}$Sn,
\\&
        $^{204-208}$Pb, $^{50}$Ti, ${52}$Cr, $^{54}$Fe, $^{86}$Kr,
\\
  &
         $^{88}$Sr, $^{90}$Zr, $^{92}$Mo, $^{138}$Ba, $^{142}$Nd, 
\\
$\sigma$: & $^{16}$O, $^{40-44,48}$Ca, $^{60-64}$Ni,
$^{118,122-124}$Sn,
\\&     $^{204-208}$Pb,  $^{50}$Ti, $^{86}$Kr,  $^{88}$Sr, $^{90}$Zr, 
\\
   &  $^{92}$Mo, $^{138}$Ba, $^{142}$Nd, 
\\
$r_\mathrm{rms}$: 
    &  $^{16}$O, $^{40-48}$Ca, $^{108,118-124}$Sn,  $^{200-214}$Pb,
\\&
       $^{36}S$, $^{38}Ar$, $^{50}$Ti, ${52}$Cr, $^{54}$Fe,
       $^{86}$Kr, $^{88}$Sr,
\\
    &   $^{90}$Zr, $^{92}$Mo,
       $^{136}$Xe, $^{138}$Ba, $^{140}$Ce,
\\&
       $^{142}$Nd, 
       $^{144}$Sm, $^{146}$Gd, $^{148}$Dy, $^{150}$Er,
\\
    &
      $^{206}$Hg, $^{210}$Po, $^{212}$Rn, $^{214}$Ra, 
\\
$\Delta^{(3)}$ &
      isotopes Z=50,82 and isotones N=50,82
\\[2pt]
l*s:  & $^{16}$O($1p_n,1p_p$) $^{132}$Sn($2p_p,2d_n$), 
\\&
     $^{208}$Pb($2d_p,1f_n,3p_n$) 
\\[2pt]
$\delta r^2$: & isotope shift $r^2(^{214}\mathrm{Pb})-r^2(^{208}\mathrm{Pb})$
\\[2pt]
NMP:  & $K$, $m^*/m$, $\kappa_\mathrm{TRK}$, $a_\mathrm{sym}$ 
   \hfill(only SV-bas)
\\
\hline
\end{tabular}
\end{center}
\caption{\label{tab:fits2} Compilation of phenomenological input from
  \cite{Kluepfel_2009} used for the fits of the various
  parametrizations presented in this paper.  Abbreviations mean: $E$
  = binding energy, $r$ = charge r.m.s. radius, $R$ = charge
  diffraction radius, $\sigma$ = charge surface thickness, $l*s$ =
  spin-orbit splitting of selected single-particle states, $\delta
  r^2$ = isotopic shift of charge radii, and NMP = nuclear matter
  properties (in symmetric matter).  }
\end{table}
Table \ref{tab:fits2} summarizes the pool of data as defined in
\cite{Kluepfel_2009}. These data were used to determine SV-min, SV-bas
and the series of forces with systematically varied NMP (used here,
e.g., in section \ref{sec:varyNMP}). They are used for the new fits in
this paper, see e.g. sections \ref{sec:groups} or \ref{sec:abinit}.
The detailed values for the listed observables are found in the
original paper \cite{Kluepfel_2009}.

\bigskip

\section*{References}

\bibliographystyle{elsarticle-num}

\bibliography{error-review}

\end{document}